\documentstyle[epsfig,natbib,natbibmnfix]{mn}

\newcommand{\be}{\begin{equation}}
\newcommand{\ee}{\end{equation}}
\newcommand{\bea}{\begin{eqnarray}}
\newcommand{\eea}{\end{eqnarray}}
\newcommand{\bc}{\begin{center}}
\newcommand{\ec}{\end{center}}

\def\gsim{ \lower .75ex \hbox{$\sim$} \llap{\raise .27ex \hbox{$>$}} }
\def\lsim{ \lower .75ex \hbox{$\sim$} \llap{\raise .27ex \hbox{$<$}} }

\renewcommand{\thefootnote}{\fnsymbol{footnote}}

\title{A unified model for AGN feedback in cosmological simulations of
  structure formation}

\author[Sijacki et al.]
       {\parbox{18cm}{Debora~Sijacki$^{1}$\footnotemark[1],
       Volker~Springel$^{1}$, Tiziana Di Matteo$^{2}$ and Lars
       Hernquist$^{3}$}\vspace{0.3cm}\\ 
       $^1$Max-Planck-Institut f\"{u}r Astrophysik,
       Karl-Schwarzschild-Stra\ss{}e 1, 85740 Garching bei
       M\"{u}nchen, Germany\\ $^2$Dept. of Physics, Carnegie-Mellon
       University, 5000 Forbes Ave., Pittsburgh, PA 15213, USA \\
       $^3$Harvard-Smithsonian Center for Astrophysics, 60 Garden
       Street, Cambridge, MA 02138, USA}

\begin{document}

\maketitle
\begin{abstract}
  We discuss a numerical model for
  black hole growth and its associated feedback processes that for the
  first time allows cosmological simulations of structure formation to
  self-consistently follow the build up of the cosmic population of
  galaxies and active galactic nuclei. Our model assumes that seed
  black holes are  present at early cosmic epochs at the centres of
  forming halos. We then track their growth from gas accretion and
  mergers with other black holes in the course of cosmic time. For
  black holes that are active, we distinguish between two distinct
  modes of feedback, depending on the black hole accretion rate
  itself. Black holes that accrete at high rates are assumed to be in
  a `quasar regime', where we model their feedback by thermally
  coupling a small fraction of their bolometric luminosity to the
  surrounding gas. The quasar activity requires high densities of
  relatively cold gas around the black hole, as it is achieved through
  large-scale inflows triggered by galaxy mergers.  For black holes
  with low accretion rates, we conjecture that most of their feedback
  occurs in mechanical form, where AGN-driven bubbles are injected
  into a gaseous environment. This regime of activity, which is
  subdominant in terms of total black hole mass growth, can be
  identified with radio galaxies in clusters of galaxies, and can
  suppress cluster cooling flows without the requirement of a
  triggering by mergers. Using our new model, we carry out TreeSPH
  cosmological simulations on the scales of individual galaxies to
  those of massive galaxy clusters, both for isolated systems and for
  cosmological boxes. We demonstrate that our model produces results
  for the black hole and stellar mass densities in broad agreement
  with observational constraints. We find that the black holes
  significantly influence the evolution of their host galaxies,
  changing their star formation history, their amount of cold gas, and
  their colours. Also, the properties of intracluster gas are affected
  strongly by the presence of massive black holes in the cores of
  galaxy clusters, leading to shallower metallicity and entropy
  profiles, and to a suppression of strong cooling flows. Our results
  support the notion that active galactic nuclei are a key ingredient
  in cosmological structure formation. They lead to a self-regulated
  growth of black holes and bring the simulated properties of their
  host galaxies into much better agreement with observations.
\end{abstract}


\section{Introduction}

\renewcommand{\thefootnote}{\fnsymbol{footnote}}
\footnotetext[1]{E-mail: deboras@mpa-garching.mpg.de }

It is now widely believed that most if not all galaxies with a spheroidal
component harbour a supermassive black hole (BH) in their centres.
Interestingly, the masses of these central BHs are found to be tightly linked
with the stellar properties of their host galaxies, as expressed, e.g., in the
correlation of bulge velocity dispersion with BH mass
\citep{Ferrarese2000, Gebhardt2000, Tremaine2002}, or in the relation of bulge
stellar mass \citep{Kormendy1995, Magorrian1998, Marconi2003, Haring2004} with BH
mass. The existence of these relationships indicates that the formation and
evolution of galaxies is fundamentally influenced by the presence of BHs, and
vice versa. We thus also expect that the environment and the cosmological
evolution of galaxies will affect the way BHs grow.

In fact, there is a plethora of observational and theoretical studies that
suggest that several different channels for interaction of BHs with their
surroundings exist. At high redshift, mergers of gas-rich galaxies occur
frequently and funnel copious amounts of cold gas towards the central regions
of galaxies, such that the embedded BHs can reach high gas accretion rates.
The radiation energy associated with the accretion can support the enormous
luminosities of powerful quasars. Theoretically it has been hypothesized
\citep{Silk1998, Fabian1999, King2003} that quasars produce high velocity
winds, which are expected to affect the properties of the host galaxy. The
presence of quasar induced outflows has been observationally confirmed in a
number of cases \citep[e.g.][]{Chartas2003, Crenshaw2003, Pounds2003}, and has
first been demonstrated in simulations of merging galaxy pairs by
\cite{DiMatteo2005}. Several numerical studies dealing with BH microphysics
\citep[e.g.][]{Proga2003, McKinney2006} also predict existence of
quasar outflows. 
Moreover, it appears that tidally disrupted galaxies are preferentially
associated with AGN activity \citep[for a review see][]{Barnes1992}.  Quasar
activity hence appears to be directly linked to mergers of galaxies, and
should represent the dominant mode of mass growth in the BH population.

Indeed, using semi-analytic models of galaxy formation,
\citet{Kauffmann2000} have demonstrated that BH growth associated with
mergers in CDM models can reproduce many properties of the observed
quasar population as well as the inferred BH mass density today
\citep[see also][]{Volonteri2003}. Based on the detailed
hydrodynamical simulations of BH growth in galaxy mergers
\citep{DiMatteo2005, Springel2005a, Robertson2006a, Robertson2006b,
Cox2006b, Cox2006a} that have recently become available,
\citet{Hopkins2005d,Hopkins2006, Hopkinsmerger2006} have proposed a
comprehensive picture of a unified, merger-driven model for the origin
of quasars and their relation to spheroid formation, which also
implies the existence of a `black hole fundamental plane'
\citep{Hopkins2007}.  Also, rapid merging of the gas-rich progenitor
systems of rare, massive galaxy clusters has been shown \citep{Li2006}
to be a viable formation path for supermassive BHs that are as massive
at $z\sim 6$ as those seen in luminous high-redshift SDSS quasars
\citep{Fan2001}.

However, there also appears to exist another channel of BH interaction with
host galaxies, which is neither related to powerful quasar activity nor
associated with galaxy mergers. Evidence for this interaction can be seen in
a number of local elliptical galaxies and central cluster galaxies, which
contain X-ray cavities filled with relativistic plasma \citep{Birzan2004,
  McNamara2005, Forman2006, Fabian2006} while harbouring seemingly ``dormant''
BHs. These X--ray depressions, often referred to as `bubbles', are thought to
be inflated by relativistic jets launched from the central BH. Even though
the radiative output from the central BH is not significant, the associated
mechanical luminosity can be very important in these systems.
There has been considerable effort from the theoretical point of view
\citep[e.g.][]{Binney1995, Churazov2001, Churazov2002, Quilis2001,
  Ruszkowski2002, DallaVecchia2004, Kawata2005, Sijacki2006a,
  Thacker2006, Okamoto2007} to understand the relevance 
of this mode of AGN feedback, which is widely considered to be leading
candidate for resolving the cooling flow problem in groups and clusters of
galaxies.

Given the complex physics of AGN-galaxy interactions, is it possible
to construct a simple unified model that accounts for the different
modes of BH feedback in a cosmological framework? First attempts in
this direction have already been made \citep{Churazov2005, Croton2006,
Merloni2006}, motivated by the observational findings of X-ray
binaries \citep{Fender1999, Gallo2003}. In particular, it has been
shown that X--ray binaries switch between two states: in the so-called
``low/hard'' state, a steady radio jet is present and the hard X--ray
spectrum is observed, while in the ``high/soft'' state, the jet
vanishes and the X--ray spectrum shows a soft, thermal component. The
transition between these two states is regulated by the accretion rate
onto the BH itself, where the threshold value is of the order of
$10^{-2}-10^{-1} \dot M_{\rm Edd}$. The ``high/soft'' state can be
explained by the standard, optically thick and geometrically thin
accretion disc \citep{Shakura1973}, with BH accretion occurring at
high rates and in a radiatively efficient mode. The ``low/hard'' state
on the other hand corresponds to optically thin, geometrically thick,
and radiatively inefficient accretion, as described by the theoretical
ADAF and ADIOS solutions \citep{Narayan1994, Blandford1999}.

The above suggests a rather simple, yet attractive scenario for distinguishing
between different modes of BH feedback in models for the cosmological
evolution of active galactic nuclei (AGN): We shall assume that for high
accretion rates, a `quasar-like' feedback occurs, while for states of low
accretion, mechanical bubble feedback applies.  It is clear that the
simplicity of this model will not allow it to explain all kinds of AGN
feedback phenomena, e.g.~powerful radio galaxies that accrete at very high
rates, as found in some proto-cluster environments, are not well represented
in this simple scheme.  Moreover, even though the physics of X-ray binaries is
expected to be quite similar to the one of AGN \citep{Heinz2005}, an analogous
transition between ``high/soft'' and ``low/hard'' states in AGN is
observationally still not well established, although there are some
encouraging observational findings \citep{Maccarone2003, Kording2006} in this
direction.

With these caveats in mind, we explore in this study a new `two-mode'
AGN feedback scenario in fully self-consistent cosmological
simulations of structure formation. This complements and extends our
study of cosmological simulations with quasar feedback in 
\citet{DiMatteo2007}, where we did not account for the ``radio'' mode.
Within our numerical model we represent BHs with collisionless
``sink'' particles and we adopt subresolution methods to compute their
gas accretion rate, estimated from a Bondi prescription. Also, we
allow BHs to grow via mergers with other BHs that happen to be in
their vicinity and that have sufficiently low relative speeds. We
assume that BH feedback is composed of two modes, as motived above,
and we track their growth and feedback with cosmic time. We consider a
vast range of objects harbouring BHs, from galaxies with $\sim 10^8
{\rm M}_\odot$ stellar mass to massive galaxy clusters with total mass
$\sim 10^{15}{\rm M}_\odot$. Even though our BH growth model is rather
simple and crude due to the inevitable numerical limitations, it
represents the first attempt to include AGN feedback effects in
cosmological simulations of structure formation. As we shall see, this
approach produces significant improvements in the properties of
simulated galaxies.

The outline of this paper is as follows. In Section~\ref{Methodology}, we
describe our numerical method for incorporating the BH growth and feedback in
cosmological hydrodynamical simulations. In Section~\ref{Isolated}, we test
our model in simulations of isolated galaxy clusters, and explore the
numerical parameter space. We then discuss cosmological simulations of galaxy
cluster formation and evolution, subject to AGN feedback, in
Section~\ref{Cosmological}, while in Section~\ref{Galform} we consider
cosmological simulations of homogeneously sampled periodic boxes. Finally, we
summarize and discuss our results in Section~\ref{Discussion}.

\section{Methodology} \label{Methodology}

We use a novel version of the parallel TreeSPH-code {\small GADGET-2}
\citep{Gadget2, Springel2001} in this study, which employs an
entropy-conserving formulation of smoothed particles hydrodynamics
\citep{SH2002}. Besides following gravitational and non-radiative
hydrodynamical processes, the code includes a treatment of radiative cooling
for a primordial mixture of hydrogen and helium, and heating by a spatially
uniform, time-dependent UV background \citep[as in][]{Katz1996}.  Star
formation and associated supernovae feedback processes are calculated in terms
of a subresolution multiphase model for the ISM \citep{S&H2003}.  We use a
simple prescription for metal enrichment, and optionally incorporate galactic
winds powered by supernovae, as implemented by \citet{S&H2003}.  Furthermore,
we follow the growth and feedback of BHs based on a new model
that combines the prescriptions outlined in \citet{Springel2005b} and
\citet{Sijacki2006a}.

In the following, we briefly summarize the main features of the BH model
relevant for this study \citep[see][for further details]{Springel2005b},
focusing on extensions that permit us to follow BH growth and feedback in a
cosmological simulation of structure formation \citep[see
also][]{DiMatteo2007}. We then discuss a new BH feedback prescription at low
accretion rates, based on a `bubble' heating scenario for representing AGN
heating in the radiatively inefficient regime of accretion.

\subsection{Black hole formation and growth}

The BHs in the code are represented by collisionless sink particles, which may
accrete gas from their surroundings based on a prescribed estimate for the
accretion rate. Two BH particles are also allowed to merge if they fall
within their local SPH smoothing lengths and if their relative velocities are
smaller than the local sound speed.

In our cosmological simulations of structure formation, we assume that low-mass
seed BHs are produced sufficiently frequently such that any halo above
a certain threshold-mass contains one such BH at its centre. Whether these
seed BHs originate in exploding pop-III stars, in the collapse of star
clusters, or are of primordial origin is not important for our analysis, but
there needs to be a process that produces initial seed BHs which can then grow
to the masses of supermassive BHs by gas accretion in the course of our
simulations.  For definiteness, in most of our simulations we adopt a seed BH
mass of $10^{5}\,h^{-1} {\rm M}_\odot$ and endow all halos with a mass larger
than $5 \times 10^{10}\,h^{-1} {\rm M}_\odot$ with a seed if they do not
contain any BH already. We identify halos without BHs on the
fly during a simulation by frequently calling a fast, parallel
friends-of-friends algorithm that is built into our simulation code.

State-of-the-art cosmological simulations of structure formation reach mass
resolutions that are at best of order of our initial BH seed mass, while the
resolution typically reached is still considerably coarser than that. This
would mean that the initial growth of the BH mass could be significantly
affected by numerical discreteness effects if the sink particle can only
swallow full gas particles, as is the case in our scheme.  In order to avoid
that the BH growth and the accretion rate estimate is strongly affected by
this numerical discreteness limitation, we treat the BH mass $M_{\rm BH}$ as
an internal degree of freedom of the sink particle. In the beginning, $M_{\rm
BH}$ may differ from the dynamical mass $M_{\rm dyn}$ of the sink particle
itself. The variable $M_{\rm BH}$ is integrated smoothly in time based on the
estimated accretion rate onto the BH, while $M_{\rm dyn}$ increases in
discrete steps when the sink particle swallows a neighbouring gas
particle. The latter process is modelled stochastically such that $M_{\rm
dyn}$ tracks $M_{\rm BH}$ in the mean, with small oscillations around it.
With this prescription for the BH mass, we can follow the early growth of BHs
accurately in a sub-resolution fashion even when their mass may be smaller
than the mass resolution of the simulation, while at late times (or in the
limit of very good mass resolution) the two mass variables coincide. Of
course, in the event of a merger of two BH sink particles, both $M_{\rm BH}$
and $M_{\rm dyn}$ are added together.

Following \cite{DiMatteo2005}, we estimate the accretion rate onto a BH
particle according to the Bondi-Hoyle-Lyttleton formula \citep{Hoyle1939,
  Bondi1944, Bondi1952} \be \dot M_{\rm BH}\,=\, \frac{4\,\pi \,\alpha\, G^2
  M_{\rm BH}^2 \,\rho}{\big(c_s^2 + v^2\big)^{3/2}}\,, \ee where $\alpha$ is a
dimensionless parameter, $\rho$ is the density, $c_s$ the sound speed of the
gas, and $v$ is the velocity of the BH relative to the gas.  We account for
the possibility that the BH accretion has an upper limit given by the
Eddington rate \be \dot M_{\rm Edd} \,=\, \frac{4\,\pi\,G\,M_{\rm BH}\,m_{\rm
    p}}{\epsilon_{\rm r}\,\sigma_{\rm T}\,c}\,, \ee where $m_p$ is the proton
mass, $\sigma_{\rm T}$ is the Thompson cross-section and $\epsilon_{\rm r}$ is
the radiative efficiency, that we assume to be 0.1, which is the mean value
for the radiatively efficient \cite{Shakura1973} accretion onto a
Schwarzschild BH. In some of our numerical models we specifically explore the
imprints of the imposed Eddington limit on the BH properties.

\subsection{Black hole feedback}\label{Methodology_feedback}

In the model of \citet{Springel2005b}, it is assumed that a fixed fraction of
the BH bolometric luminosity couples thermally to
the local gas, independent of the accretion rate
and environment.  In our model we extend this BH feedback prescription in
order to obtain a physically refined model for AGN heating both at high and at
low accretion rates. We are motivated by the growing theoretical and
observational evidence \citep{Fender1999, Gallo2003, Churazov2005, Heinz2005,
  Croton2006} that AGN feedback is composed of two modes, analogous to states
of X-ray binaries. Specifically, at high redshifts and for high accretion
rates we assume that the bulk of AGN heating is originating in the luminous
quasar activity. In this regime, BHs accrete efficiently and power luminous
quasars where only a very small fraction of their bolometric luminosity
couples thermally to the gas. On the other hand, at lower redshifts and for
BHs accreting at much lower rates than their Eddington limits, AGN heating
proceeds via radiatively inefficient feedback in a mostly mechanical form.

To model the transition between these two accretion and feedback modes, we
introduce a threshold $\chi_{\rm radio} = \dot M_{\rm BH} / \dot M_{\rm Edd} $
for the BH accretion rate (BHAR) in Eddington units, above which `quasar heating' is
operating, and below which we deal with ``radio'' mode feedback, which we model
by injecting bubbles into the host galaxy/cluster. Typically, we
adopt a value of $10^{-2}$ for $\chi_{\rm radio}$, and we impose no other
criterion to distinguish between the two modes of feedback. 

For BHAR that are higher than $\chi_{\rm radio}$, we
parameterize the feedback as in \citet{Springel2005b}, i.e.~a small fraction
of the bolometric luminosity is coupled thermally and isotropically to the
surrounding gas particles, with an amount given by \be \dot E_{\rm feed} \,=\,
\epsilon_{\rm f}\, L_{\rm r}\, = \, \epsilon_{\rm f}\, \epsilon_{\rm r}\, \dot
M_{\rm BH} \, c^2\,.  \ee Here $\epsilon_{\rm f}$ gives the efficiency of
thermal coupling. The value of $5\%$ adopted here brings the simulated $M_{\rm
  BH}-\sigma_*$ relation obtained for remnants of isolated galaxy mergers in
agreement with observations, as shown by \cite{DiMatteo2005}.
 
Below $\chi_{\rm radio}$ we assume that the accretion periodically produces an
AGN jet which inflates hot bubbles in the surrounding gas.  We clearly lack
the numerical resolution for self-consistent {\it ab initio} simulations of the
detailed physics of BH accretion and the involved relativistic MHD
that is responsible for the actual jet creation. However, we can nevertheless
try to represent the relevant heating mechanism by directly injecting the
energy contained in the AGN-inflated bubbles into the ICM.  To this end we
need to link the bubble properties, like radius, duty cycle and energy
content, directly with the BH physics.

\begin{table}
\begin{tabular}{cccccccc}
\hline \hline $M_{200}$ & $R_{200}$ & $c$ & $N_{\rm gas}$ & $m_{\rm
gas}$ & $\epsilon$ \\

{\scriptsize [$\,h^{-1}{\rm M}_\odot\,$]} & {\scriptsize
[$\,h^{-1}{\rm kpc}$]} & & & {\scriptsize [$\,h^{-1}{\rm M}_\odot\,$]}
& {\scriptsize [$\,h^{-1}{\rm kpc}\,$]}\\ \hline
$ 10^{13}$ &  444 &  8.0  & $3 \times 10^{5}$ & $4.0 \times 10^6$ &
2.0   \\ 
$ 10^{14}$ &  957 &  6.5  & $3 \times 10^{5}$ & $4.0 \times
10^7$ &  5.0   \\ 
$ 10^{15}$ & 2063 &  5.0  & $3 \times 10^{5}$ & $4.0
\times 10^8$ & 10.0   \\ 
$ 10^{15}$ & 2063 &  5.0  & $1 \times 10^{6}$
& $1.2 \times 10^8$ &  6.5 \\ \hline \hline
\end{tabular}
\caption{Numerical parameters of our simulations of isolated galaxy
  clusters. The virial masses and radii of the halos, evaluated at
  $200\, \rho_{\rm crit}$, are given in the first two columns. The
  assumed values for the concentration parameter are in the third
  column, while the number and the mass of the gas particles is shown
  in the fourth and the fifth column, respectively. The mass of the
  star particles is half that of the gas particles, because we set the
  number of generations of star particles that a gas particle may
  produce to two. Note that there are no parameters for the dark
  matter particles in these run, because we here modelled the dark
  halo with a static NFW potential.  Finally, in the last column, the
  gravitational softening length $\epsilon$ for the gas and star
  particles is given.
\label{Table_iso}}
\end{table}

Up to now there is no compelling theory that can satisfactorily explain AGN
jet formation, bubble inflation by the jet, and the duty cycle of jet
activity. On the other hand, a growing body of observational evidence
\citep[e.g.][]{Birzan2004, McNamara2005, Forman2006, Dunn2006, Fabian2006}
shows that AGN-driven bubbles are present in many systems, at different
redshifts and over a range of masses, which constrains the duty cycle to be of
the order of $10^7-10^8$ years.  However, it is still not clear whether and
how AGN activity can be influenced by properties of the host galaxy like its
mass, dynamical state, or central gas cooling rate.  In light of this
uncertainty we propose a simple model for radio feedback where we assume that
an AGN-driven bubble will be created if a BH has increased its mass by a
certain fraction $\delta_{\rm BH} \equiv \delta M_{\rm BH}/M_{\rm BH}$.  Note
that with this choice we do not constrain the possible duty cycle of the jet
itself. We only conjecture that whenever the BH increases its mass by $\delta
M_{\rm BH}$, the thermodynamical state of the surrounding gas will be affected
by the BH activity, and that this can be represented in the form of the
bubbles.

Thus, we relate the energy content of a bubble to the BH properties as
\be 
E_{\rm bub} \,=\, \epsilon_{\rm m}\, \epsilon_{\rm r} \, c^2
 \,\delta M_{\rm BH}\,, 
\ee 
where $\epsilon_{\rm m}$ is the efficiency
 of mechanical heating by the bubbles. Moreover, we link the bubble
 radius both to $\delta M_{\rm BH}$ and to  the density of the
 surrounding ICM, in the following way 
\be R_{\rm bub} = R_{{\rm
 bub},0} \bigg(\frac{E_{\rm bub}/E_{{\rm bub},0}}{\rho_{\rm
 ICM}/\rho_{{\rm ICM},0}} \bigg)^{1/5}\,,
\label{Rbub_eq}
\ee 
where $R_{{\rm bub},0}$, $E_{{\rm bub},0}$, and $\rho_{{\rm
ICM},0}$ are normalization constants for the bubble radius, energy
content and ambient density, respectively. The scaling of the bubble
radius is motivated by the solutions for the radio cocoon expansion in
a spherically symmetric case \citep{Scheuer1974, Begelman1989,
Heinz1998}. With this parameterization for $R_{\rm bub}$, we mimic a
scenario in which a more powerful jet will inflate bigger radio lobes,
and where a higher ICM density will confine the size of the buoyant
bubbles more. Finally, the spatial injection of the bubbles is random
within a sphere with radius twice the bubble radius and centred on
the BH particle in consideration.

\begin{figure*}
\centerline{ \vbox{ \hbox{
\psfig{file=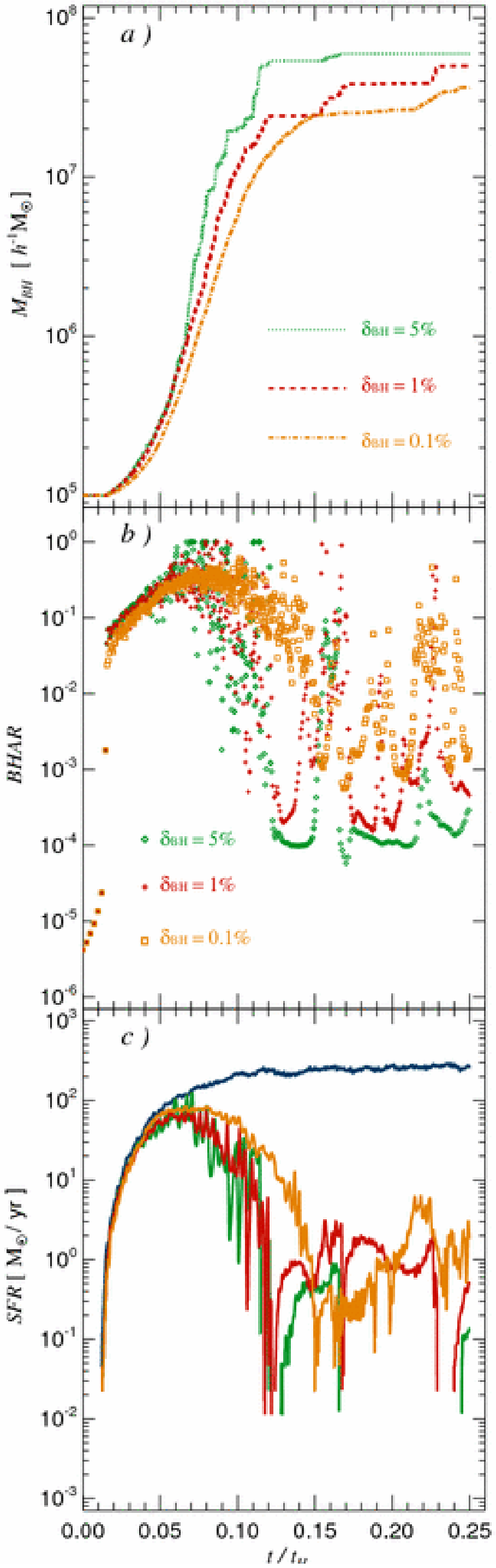,width=7.5truecm,height=21.5truecm}
\hspace{0.2truecm}
\psfig{file=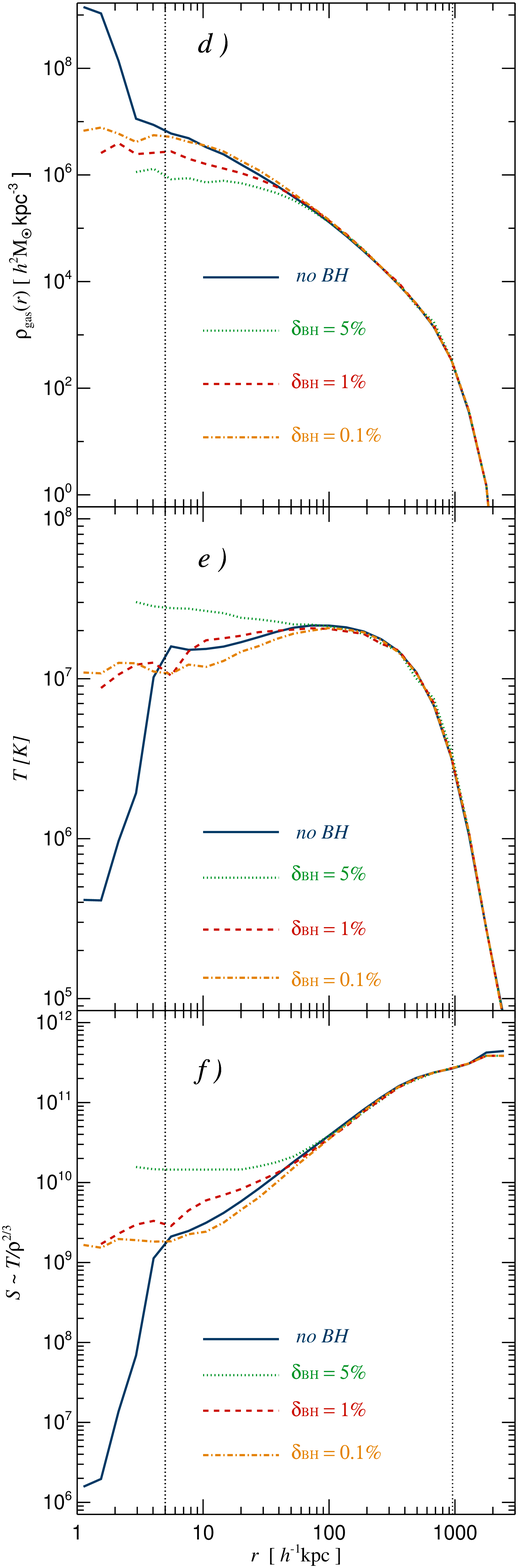,width=7.5truecm,height=21.5truecm} }}}
\caption{Panels $(a)$ and $(b)$ show the time evolution of the BH mass
and the accretion rate in Eddington units for a BH that has been
introduced in the centre of an isolated cluster of mass
$10^{14}\,h^{-1} {\rm M}_\odot$. The different curves and symbols give
results for increasing $\delta_{\rm BH}$, as labelled in
the panels. Panel $(c)$ illustrates the SFR as a function of time for
the same set of simulations (same colour-coding as panel
$(d)$). Finally, panels $(d)$, $(e)$ and $(f)$ give the gas density,
mass-weighted temperature and entropy radial profiles for this cluster
at $t=0.2\,t_{\rm H}$ in runs without AGN feedback (blue continuous
lines) and with AGN heating, where the different curves are for the
increasing value of $\delta_{\rm BH}$, as indicated on
the panels. The vertical dotted lines mark the gravitational softening
length and the virial radius, respectively.}
\label{profiles_1d14cluster}
\end{figure*}

\section{Self-regulated bubble feedback in isolated halo simulations} \label{Isolated}

We have first carried out a number of simulations of isolated galaxy clusters
in order to test our BH feedback model and to explore its parameter space.
These simulations consist of a static NFW dark matter halo
\citep{Navarro1996,Navarro1997} with a gas component initially in hydrostatic
equilibrium. The initial gas density profile has a similar form as the dark
matter, but with a slightly softened core, as explained in more detail in
\cite{Sijacki2006a}. We construct initial conditions for halos with a range of
masses (see Table~\ref{Table_iso} for the details of the numerical setup) and
evolve them non-radiatively for a time $0.25\,t_{\rm Hubble}$ to damp out
possible initial transients. Then, we ``switch on'' radiative gas cooling and
star formation, and also introduce a seed BH particle in the halo centre. In
the following subsections, we describe the results for the subsequent growth
of the BHs.

\subsection{Exploring the parameter space}

In this section, we analyse the sensitivity of our model and of the
resulting BH and host halo properties with respect to the parameter
choices we have adopted. We begin by considering an isolated halo of
mass $10^{14}\,h^{-1} {\rm M}_\odot$, comparing simulations with and
without BHs. For these numerical tests, we only consider AGN feedback
in the form of bubbles, while in all full cosmological simulations we
will include both modes of AGN feedback introduced in
Section~\ref{Methodology_feedback}.

First, we vary the threshold for bubble triggering in terms of accreted mass,
$\delta_{\rm BH}$, from $0.1\%$ to $5\%$, which will affect both
the number of bubble events and their mean energy. The other two free
parameters in the BH feedback model, $\epsilon_{\rm m}$ and the normalization
value for $R_{\rm bub}$, are kept fixed in this first series of simulations in
order to facilitate the comparison. For definiteness, for most of the isolated
halo simulations we select $\epsilon_{\rm m} = 1$, $R_{{\rm bub},0} = 30
\,h^{-1}\rm{kpc}$, $E_{{\rm bub},0} = 5 \times 10^{60}$erg, and $\rho_{{\rm
    ICM},0} = 10^6 \,h^2 {\rm M}_\odot {\rm kpc}^{-3}$. Note that our choice
for these parameters will be slightly different in cosmological simulations,
as we will discuss in Section~\ref{Cosmological}. In most of our numerical
experiments we start with a small BH seed, typically equal to $10^{5}\,h^{-1}
{\rm M}_\odot$, to establish if the BH growth is self-regulated and whether it
leads to a realistic BH mass when a quasi-stationary state is reached.  In a
number of test simulations we also explore a scenario in which a massive BH is
already present at the very beginning.

In panels $(a)$ and $(b)$ of Figure~\ref{profiles_1d14cluster} we show the BH
mass and the accretion rate expressed in Eddington units as a function of
time, for three different values of $\delta_{\rm BH}=0.1\%$, $1\%$ and
$5\%$. It can be seen that the BH is initially growing rapidly starting from
the seed mass of $10^{5}\,h^{-1} {\rm M}_\odot$, and the accretion rate onto
the BH is quite high, reaching the Eddington level at $t= 0.06\,t_{\rm H}$.
However, after this initial phase of rapid growth, AGN feedback starts to
reduce the further BH growth, because at this point enough heating is supplied
to the surrounding medium, preventing it to quickly cool and sink towards the
most inner regions. Thus, the feedback provided by AGN-driven bubbles reduces
the supply of gas available for accretion by the central BH. Consequently,
$\dot M_{\rm BH}$ drops and the mass of the BH shows no significant growth
with time for a while. However, after a certain amount of time which depends
on the heating efficiency, the central cluster gas starts to cool again,
causing an increase of the BHAR and the triggering of another bubble episode.
This mechanism establishes a self-regulated cycle, in which a balance between
the gas cooling rate, the bubble heating rate and the residual BHAR is
achieved. In panel $(b)$ of Figure~\ref{profiles_1d14cluster}, this cycle of
AGN activity can be clearly seen. Here the peaks in BHAR correspond to the
bubble injection events.  Note that the higher values of $\delta_{\rm BH}$
lead to fewer and more powerful AGN outbursts that cause the accretion rate to
drop to very low values of order of $10^{-4} \dot M_{\rm Edd}$.

\begin{figure*}
\centerline{ \vbox{ \hbox{
\psfig{file=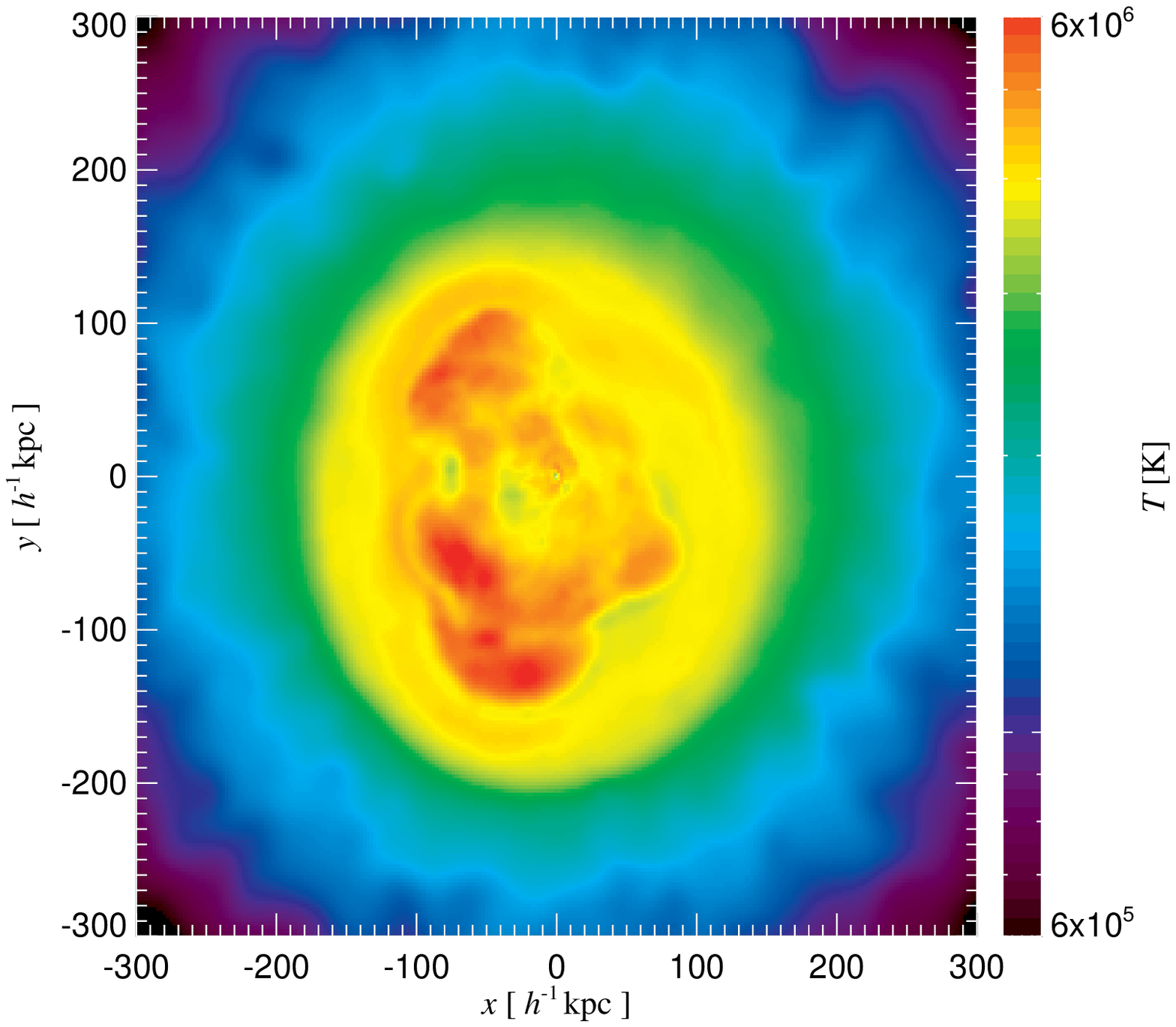,width=6.8truecm,height=6.truecm}
\hspace{-0.6truecm}
\psfig{file=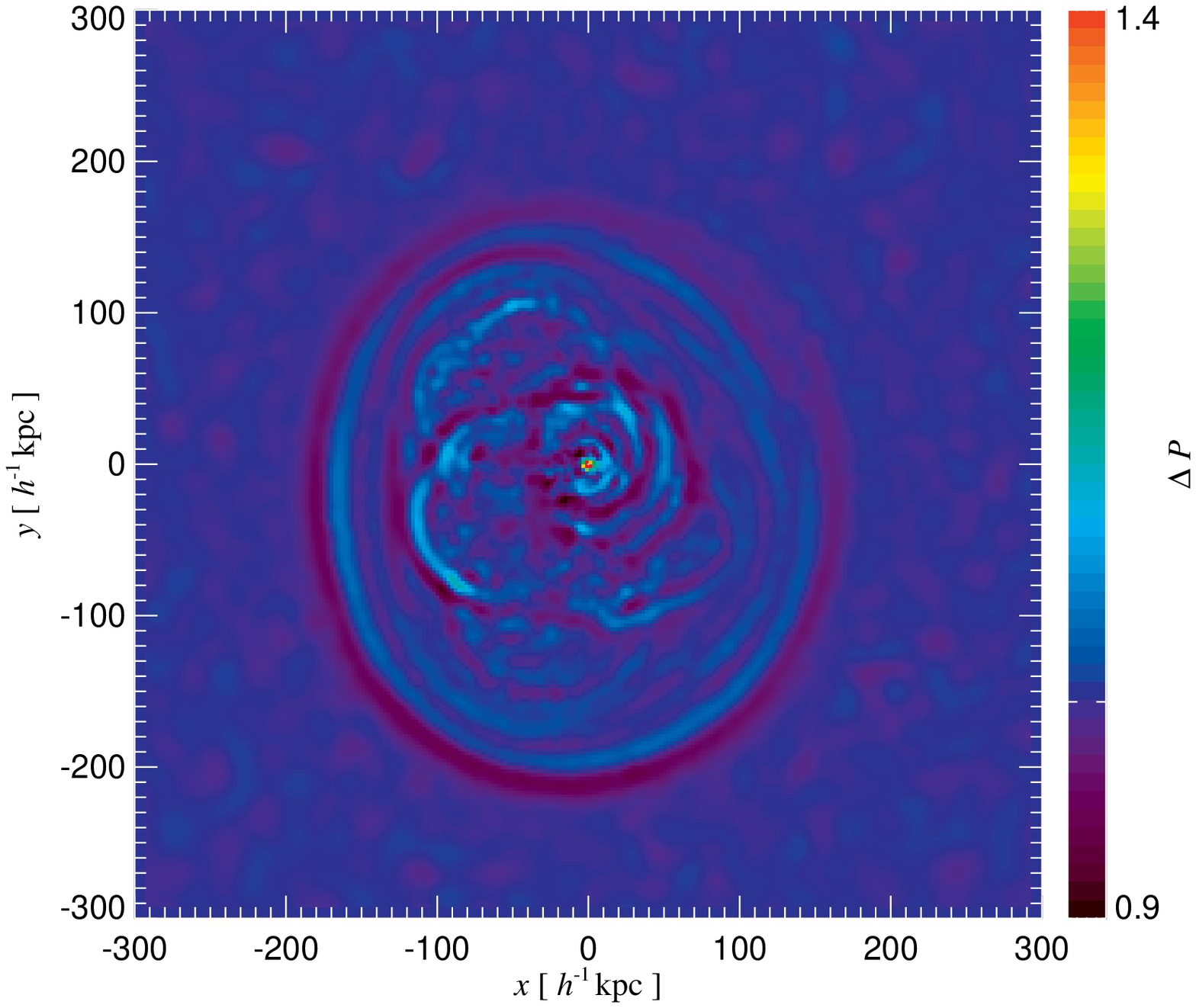,width=6.22truecm,height=6truecm}
\hspace{-0.4truecm}
\psfig{file=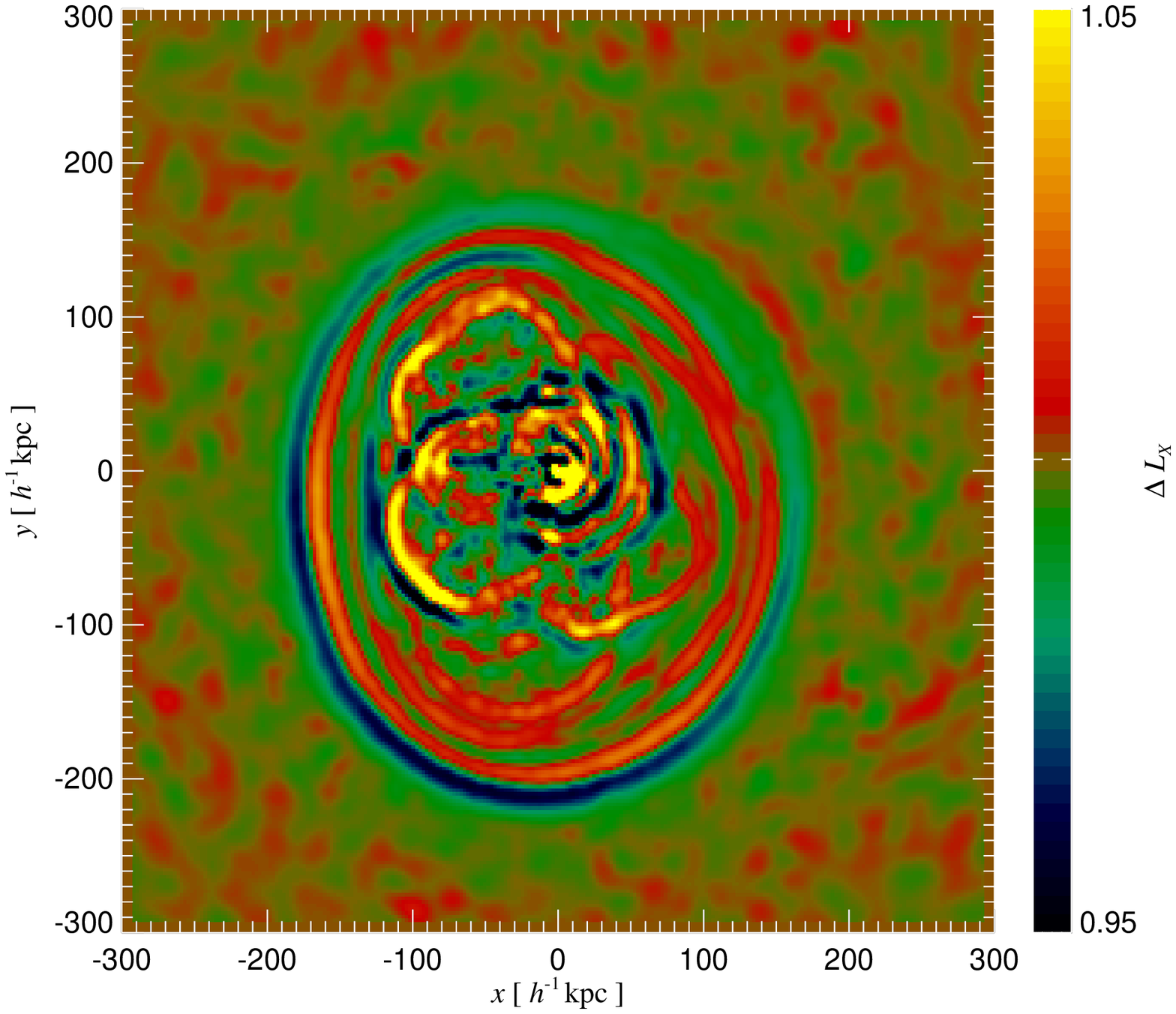,width=6.5truecm,height=6truecm} }
\vspace{-0.5truecm} \hbox{
\psfig{file=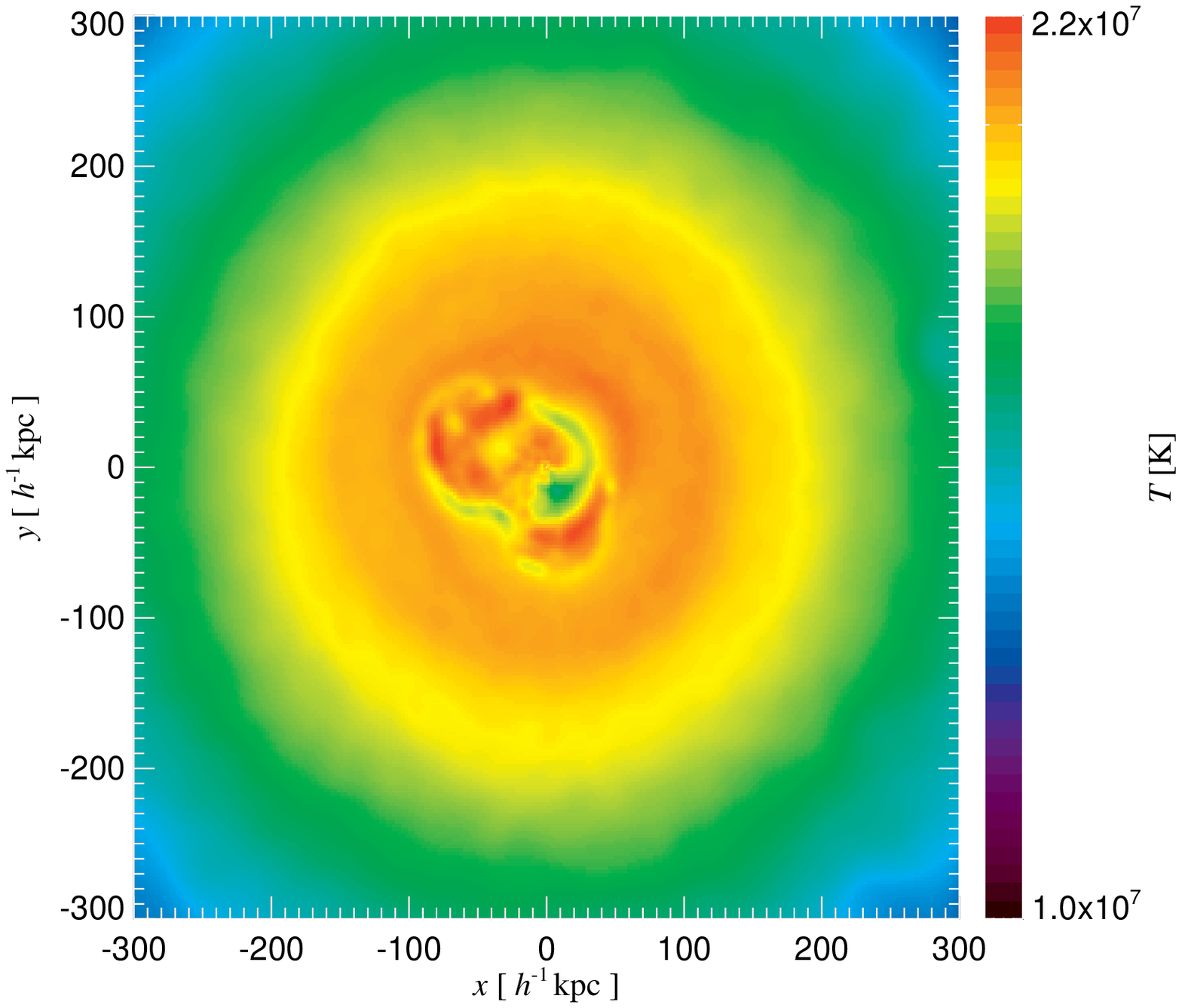,width=6.8truecm,height=6truecm}
\hspace{-0.6truecm}
\psfig{file=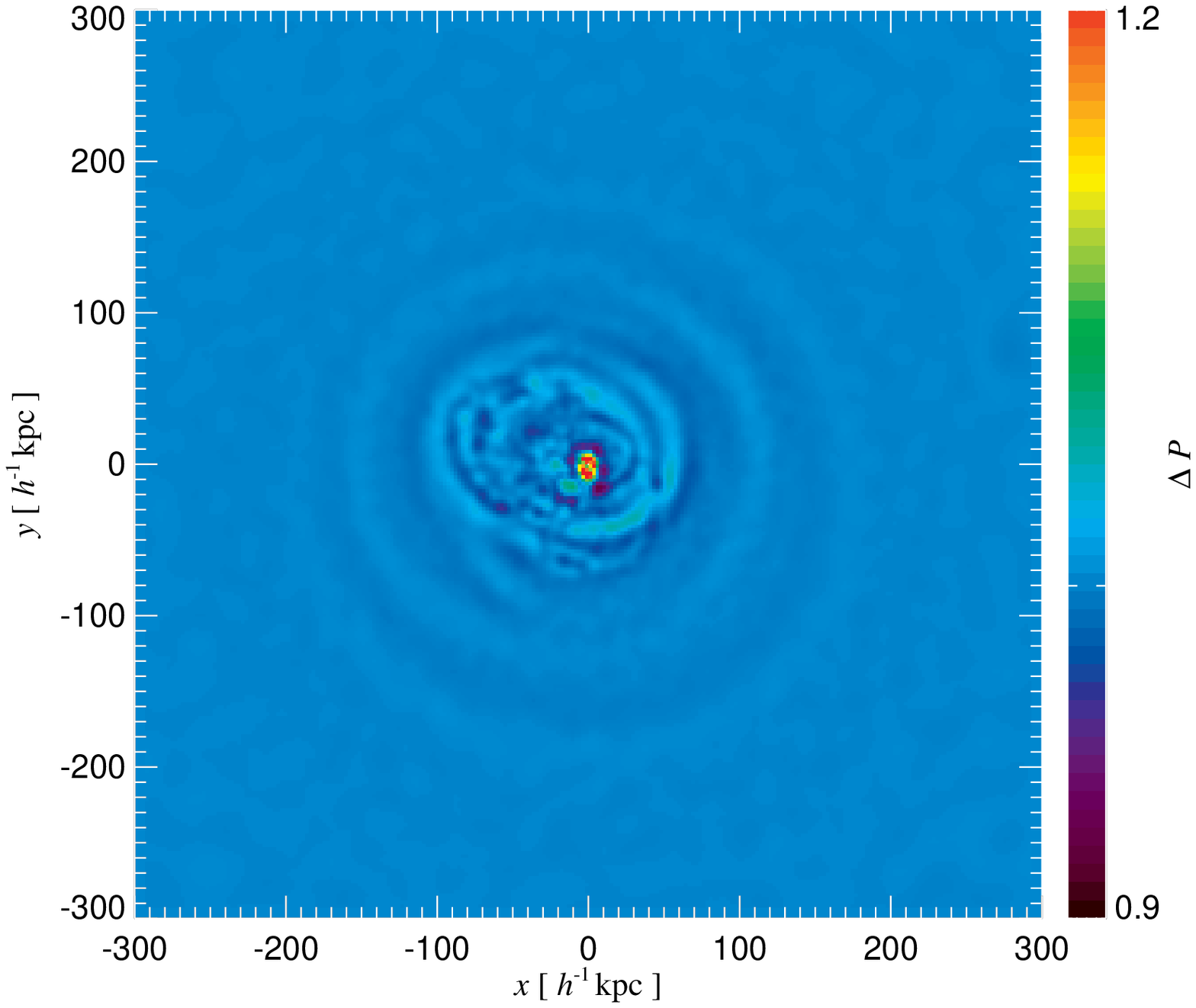,width=6.22truecm,height=6truecm}
\hspace{-0.4truecm}
\psfig{file=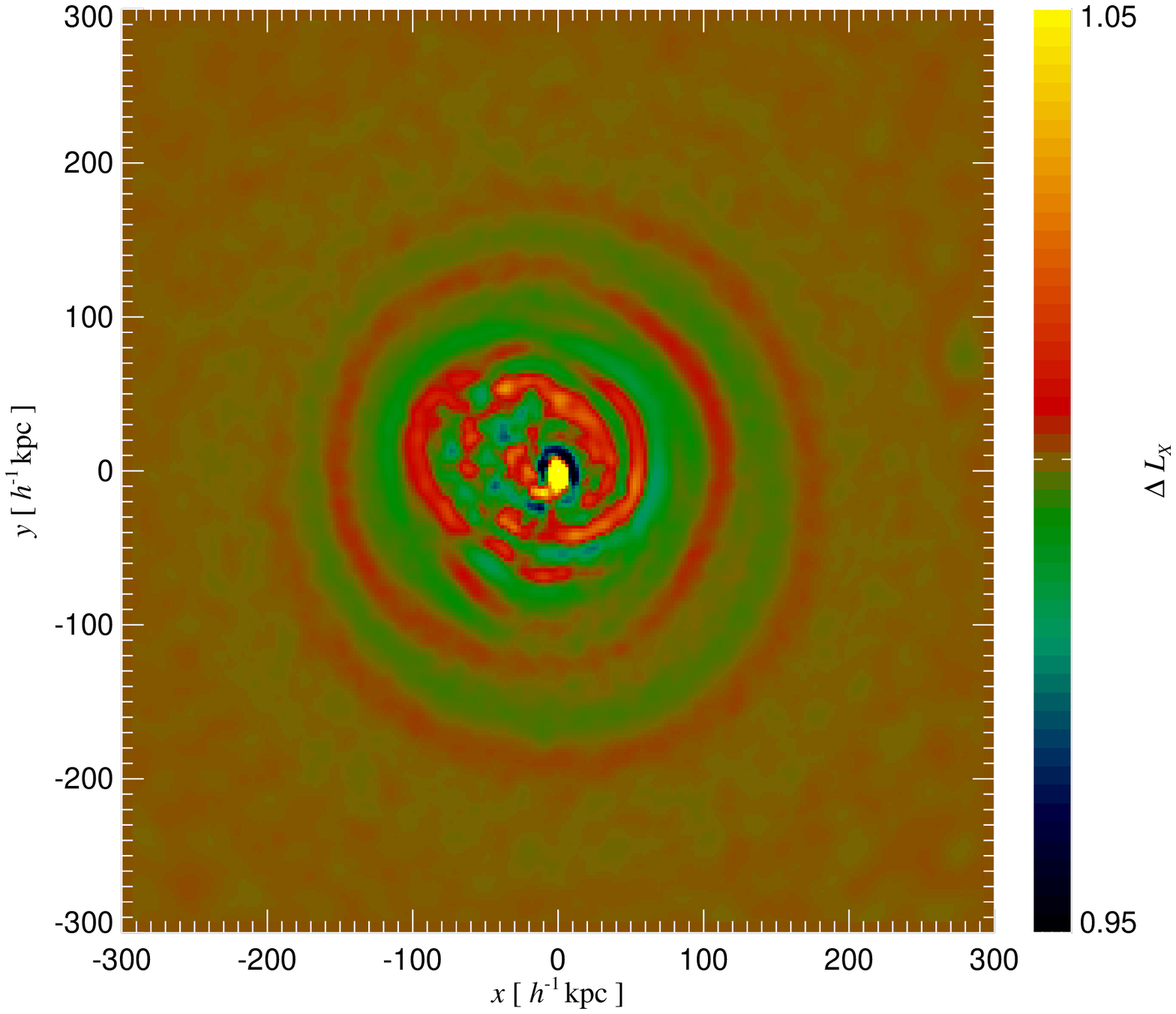,width=6.5truecm,height=6truecm} }
\vspace{-0.5truecm} \hbox{
\psfig{file=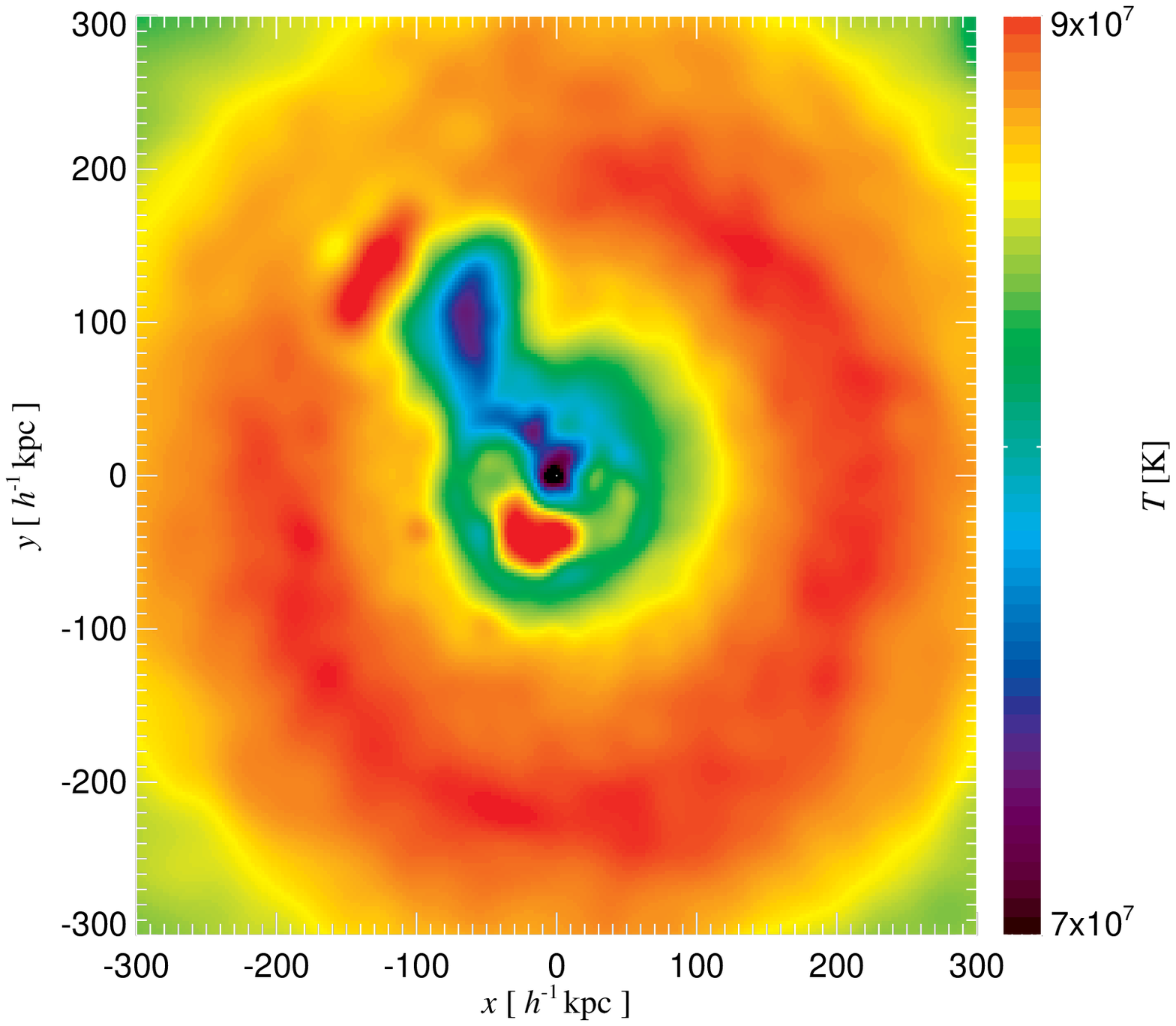,width=6.8truecm,height=6truecm}
\hspace{-0.6truecm}
\psfig{file=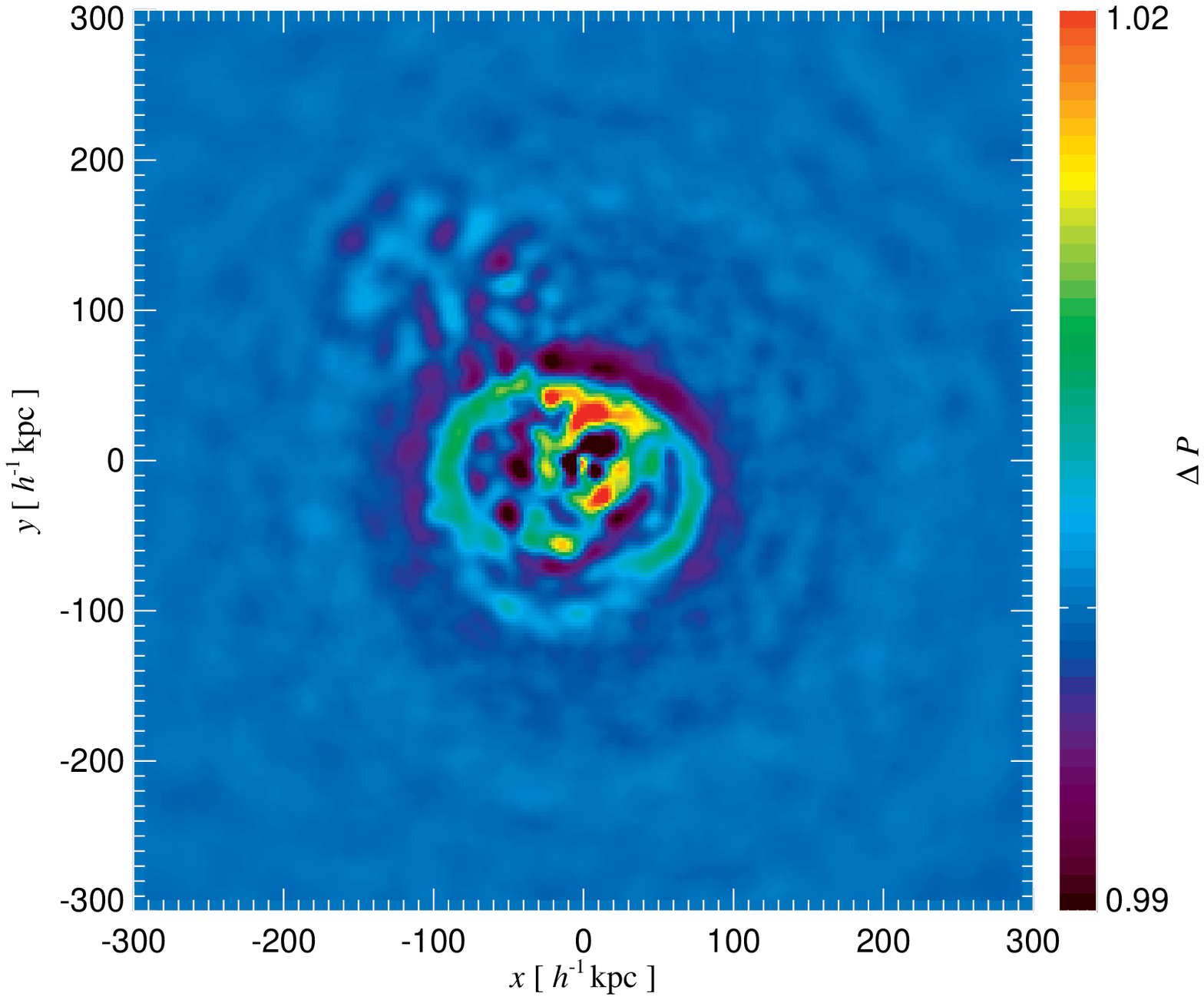,width=6.22truecm,height=6truecm}
\hspace{-0.4truecm}
\psfig{file=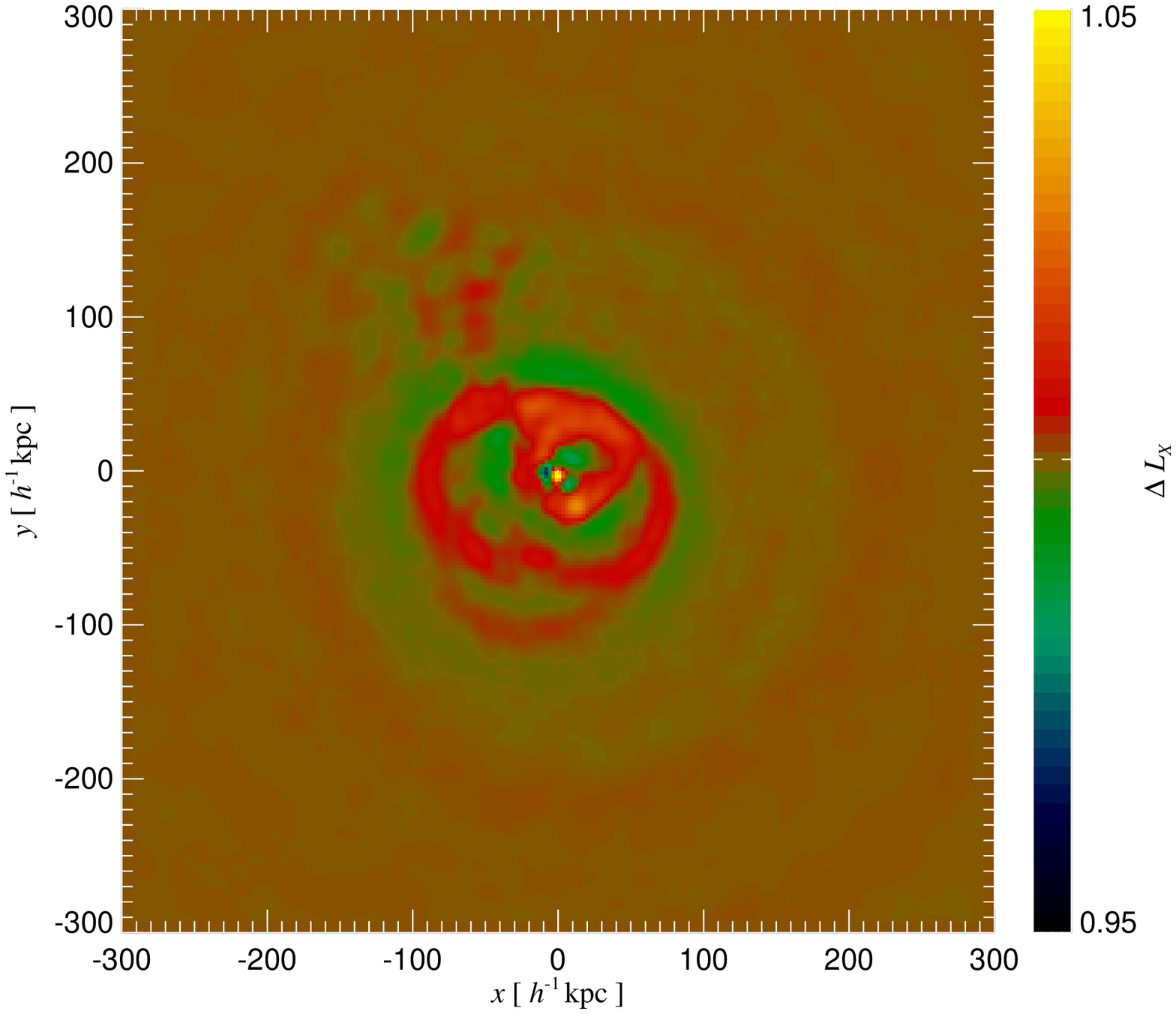,width=6.5truecm,height=6truecm} }
}}
\caption{The panels on the left-hand side show mass-weighted projected
  temperature maps of three isolated clusters with increasing mass:
  $10^{13}\,h^{-1} {\rm M}_\odot$ (upper-most panel), $10^{14}\,h^{-1}
  {\rm M}_\odot$ (central panel) and $10^{15}\,h^{-1} {\rm M}_\odot$
  (lower-most panel). It can be seen that the bubbles injected in the
  central cluster regions have mushroom-like morphologies and are
  uplifting residual cool material from the centre (lower-most
  panel). The middle and right-hand panels give projected pressure and
  X-ray emission maps, divided by the corresponding map smoothed on a
  scale of $16 \,h^{-1}{\rm kpc}$.  A number of bubble-induced sound
  waves and weak shocks are clearly visible, and in the case of the
  $10^{15}\,h^{-1} {\rm M}_\odot$ cluster, NW from the cluster centre
  a mushroom-shaped bubble in the pressure map is clearly visible
  which corresponds to the red blob in the temperature map.}
\label{iso_maps}
\end{figure*}

\begin{figure}
\psfig{file=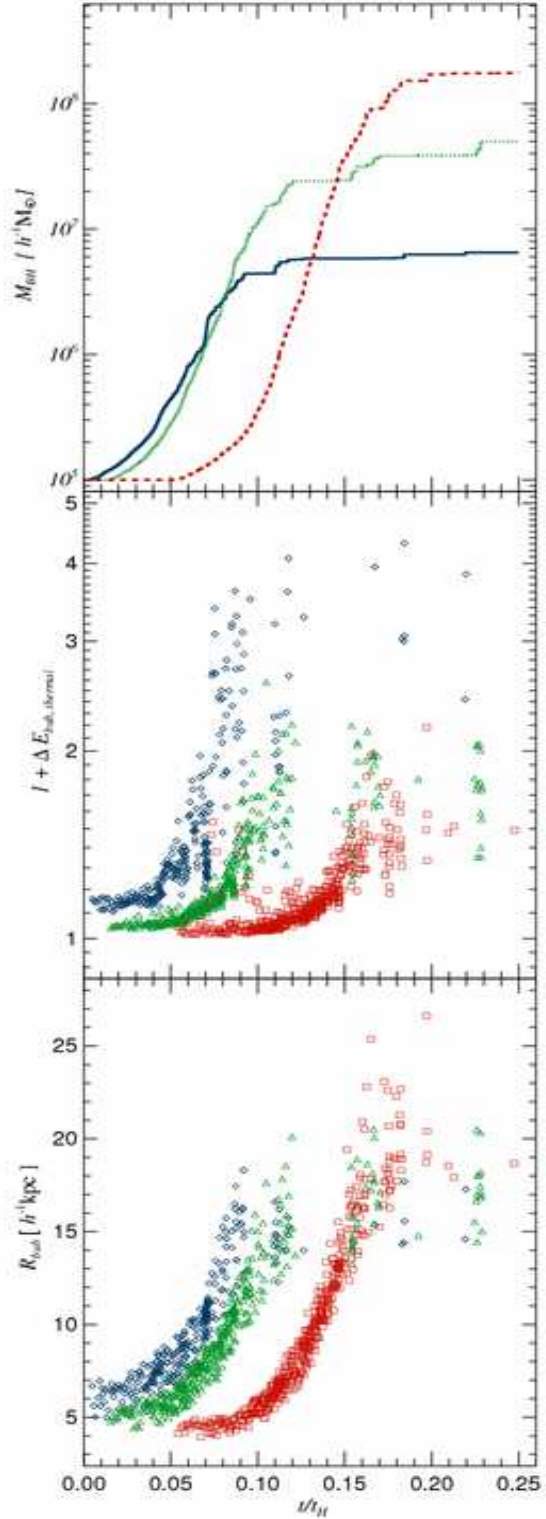,width=8.truecm,height=20.8truecm}
\caption{Growth of the BH mass with time (top panel)
  for three clusters of different mass. The middle panel shows how the
  thermal energy contrast, $\Delta {\rm E_{bub, \,thermal}}$, depends
  on time and the BH mass in our cluster simulations. The bubble
  radius as a function of time is shown in the bottom panel, for the
  same set of runs. The blue, continuous line and the diamonds are for
  a cluster of mass $10^{13}\,h^{-1} {\rm M}_\odot$, the green, dotted
  line and triangles are for a $10^{14}\,h^{-1} {\rm M}_\odot$ halo,
  while the red, dashed line and the squares denote results for a
  $10^{15}\,h^{-1} {\rm M}_\odot$ cluster.}
\label{Ebub_Mbh}
\end{figure}

\begin{figure}
\psfig{file=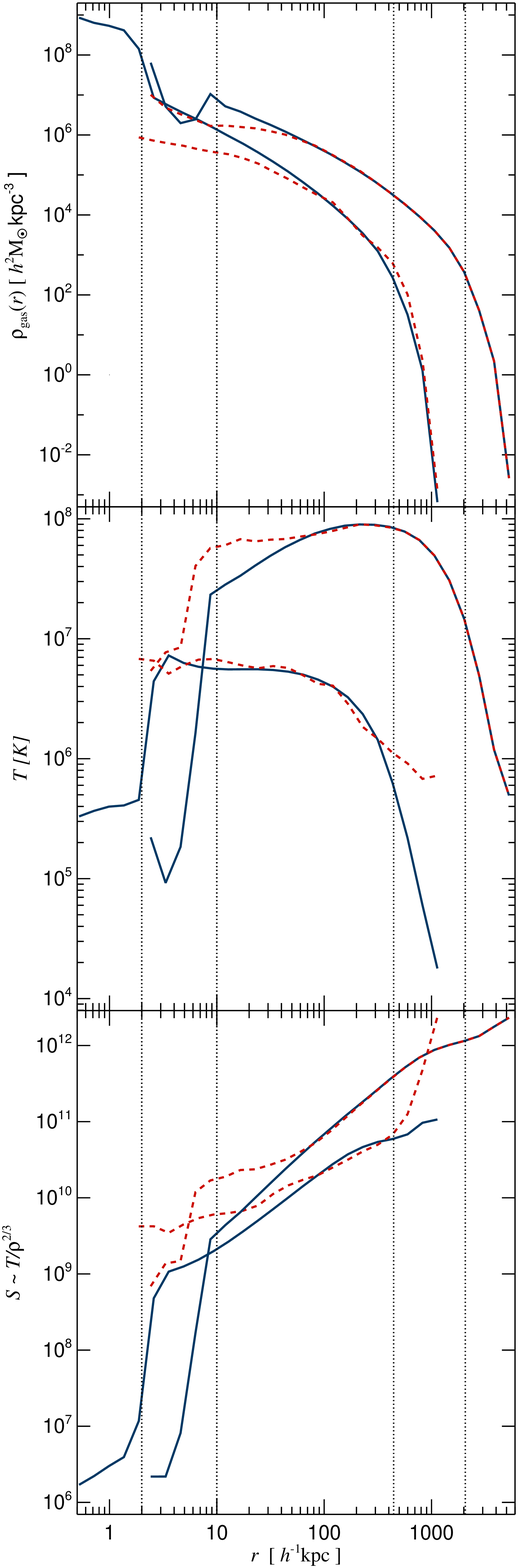,width=8truecm,height=21truecm}
\caption{Radial profiles of gas density (upper-most panel),
  temperature (central panel) and entropy (lower-most panel) for
  $10^{13}\,h^{-1} {\rm M}_\odot$ and $10^{15}\,h^{-1} {\rm M}_\odot$
  galaxy clusters at $t=0.2\,t_{\rm H}$. The blue, continuous lines
  correspond to the simulations without AGN heating, while red, dashed
  lines give the case where AGN feedback is present. On every panel,
  the upper set of profiles are for the more massive cluster. Vertical
  dotted lines mark the gravitational softening length and the virial
  radii of these clusters.}
\label{profiles_diffcluster}
\end{figure}

The $\delta_{\rm BH}$ parameter also affects the properties of
the ICM. In panels $(d)$, $(e)$ and $(f)$ of Figure~\ref{profiles_1d14cluster} we
illustrate this effect by plotting the radial profiles of gas density,
temperature and entropy, for the different values of $\delta_{\rm BH}$
assumed in the simulations, as indicated in the panels. We also show 
the corresponding profiles for the run without AGN heating, denoted by the
continuous blue lines. Clearly, there is a trend in the simulated profiles
with $\delta_{\rm BH}$: more frequent and more gentle bubble
heating events increase the central cluster entropy less than more powerful
but rarer feedback episodes.  A similar result has been found in the
simulations of \cite{Omma2004}, where a higher jet power was heating the
cluster for a longer time interval.  However, regardless of the exact value of
the $\delta_{\rm BH}$ parameter, the central cooling catastrophe
is prevented in our model cluster in all simulated cases with feedback. Once
the feedback becomes effective, the BH mass stops growing when it reaches $3-6
\times 10^7 \,h^{-1} {\rm M}_\odot$.  Note however that the final value of the
BH mass depends also on the $\epsilon_m$ parameter, which we have taken to be
as high as possible in these test runs.  Consequently, the BH masses obtained
here represent a lower limit for the actual BH mass that one would expect in a
cluster of this size.

In panel $(c)$ of Figure~\ref{profiles_1d14cluster}, we show how the star
formation rate (SFR) is affected by AGN-driven bubbles. Without AGN heating, the SFR
of the $10^{14}\,h^{-1} {\rm M}_\odot$ isolated cluster grows with time to
values as high as $\sim 250 {\rm M}_\odot/{\rm yr}$. This high SFR results
from the large amounts of cool gas that accumulate in the central regions,
forming a reservoir for intense star formation activity. However,
in the models with AGN feedback, the SFR drops dramatically to values
of order of $\sim 1{\rm M}_\odot/{\rm yr}$. For these runs it can be
also seen that the SFR shows a
number of short spikes of somewhat enhanced activity, and these
peaks are directly related with episodes of increased BHAR which
lead to a triggering of bubbles and a successive reduction of the SFR.

We have also tested to which extent the AGN feedback is affected by changes in
the normalization value for the bubble radius. As before, we adopt as default
values for bubble energy and ICM density $5 \times 10^{60}$erg and $10^6 \,
h^2 {\rm M}_\odot {\rm kpc}^{-3}$, respectively, but we vary $R_{\rm bub,0}$
for different runs. In the range of $15\,h^{-1} {\rm kpc}$ to $30\,h^{-1} {\rm
  kpc}$ for $R_{\rm bub,0}$, our results are not altered significantly by
modifications of this parameter.  However, for substantially larger values of
$R_{\rm bub,0}$, e.g. $60\,h^{-1} {\rm kpc}$, the heating energy per particle
drops to considerably smaller values, changing the initial rapid
growth phase of the BH. For large values of $R_{\rm
  bub,0}$, the AGN feedback is less efficient in the early phase of growth
such that the BH can grow to a somewhat larger mass before a self-regulated
heating loop is established.  We will further discuss the importance of the
bubble radius when we consider halos of different mass in
Section~\ref{Different_mass}.

Finally, we have numerically explored scenarios where an already massive BH is
introduced at the very beginning of a simulation. For $\delta_{\rm
  BH}$ equal to $0.1\%$, the BH mass is found to increase from its
initial value of $5 \times 10^7 \,h^{-1} {\rm M}_\odot$ to $9 \times 10^7
\,h^{-1} {\rm M}_\odot$ over a time span equal to one quarter of the Hubble
time.  Most of this BH growth, however, can be attributed to the intermittent
nature of bubble heating, while in the continuous AGN feedback regime,
the BH mass increases by less then $1\%$ over the same simulated time
span. Thus, our numerical scheme indeed leads to a stable galaxy
cluster solution where overcooling in the central parts is prevented
and the BH mass grows to a self-regulated value.

\subsection{AGN heating in halos of different mass}\label{Different_mass}

In this section we examine AGN feedback effects in cluster halos spanning a
range in mass, from $10^{13}\,h^{-1} {\rm M}_\odot$ to $10^{15}\,h^{-1} {\rm
  M}_\odot$. We adopt one set of feedback parameters for all simulations,
namely $\epsilon_{\rm m}$, $R_{{\rm bub},0}$, $E_{{\rm bub},0}$, and
$\rho_{{\rm ICM},0}$, as in the previous section, and for $\delta_{\rm BH}$ we choose the intermediate value of $1\%$.  We are
interested in the question whether for a fixed set of parameters our model
produces satisfactory solutions for halos of very different masses.

Before we delve into a detailed quantitative analysis of the cluster
simulations, we illustrate the visual morphology of simulated AGN-inflated
bubbles in Figure~\ref{iso_maps}. For this purpose, we show projected
mass-weighted temperature maps (first column), pressure fluctuation maps
(middle column) and fluctuation maps of the X--ray luminosity (last column).
Different rows are for clusters of increasing mass, from top to bottom. We
call the maps in the second and the third columns ``fluctuation'' maps because
they have been constructed by dividing the original map by a version that has
been smoothed on a scale of $16 \,h^{-1}{\rm kpc}$. This highlights local
departures from the mean ICM properties, similar to what is known as unsharp
masking technique in observational analysis.

A number of interesting features can be seen from the panels in
Figure~\ref{iso_maps}.  First, AGN-driven bubbles have characteristic
mushroom-like and cap-like morphologies. These are particularly evident for
the $10^{13}\,h^{-1} {\rm M}_\odot$ halo, where several bubbles injected
briefly after one another can be noticed. Note that the reason why there are
so many bubbles in this particular map is in part due to the selected time for
the image. At this moment $M_{\rm BH}$ happens to grow rapidly, thus the bubble duty
cycle is rather short. In the last row of Figure~\ref{iso_maps}, two bubbles
can be seen in the central region of the $10^{15}\,h^{-1} {\rm M}_\odot$
cluster, one SW from the cluster centre and the other NW and more distant. The
latter one is uplifting some residual cool gas from the cluster centre, which
forms a filamentary structure in the wake of the bubble. Looking at the
pressure map of this cluster a nicely defined mushroom-like bubble can be seen
that corresponds to the NW red blob in the temperature map.  In general, the
pressure and X-ray luminosity maps show numerous irregularities, sound waves
and weak shocks that have been generated by the bubbles. The ripples in these
maps become progressively weaker for more massive cluster, which is a
consequence of the parameterization adopted in our model, as we will clarify
in the following paragraphs.

In the upper-most panel of Figure~\ref{Ebub_Mbh}, we show the mass growth of
the central BH as a function of time for the different cluster simulations. It
can be seen that halos of increasing mass exhibit a qualitatively similar
behaviour for their BH growth: an initial rapid growth phase is followed by
self-regulated stagnation. However, the starting point of the rapid growth is
dictated by the gas cooling time, which is shortest for the smallest cluster
mass considered. Therefore, for the $10^{13}\,h^{-1} {\rm M}_\odot$ halo, the
BH starts growing first, but it is also the first one to reach stagnation,
while the BH sitting in the centre of the $10^{15}\,h^{-1} {\rm M}_\odot$
cluster waits for $\sim 0.05\,t/t_{\rm H}$ before increasing its mass
significantly. At the end of the simulated time-span, all three BHs reach
equilibrium states and their final masses are: $\sim 6.5 \times 10^{6}\,h^{-1}
{\rm M}_\odot$, $\sim 5 \times 10^{7}\,h^{-1} {\rm M}_\odot$, and $\sim 1.8
\times 10^{8}\,h^{-1} {\rm M}_\odot$, respectively.

In the other two panels of Figure~\ref{Ebub_Mbh}, we show how certain bubble
properties evolve with time for different cluster simulations. In the middle
panel, $\Delta {\rm E_{bub, \,thermal}}$ represents the thermal energy
contrast of the bubble just before and after the energy was injected into it.
$\Delta {\rm E_{bub, \,thermal}}$ is reflecting mainly the mass growth of the
BH, since it determines how much energy will be thermally coupled to the
bubbles. It can be seen that  $\Delta {\rm E_{bub, \,thermal}}$ initially grows
significantly with time, but when the BH growth saturates, it scatters
around a constant value. Also, there is a systematic trend of $\Delta {\rm
  E_{bub, \,thermal}}$ with cluster mass, with smaller mass clusters showing a
larger energy contrast for the bubbles. This is primarily due to the smaller
radius of bubbles in less massive clusters, as can be seen in the bottom
panel of Figure~\ref{Ebub_Mbh}. Given that we have adopted the same set of
parameters for our different cluster simulations, it follows from
Eqn.~(\ref{Rbub_eq}) that BHs of smaller mass will generate smaller bubbles.
This explains why $R_{\rm bub}$ is growing in time and why it is bigger for
more massive halos that harbour larger BHs. This consequence of our model has
also repercussions for the presence and intensity of sound waves generated by
the inflation of bubbles, as seen in the maps of Figure~\ref{iso_maps}.
Smaller bubbles with a higher energy contrast relative to the surrounding ICM
are more efficient in generating weak shocks and sound waves than large
bubbles with low $\Delta {\rm E_{bub, \,thermal}}$.

In Figure~\ref{profiles_diffcluster}, we show radial profiles of gas density,
temperature and entropy for our isolated clusters of mass $10^{13}\,h^{-1}
{\rm M}_\odot$ and $10^{15}\,h^{-1} {\rm M}_\odot$. The radial profiles of the
$10^{14}\,h^{-1} {\rm M}_\odot$ cluster can be found in
Figure~\ref{profiles_1d14cluster} with the same colour-coding. The blue,
continuous lines denote the runs without AGN, while the red, dashed curves are
for the simulations with AGN heating. The upper set of profiles in each panel
are for the more massive cluster. The results show that for a fixed choice of
parameters in our AGN heating model the ICM properties of clusters over a
range of masses appear satisfactory: (i) the gas density is somewhat
suppressed in central regions; (ii) very cool gas in the innermost regions of
clusters is absent, and at the same time the AGN heating is not increasing the
ICM temperature too much; (iii) the gas entropy is boosted in central regions,
but without generating entropy inversions that are typically not observed in
real systems. Together with realistic BH masses and BHAR, these
promising features of the simulated ICM properties encourage us to proceed
studying our model using fully cosmological simulations of galaxy cluster
formation, which we consider in the next section.

\section{Cosmological simulations of AGN feedback in galaxy
  clusters} \label{Cosmological}

We now turn to self-consistent cosmological simulations of galaxy cluster
formation, with the aim to understand how AGN feedback influences these
objects at different cosmological epochs and how this influence depends on the
mass and the evolutionary history of clusters. We focus our analysis on two
galaxy cluster simulations that have quite different merging histories and
different present-day masses. The clusters have been selected from a
cosmological $\Lambda$CDM model with a box size of $479\,h^{-1}{\rm Mpc}$
\citep{Yoshida2001, Jenkins2001}, and were prepared by \citet{Dolag2004} for
resimulation at higher resolution using the Zoomed Initial Condition (ZIC)
technique \citep{Tormen1997}. The cosmological parameters of the simulations
are those of a $\Lambda$CDM concordance cosmology with $\Omega_{\rm m}=0.3$,
$\Omega_\Lambda=0.7$, $\Omega_{\rm b}=0.04$, $\sigma_8=0.9$, and Hubble
constant $H_0=70 \, {\rm km\,s}^{-1}{\rm Mpc}^{-1}$ at the present epoch. In
respect to the isolated halos simulations we have adopted somewhat different
parameters of the BH feedback model, with the normalization values for $E_{\rm
bub,0} = 10^{55}$erg, and $\rho_{\rm ICM,0}=10^4\,h^2\,{\rm M_\odot\,
kpc^{-3}}$ being lower, with $\delta_{\rm BH}= 0.01\%$, and with a more
realistic mechanical feedback efficiency of $\epsilon_m=0.20$, while the other
parameters were kept exactly the same.

The primary numerical parameters of our simulations are summarized in
Table~\ref{tab_simpar}, and the main physical properties at $z=0$ of the
formed clusters are listed in Table~\ref{tab_GCpar}. For both simulated galaxy
clusters, the smaller one (``g676'') and the more massive one (``g1''), the
virial radius, virial mass, total gas mass and stellar mass have been measured in
runs without AGN feedback (labelled ``csf'') and when it is included (labelled
``csfbh'').  Additionally, the mass-- and emission--weighted temperature,
X--ray luminosity and total SFR are given in the 6th to the 9th column of
Table~\ref{tab_GCpar}, respectively.  Looking at the gas properties listed in
Table~\ref{tab_GCpar} it can be seen that AGN feedback does not significantly
alter the global gas temperature, or the X--ray luminosity.  However, there are
still important local changes of these properties when AGN heating is
included, as we will discuss in more detail in Section~\ref{ICMproperties}. On
the other hand, the global stellar properties are strongly affected by AGN
feedback: the total stellar mass is reduced considerably, as well as the total
SFR, while the total gas mass is increased.  This implies that the relative
amount of gas versus stars is a function of AGN heating efficiency.

\begin{table*}
\bc
\begin{tabular}{crrccccc}
\hline
\hline
Simulation & $N_{\rm HR}$ & $N_{\rm gas}$ & $m_{\rm DM}$
[$\,h^{-1}{\rm M}_\odot\,$] & $m_{\rm gas}$  [$\,h^{-1}{\rm
    M}_\odot\,$] & $z_{\rm start}$ & $z_{\rm end}$ & $\epsilon$
[$\,h^{-1}{\rm kpc}\,$] \\
\hline
g676 & $314518$  & $314518$  & $1.13\times 10^9$ & $0.17\times 10^9$ &
$60$ & $0$ & $5.0$ \\
g1   & $4937886$ & $4937886$ & $1.13\times 10^9$ & $0.17\times 10^9$ &
$60$ & $0$ & $5.0$ \\

\hline
\hline
\end{tabular}
\caption{Numerical parameters of the cosmological galaxy cluster
  simulations analysed in this study. The values listed from the
  second to the fifth column refer to the number and to the mass of
  high resolution dark matter particles and of gas particles. Note
  that the actual values of $N_{\rm gas}$ and $m_{\rm gas}$ vary in
  time due to star formation. The last three columns give the initial
  and final redshifts of the runs, and the gravitational softening
  length $\epsilon$.
\label{tab_simpar}}
\ec
\end{table*}
\begin{table*}
\bc
\begin{tabular}{lcccccccc}
\hline
\hline
Cluster & $R_{\rm 200}$ & $M_{\rm 200}$ &$M_{\rm gas,200}$ & $M_{\rm
  stars,200}$ & $T_{\rm mw}$ & $T_{\rm ew}$ & $L_{\rm X}$ & SFR \\
 & [$\,h^{-1}{\rm kpc}\,$] & [$\,h^{-1}{\rm M}_\odot\,$] &
[$\,h^{-1}{\rm M}_\odot\,$] & [$\,h^{-1}{\rm M}_\odot\,$] & [$K$] &
[$K$] & [$\,\rm ergs^{-1}\,$] & [$\,{\rm yr}^{-1}{\rm M}_\odot\,$]    \\
\hline \hline
g676\_csf   & $1176$ & $1.13\times10^{14}$ & $9.3\times10^{12}$ &
$4.7\times10^{12}$ & $1.4\times10^7$ & $2.6\times10^7$ &
$1.6\times10^{43}$ & 51 \\ 
g676\_csfbh & $1165$ & $1.10\times10^{14}$ & $1.1\times10^{13}$ &
$1.4\times10^{12}$ & $1.3\times10^7$ & $2.4\times10^7$ &
$1.0\times10^{43}$ & 1 \\ \hline 
g1\_csf  & $2857$ & $1.63\times10^{15}$ & $1.4\times10^{14}$ &
$6.3\times10^{13}$ & $7.3\times10^7$ &
$1.3\times10^8$ & $1.0\times10^{45}$ & 742\\
g1\_csfbh  & $2859$ & $1.63\times10^{15}$ & $1.7\times10^{14}$ &
$2.5\times10^{13}$ & $7.3\times10^7$ & $1.3\times10^8$ &
$1.0\times10^{45}$ & 144\\ 
\hline
\hline
\end{tabular}
\caption{Physical properties of our sample of simulated galaxy
  clusters at $z=0$, selected with a mean overdensity of
  $200\rho_c$. For two different galaxy clusters, as labelled in the
  first column, the cluster radius, total mass, gas mass and stellar
  mass are given. Also, the mass-- and emission--weighted gas
  temperatures, X--ray luminosity and total SFR are listed, in the 6th
  to 9th column, respectively. Note that the values give results both
  for the simulations with cooling and star formation only (denoted
  with ``csf''), and for runs performed with AGN heating (labelled
  ``csfbh'').
\label{tab_GCpar}}
\ec
\end{table*}

\subsection{Black hole growth in clusters}\label{BHgrowth}

\begin{figure}
\psfig{file=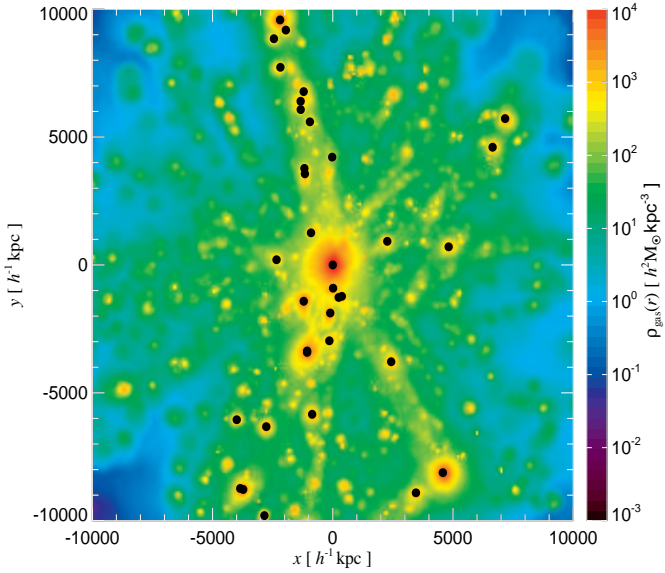,width=9truecm,height=8.5truecm}
\caption{Projected gas density map of the g676 galaxy cluster simulation at
  $z=0$, subject to AGN heating. Black dots mark the positions of BH particles
  that are more massive than $1.5 \times 10^{7}\,h^{-1} {\rm M}_\odot$. It can
  be seen that the BHs are residing not only in the centre of the most massive
  galaxy cluster, but also in almost all smaller halos that are visible on the
  map. Interestingly, there is a concentration of BHs in a filament that is
  extending from the central cluster towards NW, due to the ongoing structure
  formation there.}
\label{g676_bhmap}
\end{figure}

In Figure~\ref{g676_bhmap}, we show a projected gas density map of the g676
galaxy cluster simulation at $z=0$, with the positions of BHs more
massive than $1.5 \times 10^{7}\,h^{-1} {\rm M}_\odot$ overlaid as black dots.
The density map is $20 \,h^{-1} {\rm Mpc}$ on a side and is centred on the
most massive cluster galaxy. Interestingly, the surrounding large-scale structure
shows many smaller halos with embedded BHs, many of them residing in a
filamentary region NW of the central object.  In fact, most halos are
harbouring a central, massive BH. However, the most massive BH among them is
sitting at the centre of the biggest halo, and this is also true at higher
redshifts.

\begin{figure}
\psfig{file=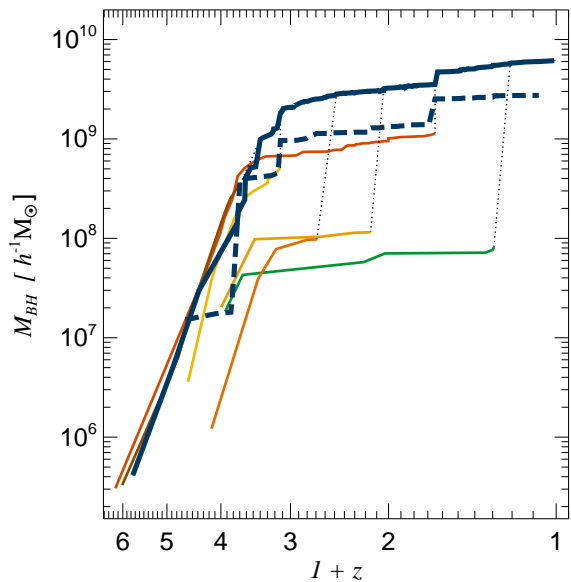,width=9truecm,height=9truecm}
\caption{Merger tree of the most massive BH at $z=0$ in the g676
  galaxy cluster simulation. The thick blue line is showing the mass
  growth of the BH's first progenitor as a function of redshift, while
  the thin lines of different colour represent the evolutionary
  history of secondary progenitors, until the merger with the main
  progenitor occurs. For clarity, only secondary progenitors that at
  the moment of the last merger have a mass greater than $5 \times
  10^{7}\,h^{-1} {\rm M}_\odot$ are shown. Vertical dotted lines
  indicate when merger events between the BHs in consideration
  occur. The dashed, thick blue line represents the cumulative mass of
  all secondary progenitors.}
\label{g676_mtree}
\end{figure}

The mass as a function of time of this most massive BH is shown in
Figure~\ref{g676_mtree}. To produce this measurement, we have first
constructed the full merger tree of all BHs formed in the simulated
volume, allowing us to extract the main progenitor trunk corresponding
to the most massive BH at $z=0$, which is plotted as a thick
continuous blue line. Also, we show the mass growth of the secondary
progenitors that at the moment of their last merger have a mass
greater than $5 \times 10^{7}\,h^{-1} {\rm M}_\odot$. It can be seen
that the most massive BH at $z=0$ is not the first one to form, but it
merges with two BHs that formed somewhat earlier and that are more
massive for $z > 3.5$. This is similar to our findings in
\citet{DiMatteo2007}, where we showed that the most massive black
holes found at high redshift in a cosmological volume need not
necessarily be the most massive ones at low redshift.  The thick,
dashed blue line in Figure~\ref{g676_mtree} represents the cumulative
mass of all secondary progenitors of the most massive BH at
$z=0$. This shows that mergers contribute up to $45\%$ to the final
mass of the BH that defines the trunk, while the remaining $55\%$ are
due to gas accretion by the BH on the trunk itself. Note however that
ultimately all the BH mass is built-up by gas accretion (the mass of
the initial seed BHs is negligible), but the part we here attributed
to mergers was built up by gas accretion in other systems.

We also note that the mass growth via mergers is more important at
higher redshifts, with the last significant merging event occurring at
$z \sim 0.7$.  The merger events of BHs are mainly driven by the
merging history of the host galaxy clusters.  Indeed, there is a quite
close correspondence between the evolution of the main progenitor of
the most massive BH and the main progenitor of the biggest halo in the
simulation. In particular, the main progenitor of the g676 galaxy
cluster undergoes a number of mergers between redshifts $3.5 < z <
2.8$, when also the BH grows significantly via mergers with other
BHs. Then at $z \sim 0.75$, the g676 cluster experiences its last
major merger, visible as a jump in the main trunk in
Figure~\ref{g676_mtree}.  At lower redshifts, the cluster becomes
fairly relaxed and isolated, such that its central BH grows mostly by
accreting gas that is cooling off from the hot cluster atmosphere. As
we will discuss in Section~\ref{Radio} this mode of accretion (at low
accretion rates) does however not contribute significantly to the
overall BH mass growth.

\begin{figure}
\psfig{file=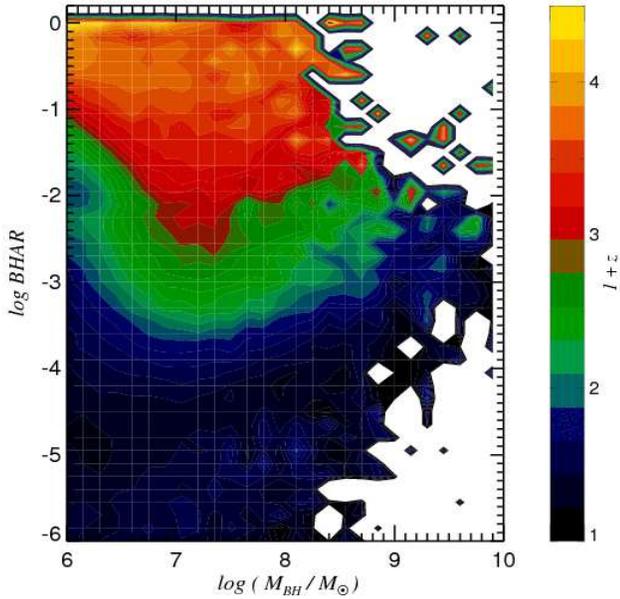,width=9.0truecm,height=8.5truecm}
\caption{BHAR in Eddington units as a function of BH
mass, for all BHs belonging to the g1 galaxy cluster simulation. The
colour-coding expresses the redshift, and the redshift interval
considered here ranges from $4.3$ to $0.0$. For a given BH mass, the
accretion rate is highest at early cosmic times. At lower redshifts,
there is an upturn in the BHAR at the massive BH end, corresponding to
BHs that are fed by gas that cools off from the atmosphere of the
cluster and its massive progenitor systems.}
\label{g1_BHAR}
\end{figure}

Now we consider in more detail the mass growth and the accretion rate,
not only of the most massive BH, but of all the BHs belonging to the
simulated volume.  Due to the better statistics provided by the g1
galaxy cluster simulation, we focus on this simulation, noting that
the results for the g676 cluster are very similar. In
Figure~\ref{g1_BHAR}, we plot the BHAR, expressed in Eddington units,
as a function of the BH mass.  The colour-coding denotes the redshift,
over the redshift interval $z = 4.3$ to $z = 0$ considered here. A
number of interesting features can be noticed in this plot: (i) at a
given BH mass, the accretion rate is decreasing with redshift; (ii)
for intermediate redshifts, $2 < z < 3$, the BHAR is lowest for
intermediate mass BHs; (iii) at still lower redshifts, the BHAR
for intermediate masses keeps falling strongly, while it stays
comparatively high for low-mass BHs as well as for very massive BHs (iv)
at the present epoch there is a large population of low-mass BHs that is
still accreting efficiently. These features are
qualitatively consistent with observational findings both from
optical and X-rays surveys \citep[e.g.][]{Steffen2003, Ueda2003,
Heckman2004, Barger2005, Hasinger2005} that indicate that BHs are
growing in a so-called `anti-hierarchical' manner.  In
Section~\ref{Galform}, we will come back to this issue, and discuss
how the BHAR depends on the BH mass in a galactic environment.

At low redshifts, the increase of the BHAR with mass at the high
BH mass end is driven by BHs that sit at the centres of massive galaxy
clusters. These BHs are fed by gas from the hot cluster atmosphere,
which would develop a strong cooling flow without the periodic heating
by the AGN feedback. Interestingly, we also find that our BHAR of
massive BHs at $z=0$ agrees very well with a recent estimate of the
Bondi accretion rate for a sample of X-ray luminous elliptical
galaxies by \citet{Allen2006}.

However, we need to emphasize that the simulations performed in this study
were not designed to address the problem of BH formation and growth in the
very early universe, for which they lack the required resolution and volume.
This becomes also apparent in the plot of Figure~\ref{g1_BHAR}, where the
upper right corner is not populated.  This means that these simulations cannot
account for the existence of supermassive BHs of mass $\sim 10^{9}\,{\rm
  M}_\odot$ already at $z=6$, as inferred from the luminosity of high redshift
quasars \citep{Fan2001}.  Using high-resolution compound galaxy models to
populate a merger tree measured from a zoom simulation of a protocluster
region, \citet{Li2006} have however recently demonstrated that at least in
principle such early supermassive BHs can grow from accretion in
gas-rich mergers at high redshift.

\subsection{Heating the cluster outskirts}\label{Outskirts}

\begin{figure}
\psfig{file=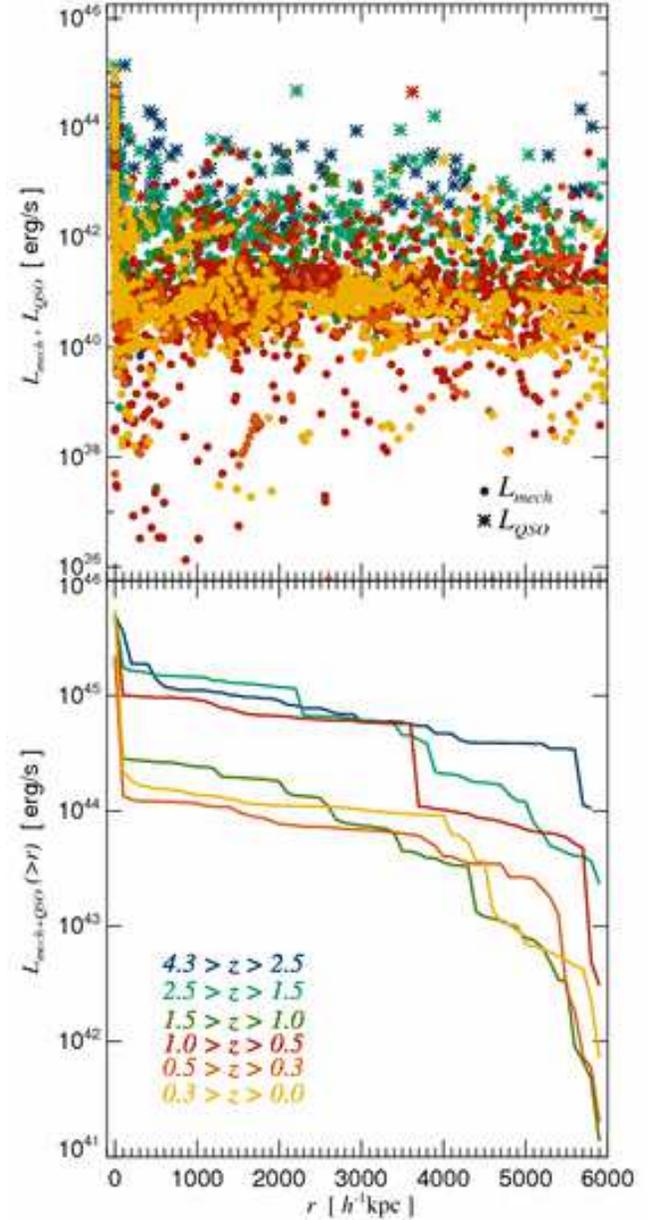,width=8.5truecm,height=17truecm}
\caption{In the upper panel, the mechanical and quasar AGN
  luminosities are plotted as a function of distance from the most
  massive halo in the g676 galaxy cluster simulation. Depending on the
  BHAR, a BH is assumed to be either in a ``quasar'' or
  in a ``radio'' mode, and the BH luminosity is calculated accordingly
  and denoted with different symbols. The colour-coding is according to
  redshift bins, as indicated in the lower panel. The total AGN
  luminosity outside a given radius, regardless of the feedback mode,
  is shown in the lower panel. It can be seen that the highest AGN
  luminosities always correspond to BHs sitting in the centre of the
  most massive clusters, and that at higher redshifts (especially
  $1.5<z<4.3$) heating from the quasars in cluster outskirts can be
  important.}
\label{g676_Lmech}
\end{figure}

In Figure~\ref{g676_Lmech}, we examine the radial dependence of AGN heating in
our simulated clusters for different redshift bins. To this end, we evaluate
the AGN luminosity at the given epoch and plot it as a function of distance
from the centre of the most massive cluster in the simulation. Depending on
the BHAR, we decide whether the BH is in the ``quasar'' or
``radio'' mode, which we denote with star symbols or circles, respectively
(see upper panel). In the lower panel, we plot the total AGN luminosity
outside of a given radius, regardless in which mode the BHs accrete. It can be
seen that the AGN heating at all redshifts considered and regardless of the
feedback mode is most important in the cluster centre. However, at early
times, and in particular for $1.5<z<4.3$, BHs that happen to reside in cluster
outskirts during this time could provide an important additional source of ICM
heating.  These BHs are essentially all in the quasar phase where they accrete
gas efficiently. Most of these BHs reside in satellite halos that are entering
the most massive system for the first time, and thus possibly are preheating
the gas of smaller halos prior to and during the merger with the massive
cluster. These findings and the spatial distribution of AGN in galaxy
clusters, illustrated in Figure~\ref{g676_bhmap}, are in a good agreement with
observational evidence \citep[e.g.][]{Cappi2001, Cappelluti2005,
  Ruderman2005, Martini2006} for the presence of AGN in galaxy cluster
environments. At low redshifts, though, the AGN sitting
at the cluster centre appears to be the dominant source of heating, as we
confirm in Section~\ref{ICMproperties}. Interestingly, the feedback luminosity
of the central AGN, regardless of its accretion regime, peaks at the similar
value for the whole time span considered.

\begin{figure*}
\centerline{
\psfig{file=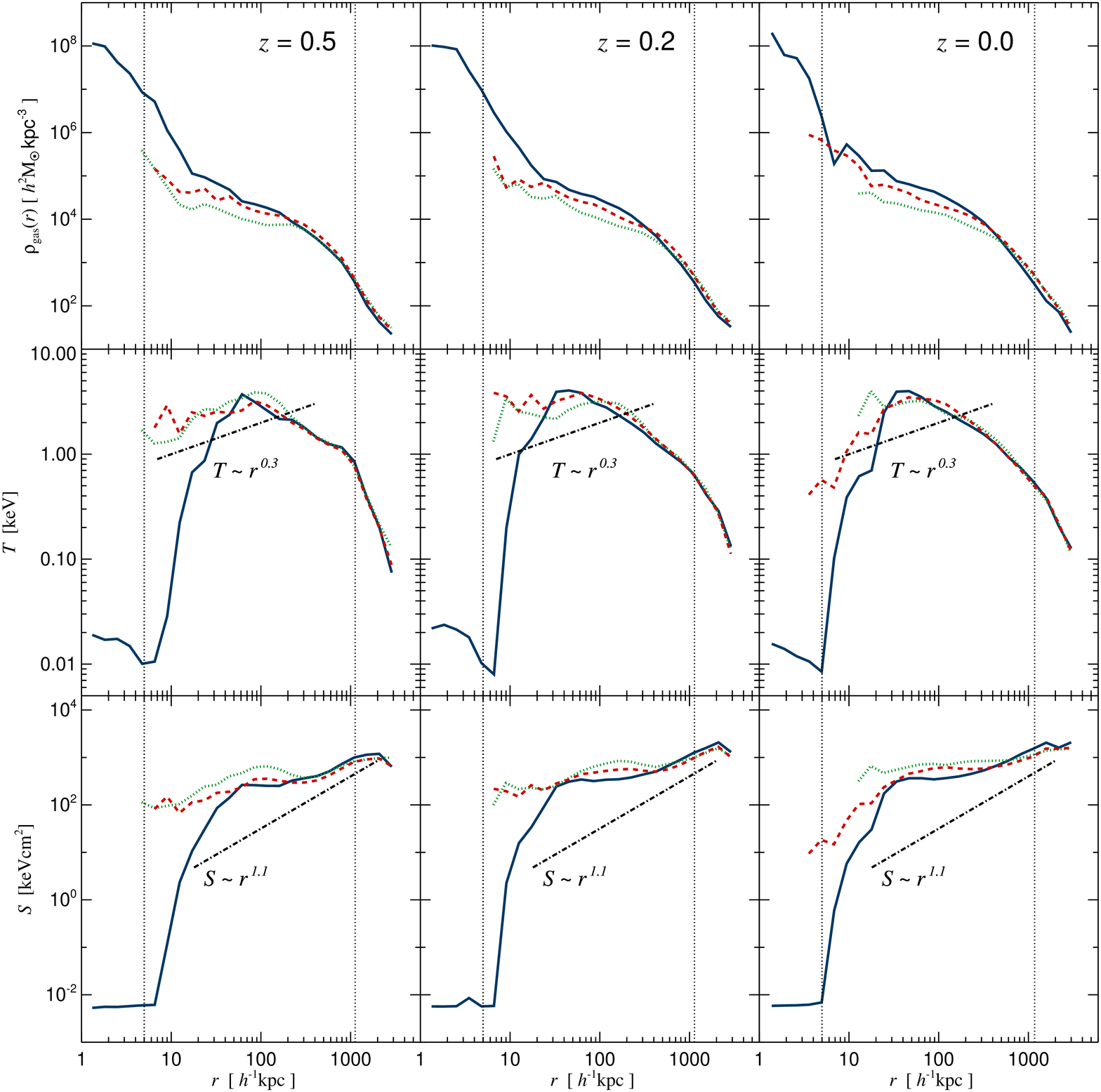,width=19truecm,height=19truecm}
}
\caption{Radial profiles of gas density (upper row), mass-weighted
  temperature (central row) and entropy (lower row) of the g676 galaxy
  cluster at $z = 0.5$, $0.2$ and $0.0$, respectively. Continuous blue
  lines illustrate the case without AGN heating, while dashed red
  lines are for simulations where AGN feedback is included. The run
  where the Eddington limit was not imposed on the BHAR is shown with
  a dotted green line. The vertical dotted lines denote the
  gravitational softening and the virial radius of this cluster.
  The dash-dotted lines in the central row of panels show the slope of
  the central temperature profiles of the cool core clusters found by
  \cite{Sanderson2006}, while the dash-dotted lines in the lower row
  illustrate the entropy scaling with radius, i.e. $S \propto
  r^{1.1}$.}
\label{g676_profiles}
\end{figure*} 

\begin{figure*}
\centerline{\vbox{
\psfig{file=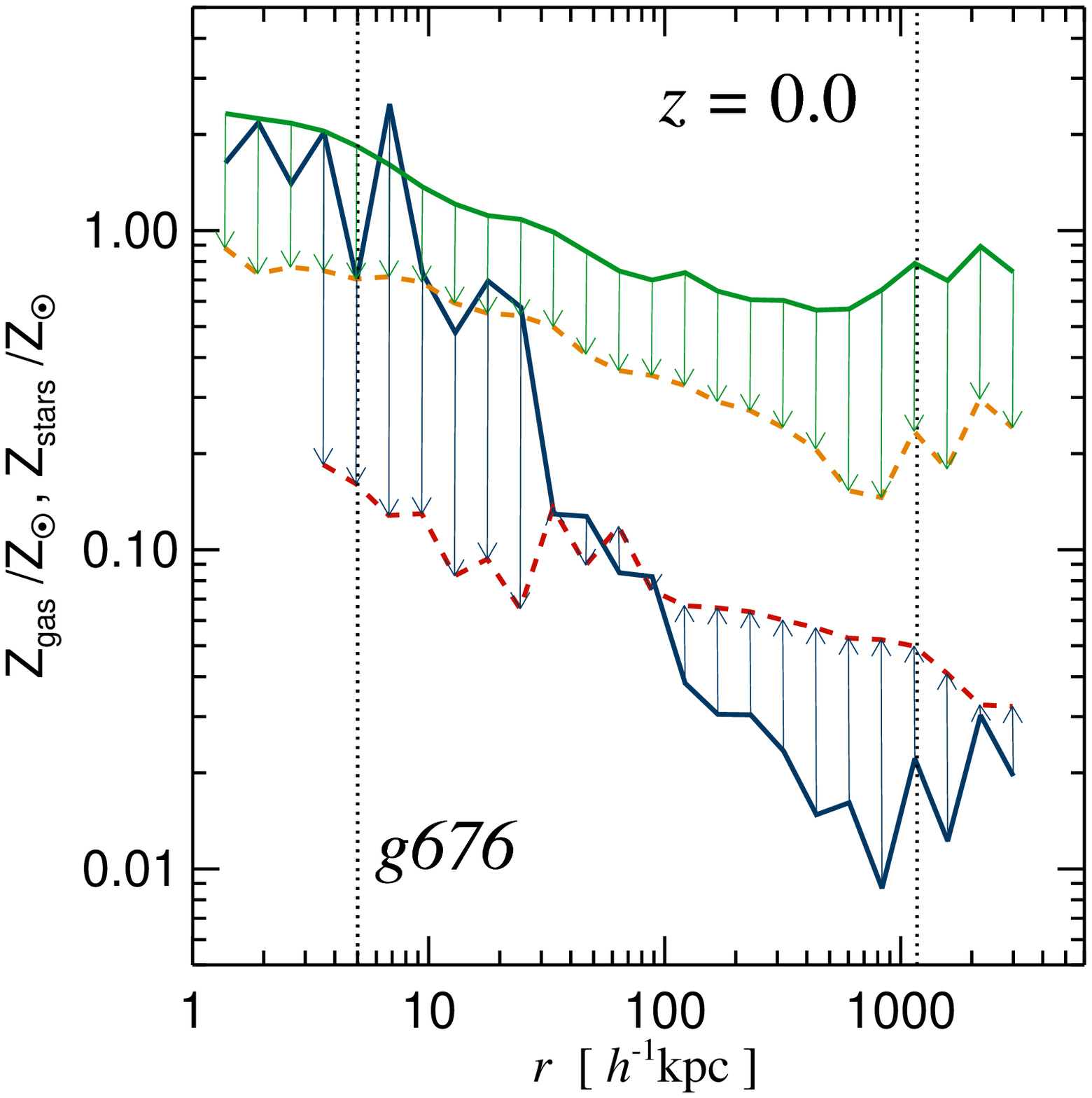,width=8.5truecm,height=8.5truecm}
\psfig{file=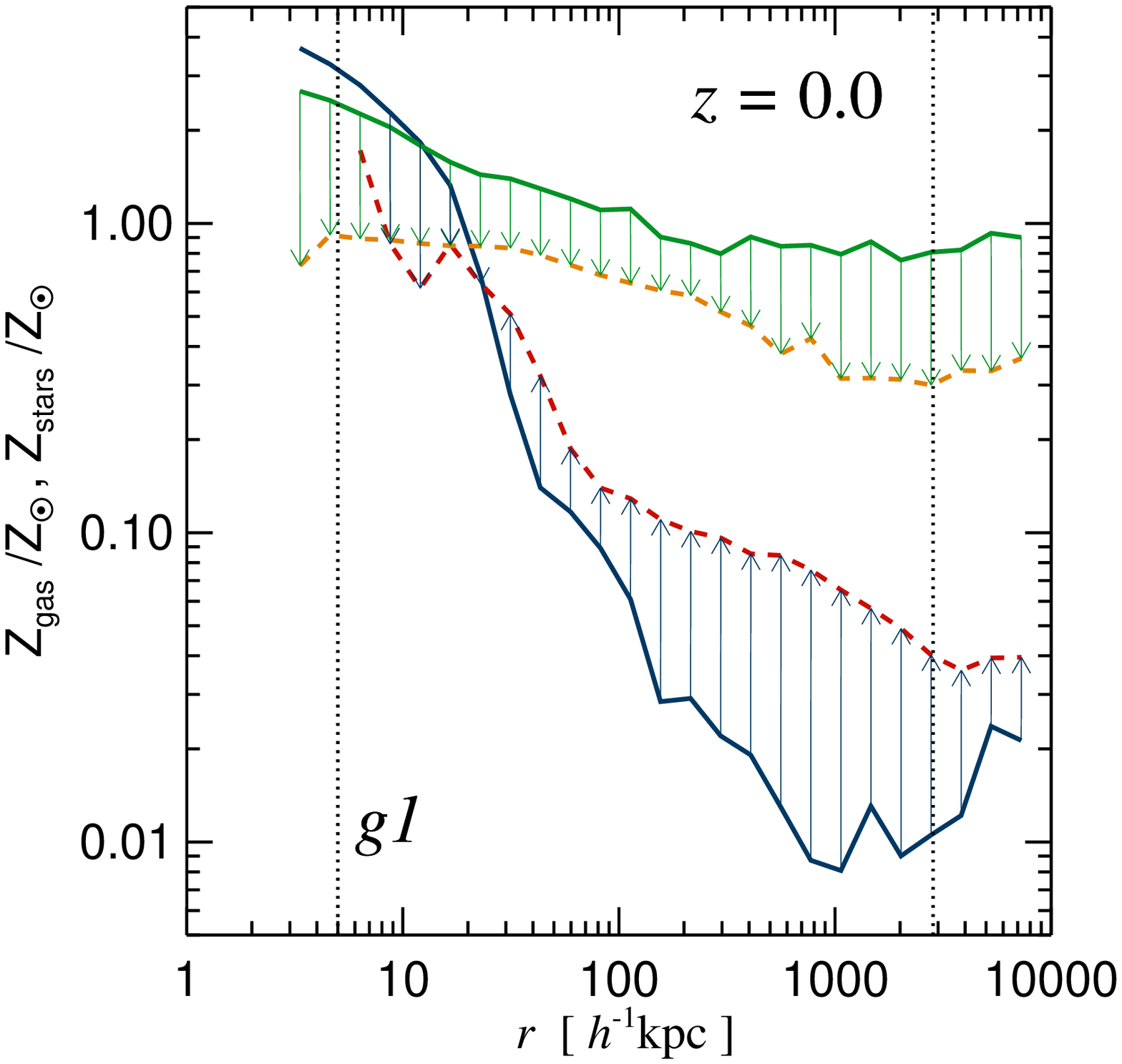,width=8.5truecm,height=8.5truecm}}}
\caption{Radial profiles of mass-weighted gas metallicity and stellar
  metallicity of the g676 (left-hand panel) and g1 (right-hand panel)
  galaxy cluster simulations at $z=0$. Continuous lines show the
  metallicity profiles without AGN heating, while dashed lines are for
  the runs with AGN feedback. The arrows illustrate how the ICM
  metallicity is affected by the BHs. The stellar metallicity is
  reduced at all radii, while the gas metallicity shows a tilt in the
  profile, decreasing in the inner regions and increasing for $r > 30
  h^{-1}{\rm kpc}$, while still exhibiting a residual gradient.}
\label{g676_Z}
\end{figure*}

\begin{figure*}
\centerline{\vbox{
\psfig{file=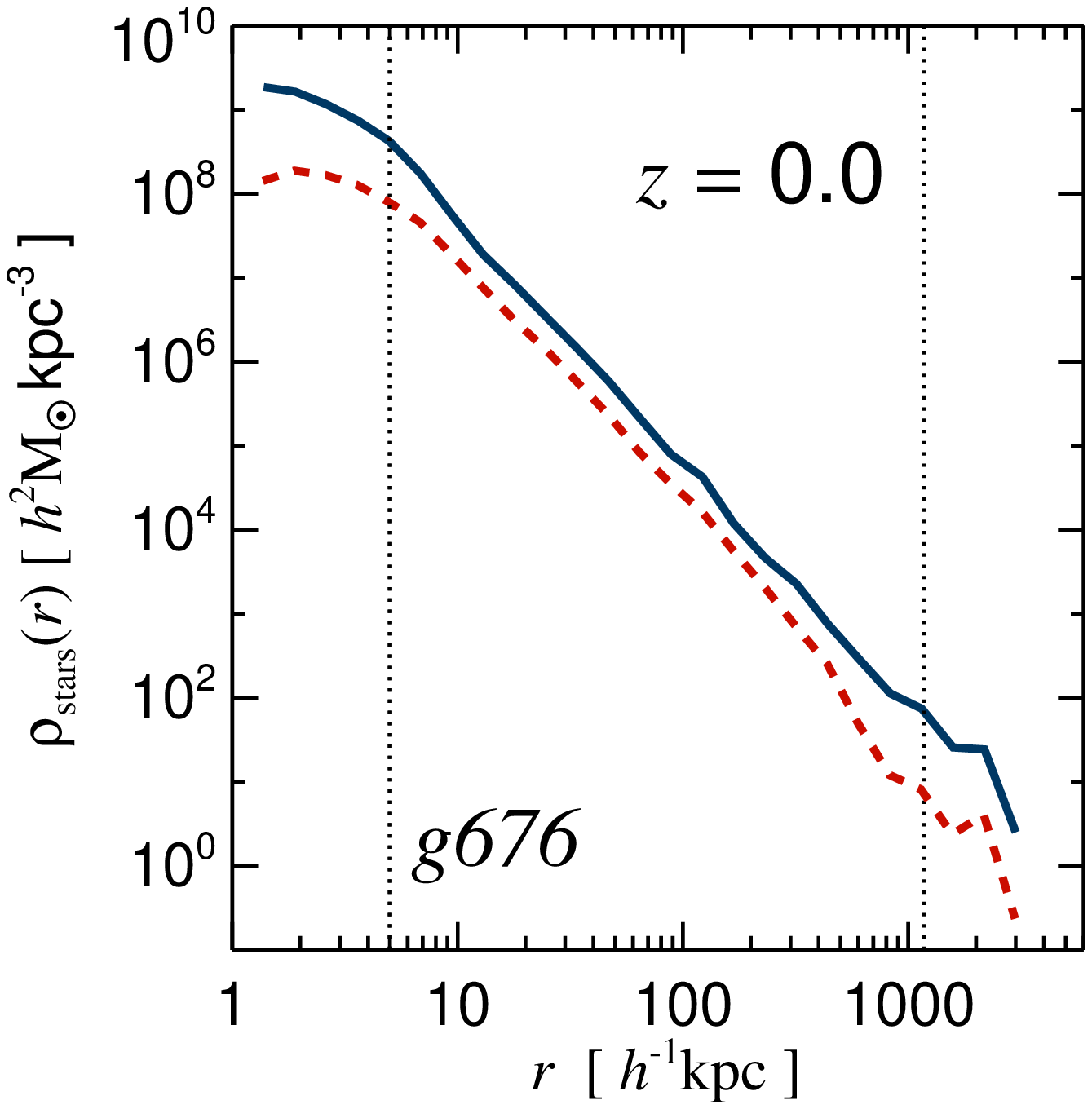,width=8.5truecm,height=8.5truecm}
\psfig{file=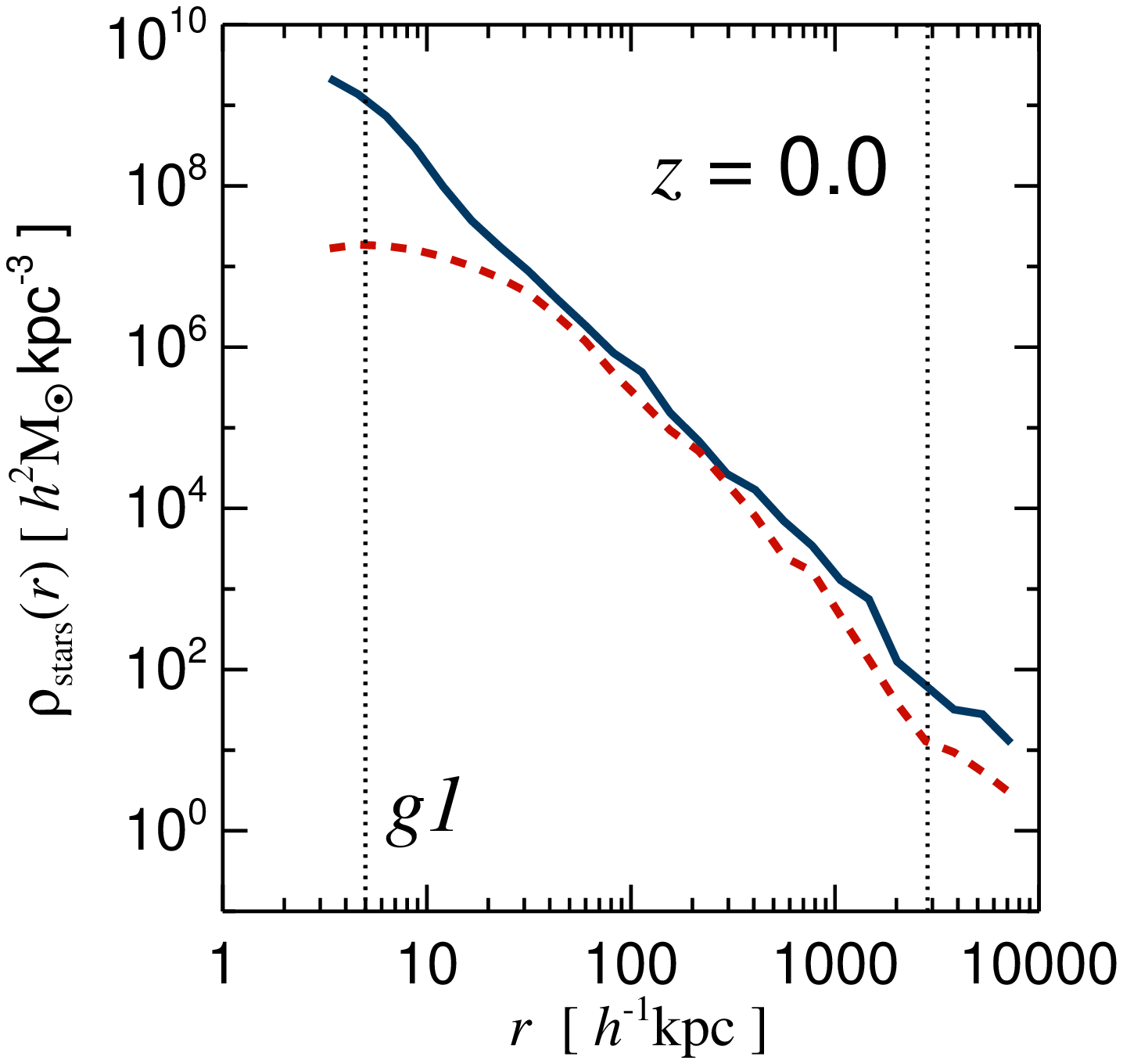,width=8.5truecm,height=8.5truecm}}}
\caption{Stellar density profile of the g676 (left-hand panel) and g1
  (right-hand panel) galaxy cluster simulations at $z=0$. Continuous
  blue lines show the case without AGN heating, while dashed red lines
  are for simulations where AGN feedback is included. It can be seen
  that the stellar density is decreased at all radii, and in the case
  of the g1 cluster, the central stellar density develops a core when
  AGN feedback is operating.}
\label{g676_rhostars}
\end{figure*} 

\subsection{The impact of AGN heating on the ICM}\label{ICMproperties} 

We here analyse how AGN feedback affects the intracluster medium and the
stellar properties of the cluster galaxies. In Figure~\ref{g676_profiles}, we
show radial profiles of gas density, temperature and entropy of the g676
cluster at three different epochs, both without (continuous blue lines) and
with AGN heating (dashed red lines). Additionally, we have explored a model in
which the BHAR was not limited to the Eddington rate, which is drawn with a
dotted green line. It can be seen that at all redshifts considered, the
central gas density is reduced significantly as a result of AGN feedback. We
also observe that the gas temperature in the central regions is fluctuating
around the ``isothermal'' value, sometimes for short time intervals increasing
towards the centre, but most of the time being roughly constant or
decreasing. These trends in gas temperature are due to the periodic nature of
the bubble feedback. If a very energetic bubble is injected in the innermost
cluster regions, the gas temperature increases for a short time, and reduces
the BHAR considerably. After some time has elapsed, the gas will begin to cool
again such that the central gas temperature will gradually start to decrease
towards the centre. When enough gas has cooled off from the hot cluster
atmosphere and has become available for accretion onto the supermassive BH, a
new bubble will be triggered, establishing a self-regulated loop for the
growth of the BH and the heating of the ICM. In some sense, this feedback loop
acts as a thermometer for the ICM, preventing it to develop strong cooling
flows.

On top of our simulated gas temperature profiles, we plot the slopes
of the central temperature profiles of the cool core clusters recently
found by \cite{Sanderson2006}, showing a very encouraging agreement. Indeed,
a number of observational studies \citep[e.g.][]{DeGrandi2002,
Vikhlinin2005, Dunn2006, Pratt2007} have shown that the central
temperature profiles of relaxed, cool core clusters decrease with
decreasing radius, while at the same time these systems require an
additional source of heating in order to avoid excessive cooling
flows. Theoretically, it is not trivial to explain this observed
feature of the central cluster temperature -- a self-regulated
feedback mechanism is apparently needed which is sensitive to the
local properties of the intracluster gas. Our model contains such a
mechanism and represents to our knowledge the first successful attempt
to simulate such a process in a cosmological environment.
    
As a result of the non-gravitational bubble heating, the ICM entropy
is increased in the central regions, but maintains a monotonically
increasing radial profile. Given the temperature of this cluster, the
overall shape and the normalization of the simulated entropy profile
are consistent with observational findings \citep[e.g.][]{Pratt2003,
Ponman2003}. Observationally, it is found that the entropy scales with
radius roughly as $S \propto r^{1.1}$ in the cluster outskirts, while
departures from this scaling are seen in the central
regions, out to  $r \sim 0.1-0.2\,R_{\rm 200}$. As can be noticed from the
bottom row of Figure~\ref{g676_profiles}, our simulated entropy
profiles depart from this scaling out to somewhat larger radii, and
this is also found for the `g1' cluster simulation which has quite
similar radial profiles. However, to decide whether this constitutes a
real discrepancy between the simulations and observations, a larger
set of simulated galaxy clusters is required, which we plan to compile
in forthcoming work.

Looking at the radial gas profiles for $r > 400\, h^{-1}{\rm
 kpc}$ it is clear that the AGN heating does not significantly change
 the ICM properties at larger radii. Thus, heating of the cluster
 outskirts does not appear to be relevant for $z < 0.5$, confirming
 our findings in Section~\ref{Outskirts}. Moreover, the radial
 profiles of our simulated galaxy cluster at $z=0.5$ do not differ
 significantly from the ones at $z=0$. This is consistent with
 observational findings \citep{Bauer2005}, where already formed cool
 core clusters are seen at $z \sim 0.15-0.4$, with similar properties
 to the local population.

\begin{table*}
\bc
\begin{tabular}{lccccccc}
\hline
\hline
Cluster & $M_{\rm stars}$ & $M_{\rm gas}$ &$M_{\rm gas,\,cold}$ & SFR & $M_{\rm r}$ & $u-r$ & $g-r$ \\
 & [$\,h^{-1}{\rm M}_\odot\,$] & [$\,h^{-1}{\rm M}_\odot\,$] &
[$\,h^{-1}{\rm M}_\odot\,$] & [$\,{\rm yr}^{-1}{\rm M}_\odot\,$] & & & \\
\hline \hline
g676\_csf   & $1.3\times10^{12}$ & $2.3\times10^{10}$ &
$1.9\times10^{10}$ & $30$ & $-24.64$ & $1.83$ & $0.59$ \\ 
g676\_csfbh & $3.4\times10^{11}$ & $3.5\times10^{9}$ &
$1.2\times10^{9}$ & $0$ & $-22.55$ & $2.68$ & $0.84$ \\ \hline 
g1\_csf & $4.2\times10^{12}$ & $8.4\times10^{10}$ &
$7.8\times10^{10}$ & $163$ & $-26.37$ & $0.76$ & $0.29$ \\ 
g1\_csfbh & $3.4\times10^{11}$ & $7.5\times10^{9}$ &
$2.5\times10^{8}$ & $0$ & $-22.71$ & $2.64$ & $0.83$ \\ 
\hline
\hline
\end{tabular}
\caption{Properties of the cD galaxy of the g676 and g1 galaxy
  clusters at $z=0$. The listed values refer both to simulations with
  cooling and star formation only (denoted with ``csf''), and to the
  runs where AGN heating was included (labelled ``csfbh''). From the
  second to the fourth column, we give the stellar mass, the gas mass
  and the cold gas mass with $T < 1\,{\rm keV}$ of the cD galaxy.  The
  star formation rate of the cD galaxy is found in the fifth column,
  while the $M_{\rm r}$ magnitude and the $u-r$ and $g-r$ colours are
  given in the last three columns, respectively.
\label{tab_cDpar}}
\ec
\end{table*}

In Figure~\ref{g676_Z}, we show radial profiles of gas and stellar
metallicity, in Solar units, for the g676 and g1 cluster simulations
at $z =0$. The continuous lines show the metallicity profiles without
AGN feedback, while the dashed lines illustrate how these profiles
change when our BH model is ``switched-on''. The stellar metallicity
is reduced at all radii, while the gas metallicity shows a tilt: it is
reduced in the innermost regions, for $r < 30\, h^{-1}{\rm kpc}$,
while it is increased for larger radii. These changes in the
metallicity gradients are caused by two effects: first, due to the AGN
feedback the total number of stars is reduced, not only in the central
regions, but all over the cluster, as illustrated in
Figure~\ref{g676_rhostars}; second, some of the metals accumulated in
dense, star-forming regions are expelled into the hot ICM. This
mechanism drives the tilt in the gas metallicity, increasing the metal
content of the hot intracluster gas.

\begin{table*}
\bc
\begin{tabular}{cccccccc}
\hline
\hline
Simulation & $L_{\rm box}$ [$\,h^{-1}{\rm Mpc}\,$] & $N_{\rm part}$ & $m_{\rm DM}$
[$\,h^{-1}{\rm M}_\odot\,$] & $m_{\rm gas}$  [$\,h^{-1}{\rm
    M}_\odot\,$] & $z_{\rm start}$ & $z_{\rm end}$ & $\epsilon$
[$\,h^{-1}{\rm kpc}\,$] \\
\hline
R1 & $25$  & $2\times176^3$  & $1.72\times 10^8$ & $0.35\times 10^8$ &
$100$ & $1$ & $3.0$ \\
R2 & $25$ & $2\times256^3$ & $5.58\times 10^7$ & $1.14\times 10^7$ &
$100$ & $1$ & $2.0$ \\
R3 & $25$ & $2\times384^3$ & $1.65\times 10^7$ & $0.34\times 10^7$ &
$100$ & $1$ & $1.3$ \\
\hline
\hline
\end{tabular}
\caption{Numerical parameters of our cosmological simulations in
  periodic boxes. The values listed in the second to the fifth column
  refer to the size of the box, to the number of gas and dark matter
  particles and to their masses. The last three columns give the
  initial and final redshifts of the runs, and their gravitational
  softening lengths $\epsilon$.
\label{tab_boxpar}}
\ec
\end{table*}

The effect of AGN heating on the metal production and mixing in
clusters brings the simulated gradients into a much better agreement
with observational results. However, the simulated gas metallicity in
the cluster outskirts appears still too low in comparison with the
metallicity levels of cool core clusters at similar radii
\citep[e.g.][]{DeGrandi2001}. This discrepancy may in part be
attributed to the fact that in these simulations we have not included
feedback effects from galactic winds and outflows powered by star
formation. As we have shown in \cite{Sijacki2006a}, galactic winds
help spreading metals throughout the cluster environment, especially
into the outer parts. Another restriction of our model for metal
enrichment lies in its highly simplified treatment of supernovae,
where we neglect the time delay of supernovae type Ia.  More
sophisticated models of chemical enrichment
\citep[e.g.][]{Tornatore2004, Scannapieco2005, Kobayashi2007} that
take this into account might therefore help as well.

We now turn to the effects of AGN feedback on the stellar properties of
the simulated clusters. In Figure~\ref{g676_rhostars}, it can be seen
that the stellar density is reduced at all radii in both clusters, as
we have anticipated above. This reduction is caused not only by the
activity of the central cluster BH, but also by other BHs that are
harboured at the centres of individual galaxies throughout the cosmic
evolution. As we will discuss in more detail in
Section~\ref{Galprop}, when a BH at the centre of a given galaxy becomes
massive enough to influence its host, it acts on the stellar
properties by reducing both the instantaneous SFR as well as the 
integrated total stellar mass by a significant amount.

In Table~\ref{tab_cDpar}, we list a number of properties we have measured
for the central cluster galaxy. For the identification of the cD
galaxy, we have simply taken its radius to be $20\, h^{-1}{\rm kpc}$, based
on the apparent break in the stellar density profile at this radius. A
number of interesting results can be noticed from the values in the
table: due to the AGN feedback, the stellar mass of the cD galaxy is
significantly reduced, as well as the central total gas mass. There is
some residual cold gas in the cD galaxy, but star formation is
completely quenched. This has an immediate and important consequence
for the photometric colours of the cD galaxies. In order to estimate
the colours, we have used the stellar population synthesis models of
\cite{Bruzual2003} and computed the rest-frame magnitudes in the SDSS
bands, assuming Solar metallicity and a Chabrier initial mass
function. Both cD galaxies in our cluster simulations become much less
luminous in the $r$, $u$ and $g$-bands as a result of AGN feedback,
such that the $u-r$ and $g-r$ colour indices are substantially
increased. This reflects the fact that the stellar populations of the
cD galaxies have become much older in our BH model, with very little
recent star formation. With these properties, the simulated cD
galaxies are in quite good agreement with observational findings
\citep[e.g.][]{Linden2006}, in terms of their stellar mass, their SFR
and their colours.

Finally, we briefly discuss whether the Eddington limit that we have imposed
on the BHAR has a significant effect on the BH growth and the feedback in
clusters. For this purpose we have rerun the g676 galaxy cluster simulation
with exactly the same feedback parameters, but relaxing the maximum accretion
rate assumption. In Figure~\ref{g676_profiles}, we have included results for
the radial profiles of gas density, temperature and entropy in this run as
dotted green lines.  It can be seen that the ICM properties are modified
slightly more strongly in this case with respect to our default model,
i.e.~the AGN feedback effects appear to be somewhat stronger.  A
closer look at the detailed mass growth and accretion rate history of
the BH as function of redshift shows that the feedback is regulating
the BHAR even without an upper limit in the form of the Eddington
rate. In fact, the total mass of all BHs at $z=0$ in the simulated
volume increases only by roughly $20\%$ if the Eddington limit is
disregarded.  The most massive BH in the cluster centre shows an
increase of its mass of the same order.  This additional growth of the
BH mass mostly comes from brief episodes of very high accretion rate
 (typically reaching a few times up to ten times $\dot
M_{\rm Edd}$), which are quickly terminated by the onset of AGN
heating. Without an explicit Eddington limit, a bit more growth can
happen before the `suicidal' AGN activity shuts off the accretion, but
the limit itself is not required for establishing a stable
self-regulation loop.  The Eddington limit therefore does not seem to
pose an important restriction on the growth of BHs in
clusters. However, this may still be quite different in the early
phases of BH accretion at high redshifts, when the BH mass is so low
that feedback effects are weak. Then the Eddington limit sets the
shortest growth timescale that can be realized, which makes it a
challenge to grow BHs quickly enough to the observationally inferred
large masses already at very high redshifts.  We plan
to address this interesting problem in full cosmological simulations
in future work.

\begin{figure*}
\centerline{\vbox{
\psfig{file=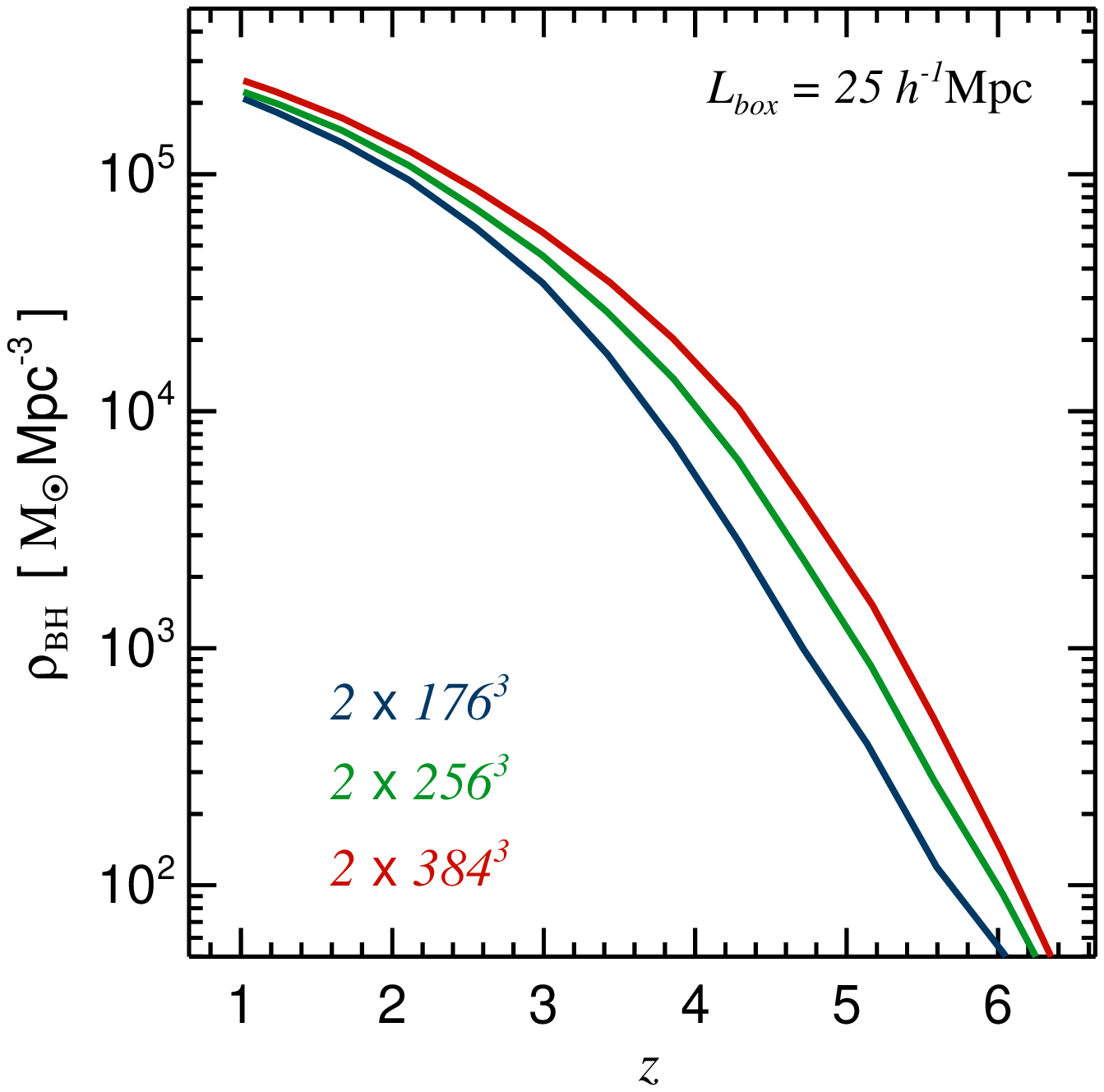,width=8.5truecm,height=8.5truecm}
\psfig{file=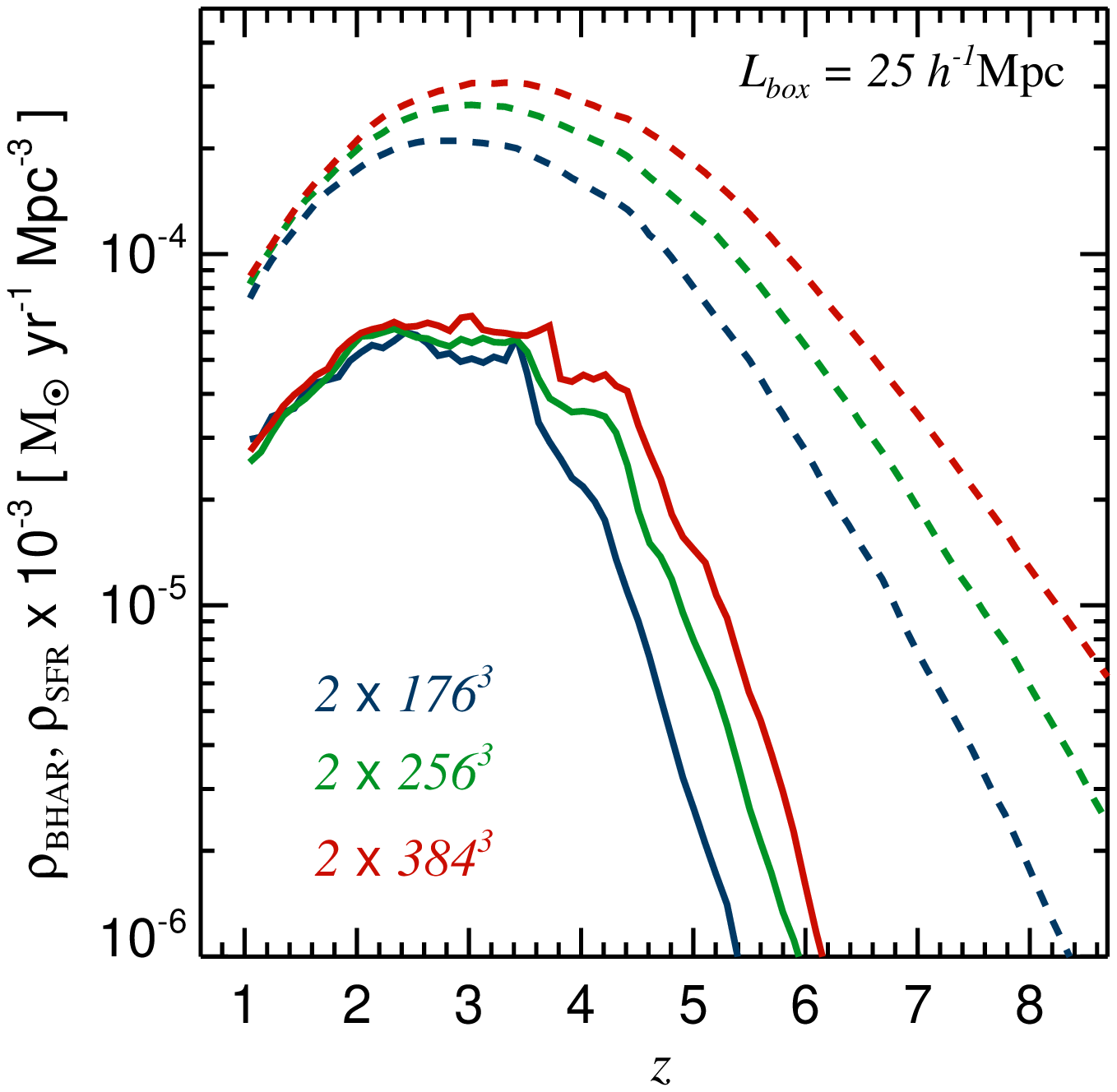,width=8.5truecm,height=8.5truecm}}}
\caption{Redshift evolution of the comoving BH mass density in a
  cosmological box of $25 \,h^{-1}{\rm Mpc}$ on a side (left-hand
  panel). The three different curves show runs with increasing
  resolution as labelled in the panel. The right-hand panel gives the
  BHAR (continuous lines) and SFR (dashed lines) densities as a
  function of redshift for the same set of runs. For plotting
  purposes, $\rho_{\rm SFR}$ has been rescaled by a factor of
  $10^3$ to fit onto the same scale. It can be seen that numerical
  convergence is achieved at low redshift, where BHAR and SFR
  densities of the different runs asymptotically approach each other
  with increasing resolution.}
\label{rhobh}
\end{figure*}

\section{Simulations of galaxy formation with AGN feedback} \label{Galform}

In this section we analyse BH growth and feedback in simulations of
cosmic structure formation in homogeneously sampled volumes.  We are
mainly interested in the question whether our numerical model for a
two-mode AGN growth produces realistic results for a range of object
masses, from the scales of isolated galaxies to the ones of massive
galaxy clusters.  In Section~\ref{Cosmological}, we have already
confirmed that the BH model works well for clusters of galaxies, where
it in fact drastically improves the properties of simulated galaxy
clusters with respect to observations, and yields at the same time
plausible evolutionary histories for the BH masses and the accretion
rates. Here, we want to see whether this success also carries over to
the scale of individual galaxies, where the ``quasar'' mode of feedback
will be dominant. In particular, we want to establish whether our
unified model for AGN feedback maintains and extends the successes we
have found in \citet{DiMatteo2007} for the high-redshift galaxy
population in a model that only accounts for the ``quasar'' mode.

We have performed hydrodynamical simulations of a cosmological box
with $25 \, h^{-1}{\rm Mpc}$ on a side, at three different mass
resolutions, as summarized in Table~\ref{tab_boxpar}. The small
box-size allows us to study comparatively small galaxies with stellar
masses in the range of $\sim 10^8 {\rm M}_\odot$ to $\sim 10^{11} {\rm
M}_\odot$.  However, since the fundamental
mode of this small box becomes non-linear at around $z=1$, we had to
stop the simulations at this redshift, since they would not be
representative any more for the low-redshift universe.  For the
initial conditions, we have used a `glass' as unperturbed particle
load and the power spectrum fit of \cite{Eisenstein1999} for imposing
initial perturbations with WMAP-3 cosmological parameters
\citep{Spergel2006}. In particular, we have adopted a $\Lambda$CDM
cosmology with $\Omega_{\rm m}=0.26$, $\Omega_{\rm b}=0.044$,
$\Omega_\Lambda=0.74$, $\sigma_8=0.75$, and $H_0=71 \, {\rm
km\,s}^{-1}{\rm Mpc}^{-1}$ at the present epoch, and a  primordial
spectral index of $n_s=0.938$. 

For all three mass resolutions R1-R3 that we considered, we have
computed two runs, one only with cooling and star formation and the
other with the BH model included as well. The parameters of the BH
model were exactly the same as the ones adopted in our cosmological
simulations of galaxy cluster formation.  
For our intermediate
resolution case R2, we have carried out additional simulations to
study the influence of further parameters.  This includes a run where
we used the same cosmological parameters as in
Section~\ref{Cosmological} in order to gauge the importance  of
the cosmological model. We have also explored the relative importance
of the bubble feedback on galactic scales relative to the quasar
feedback, that we will discuss more in detail in Section~\ref{Radio}. 
Finally, we have studied the influence of including a model
for supernova-driven galactic winds with velocity $v\sim 480\,{\rm
km/s}$, which can be especially important in low mass systems, as we
will discuss below.

\begin{figure*}
\centerline{ \hbox{
\psfig{file=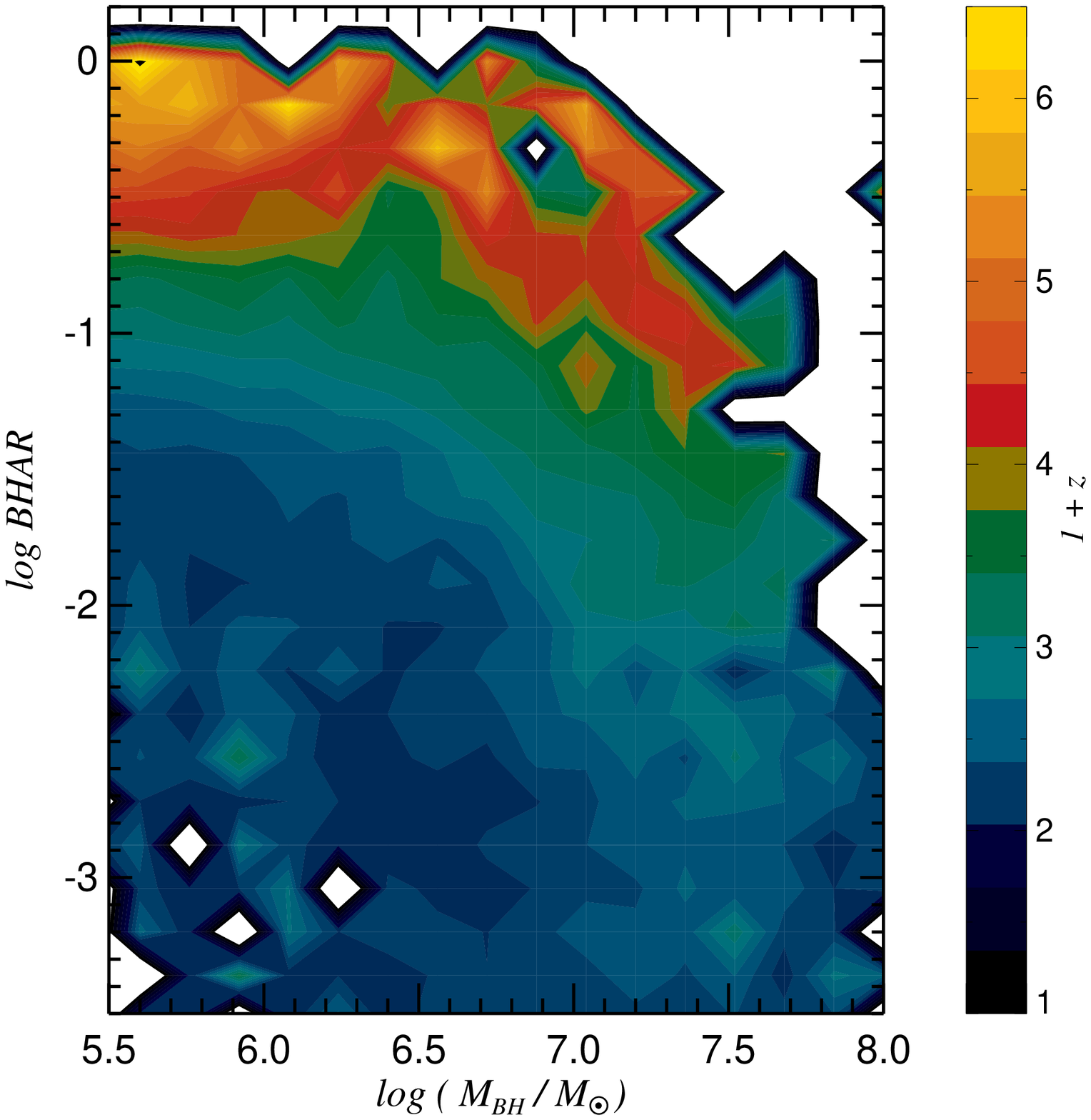,width=8.5truecm,height=8.truecm}
\psfig{file=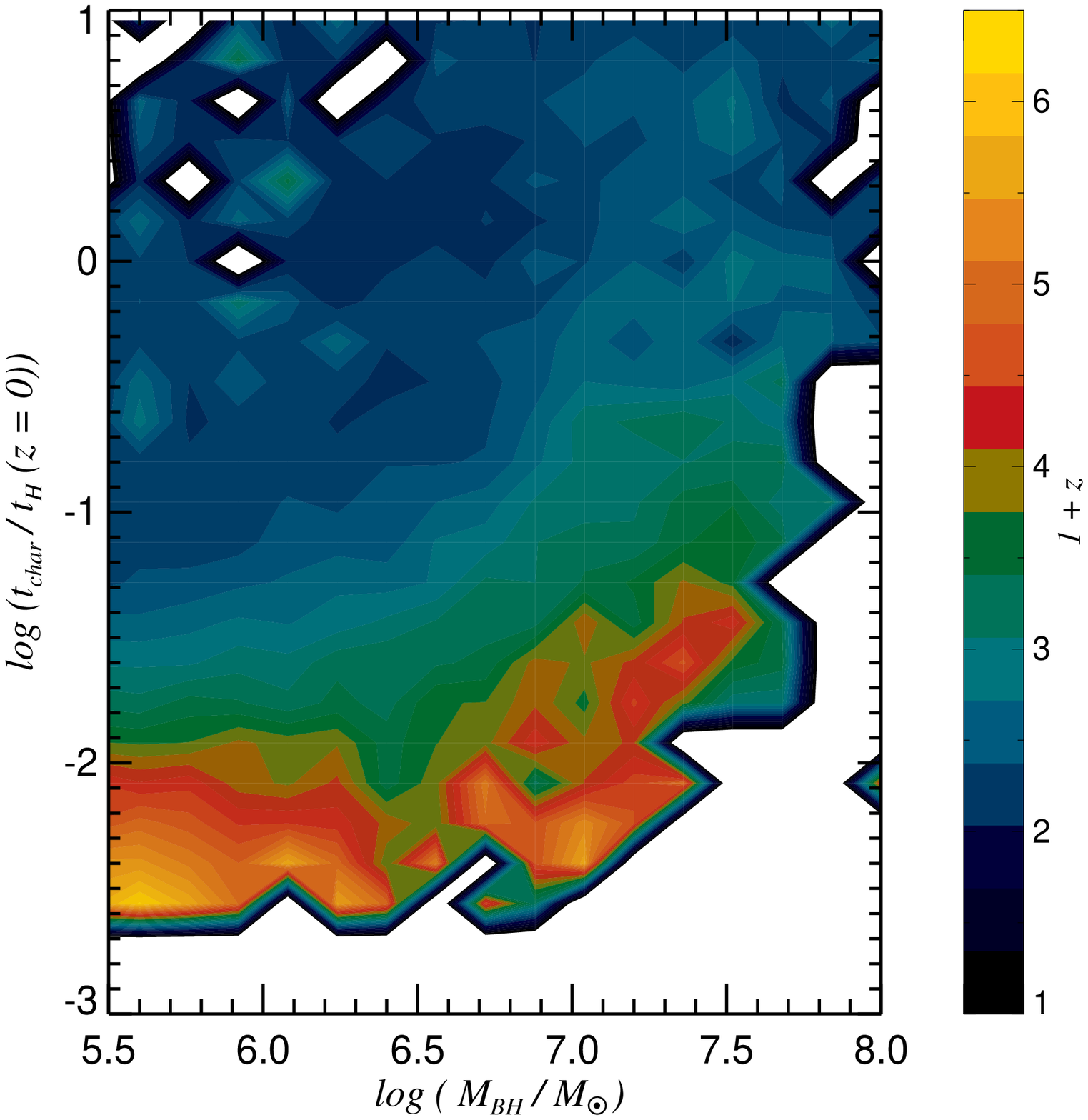,width=8.5truecm,height=8.0truecm}}}
\caption{Left-hand panel: BHAR in Eddington units as a
  function of BH mass, for all BHs belonging to the R2 cosmological
  simulation. The colour-coding expresses the redshift, and the
  redshift interval considered is from $5.0$ to $1.0$. Right-hand
  panel: Characteristic growth time of the BHs as a function of their
  mass, colour-coded according to redshift. The growth time has been
  defined as the ratio of the BH mass and the BHAR at the given epoch,
  normalized to the Hubble time at $z=0$.}
\label{box25_256}
\end{figure*}

\subsection{Black hole growth}

In Figure~\ref{rhobh}, we consider the comoving BH mass density
(left-hand panel) and BHAR density (right-hand panel) as a function of
cosmic time. For comparison, we also show the redshift evolution of
the SFR density (right-hand panel, dashed lines), which we have
rescaled by a constant factor of $10^3$ to put the curves onto the
same plot. It can be seen that the BHAR and SFR densities approach
each other asymptotically towards low redshifts when the numerical
resolution is increased. The fact that numerical convergence can be
more easily achieved at low than at high redshift is not surprising in
light of the hierarchical growth of structure. At high redshift, most
of the small mass systems that undergo star formation and BH accretion
in our highest resolution simulation are unresolved in our low
resolution simulation, which hence underpredicts the BHAR and SFR
density. However, towards later times, the mass scale where most of
the accretion happens shifts to larger mass objects, which can be
resolved in all of our simulations, such that approximate convergence
can be reached even with moderate resolution. This behaviour is also
similar to the one seen by \citet{Springel2003} in a comprehensive
simulation study of the cosmic star formation rate history.  We
consider it highly encouraging that we find such a behaviour for our
BH growth model as well, which shows that the model is numerically
well posed and can be meaningfully applied to cosmological volumes
with their comparatively poor resolution per galaxy.

Nevertheless, it is clear that the numerical resolution in our
cosmological runs remains too poor to correctly resolve the internal
structure of galaxies, in particular the thin gaseous and stellar
disks in spiral galaxies. This presumably affects the detailed time
evolution of star formation and black hole accretion in galaxy
mergers. For example, our cosmological runs will largely blend over
the time delay we see in high-resolution galaxy mergers between BH
activity and the peak of the star burst. But the final BH mass reached
in a merger is comparatively insensitive to these details and probes
primarily the depth of the gravitational potential of the forming
spheroid, which our simulations can still represent quite well. While
numerical resolution is clearly an important limitation in our
cosmological results, we therefore believe they are reasonably
accurate for the basic properties of the BH population.

Our estimated BH mass density at $z=1$ in our highest resolution run
is $2.5\times10^5{\rm M}_\odot \, {\rm Mpc}^{-3}$. This is in very
good agreement with a number of observational estimates, based both on
optical and X--ray data, which typically find values in the range of
$2-6\times10^5{\rm M}_\odot \, {\rm Mpc}^{-3}$
\citep[e.g.][]{Fabian1999, Merritt2001, Yu2002, Cowie2003,
Graham2007}. The SFR density shows a peak at $z \sim 3$, while the
BHAR density peaks at somewhat lower redshift, i.e. $z \sim 2.5$. The
BHAR and SFR then decline to lower redshifts, with the SFR density
decreasing slightly more steeply. If we extrapolate the BHAR and SFR
densities to $z=0$, the ratio of the two is of order of $2\times10^3$,
a value that is roughly in agreement with estimates from the local
population of galaxies \citep{Heckman2004}. In contrast, at high
redshift, the BHAR density is rising more steeply than the SFR
density, in broad agreement with the estimates by
\cite{MerloniRudnick2004}.

However, \cite{HopkinsRichards} find that the quasar luminosity
density declines more steeply for $z>2$ than in our simulations. It is
quite plausible that this discrepancy at high redshift is due to the
limited numerical resolution available in our cosmological
simulations, and in particular, their inability to correctly reproduce
galactic disks of the right sizes, as discussed above. While the bulk
of the BH accretion happens during major merger events in our
simulations, we expect that there is a significant contribution from
minor merger episodes or secular processes, which boost the BHAR at
high $z$ and induce a less steep rise. If the galaxies could be fully
spatially resolved, the gas would instead not be able to easily reach
the galactic centres in these minor merger events.  Nonetheless, it is
encouraging that the shapes and the peaks of the SFR and BHAR
densities we find are significantly different, clearly indicating that
the BHAR is not simply following the evolution of the SFR with
time. Finally, by comparing the simulations with and without AGN
feedback we conclude that the SFR density starts to be noticeably
reduced by BH heating for $z < 4$, and at $z=1$, this reduction is of
order $\sim 20\%$.

In \citet{DiMatteo2007}, similar results for the BH mass density
and BHAR density are obtained for a cosmological box performed at
somewhat higher numerical resolution, further supporting the trends we
find here. The peak of the BHAR density appears moderately broader
here than in \citet{DiMatteo2007}, which could be a result of our
smaller simulation box, or simply due to different choices of
cosmological parameters, or due to the absence of galactic winds in
Figure~\ref{rhobh}. As we shall discuss in detail in Section~\ref{Radio}, the
`radio mode' does not play a significant role at high redshift and
hence is unlikely to modify the shape of the BHAR evolution at the
peak of the quasar activity.

In the left-hand panel of Figure~\ref{box25_256} we show how the BHAR evolves
with redshift as a function of the BH mass. We have already considered an
analogous plot for our galaxy cluster simulations in Figure~\ref{g1_BHAR}.
While the two plots are consistent for the overlapping range of BH masses, the
most striking difference can be seen at the high BH mass end: in the galaxy
cluster simulation we saw an upturn of the BHAR at low redshifts for very
massive BHs, while here instead the BHAR keeps decreasing with increasing BH
mass at all redshifts considered. This can be understood as a consequence of
the limited size of the cosmological box. As a result of the small volume,
rare big objects are absent, and hence massive BHs residing in cluster-sized
objects are missing as well.

The characteristic time of BH growth, as a function of the BH mass, is
illustrated in the right-hand panel of Figure~\ref{box25_256}. For each BH, we
have defined the characteristic growth time as the ratio of its mass to its
accretion rate at the given redshift, normalized to the Hubble time at the
present epoch. It can be seen that the BH growth time is shortest at high
redshifts and for the smallest BHs. At $z=1$, BHs with mass of order of
$10^7{\rm M}_\odot$ reach a characteristic growth time of the order of the
Hubble time for the first time. At different redshifts, the $\log \,t_{\rm
  c}-\log \,M_{\rm BH}$ curve has a similar shape, which is consistent with
results obtained for the local galaxy population \citep{Heckman2004}.

\subsection{Effects on the intergalactic medium}\label{Galprop}

\begin{figure*}
\centerline{ \vbox{ \hbox{
\vspace{-0.5truecm}
\psfig{file=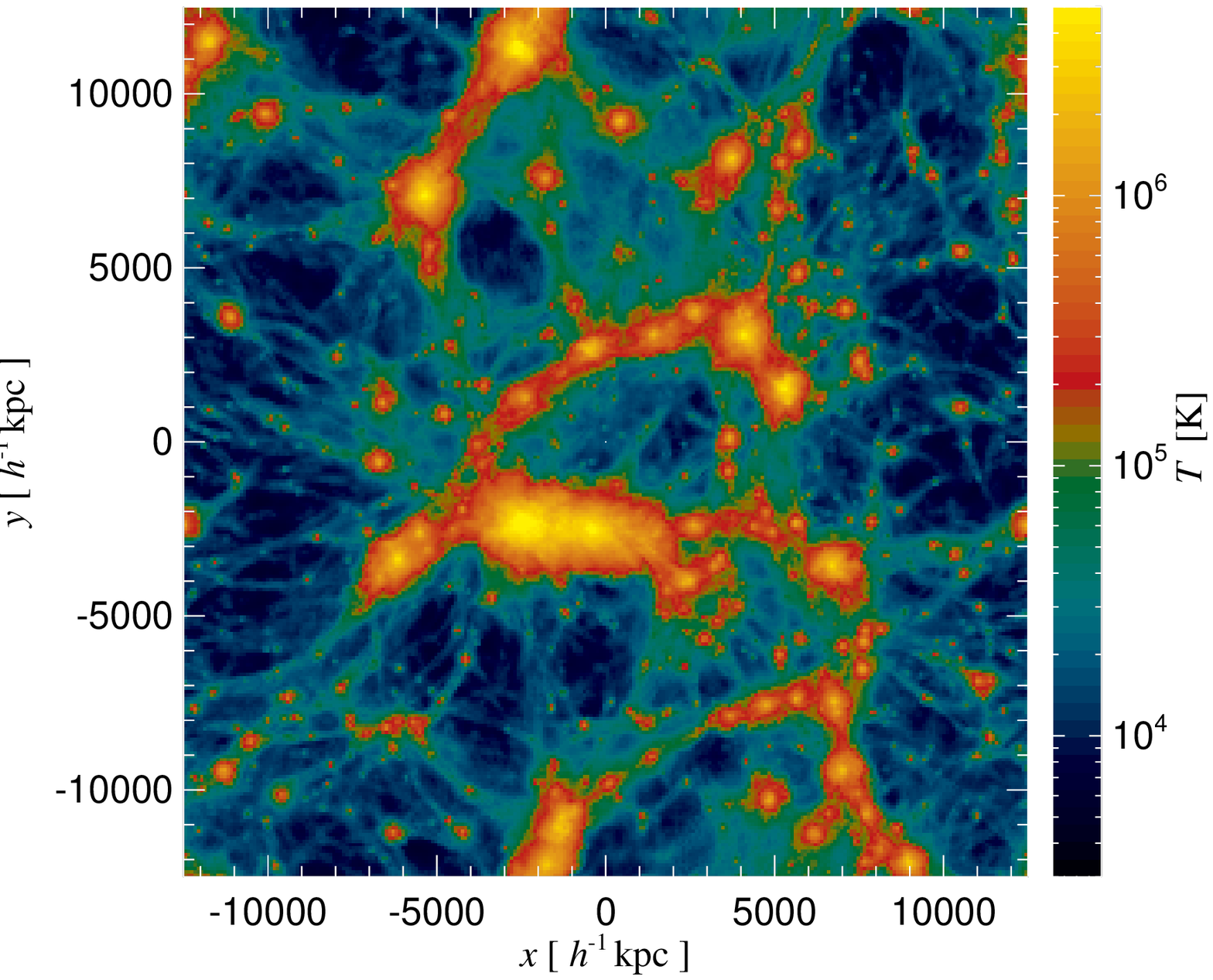,width=8.5truecm,height=8.truecm}
\psfig{file=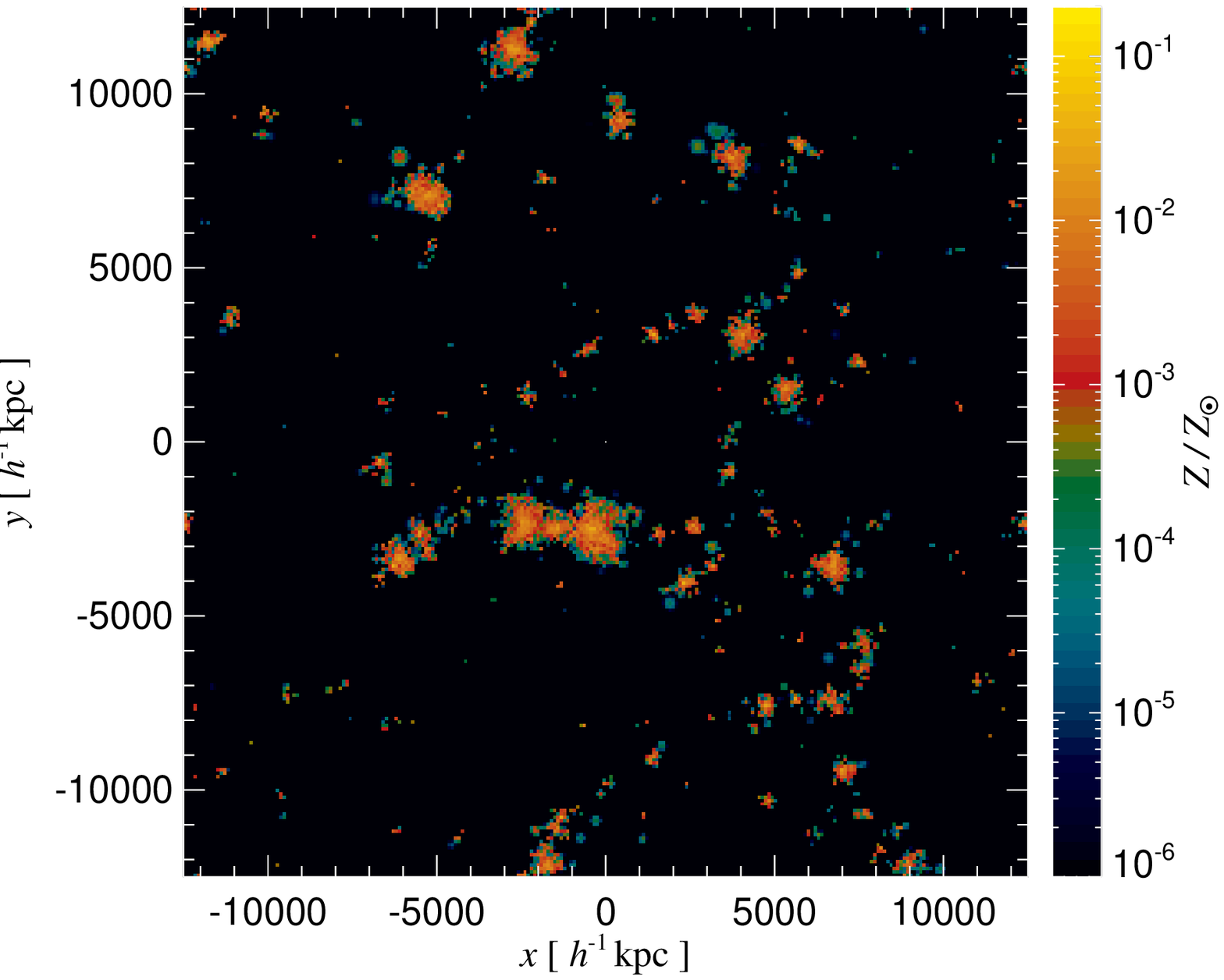,width=8.5truecm,height=8.truecm}
}
\hbox{
\vspace{-0.5truecm}
\psfig{file=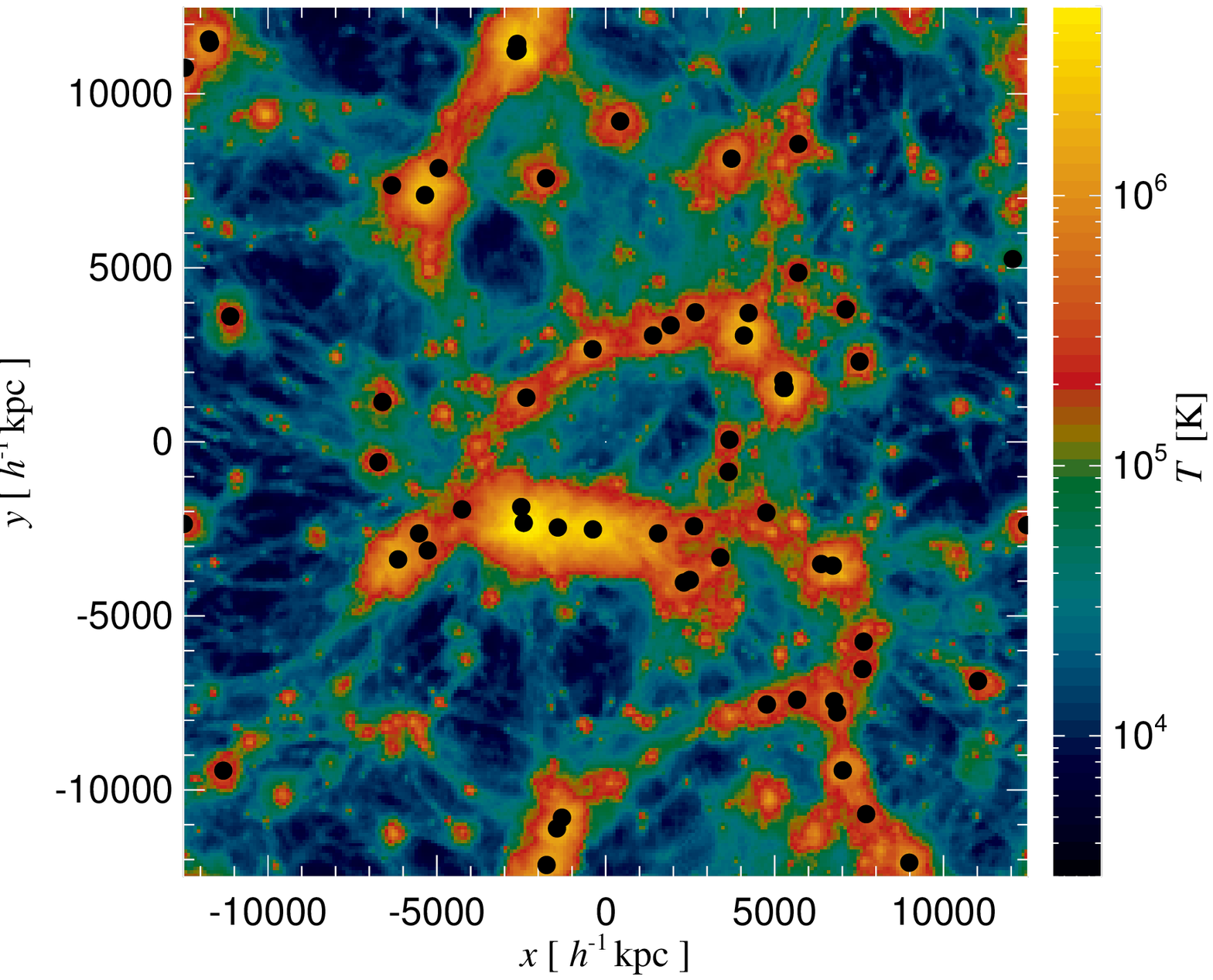,width=8.5truecm,height=8.truecm}
\psfig{file=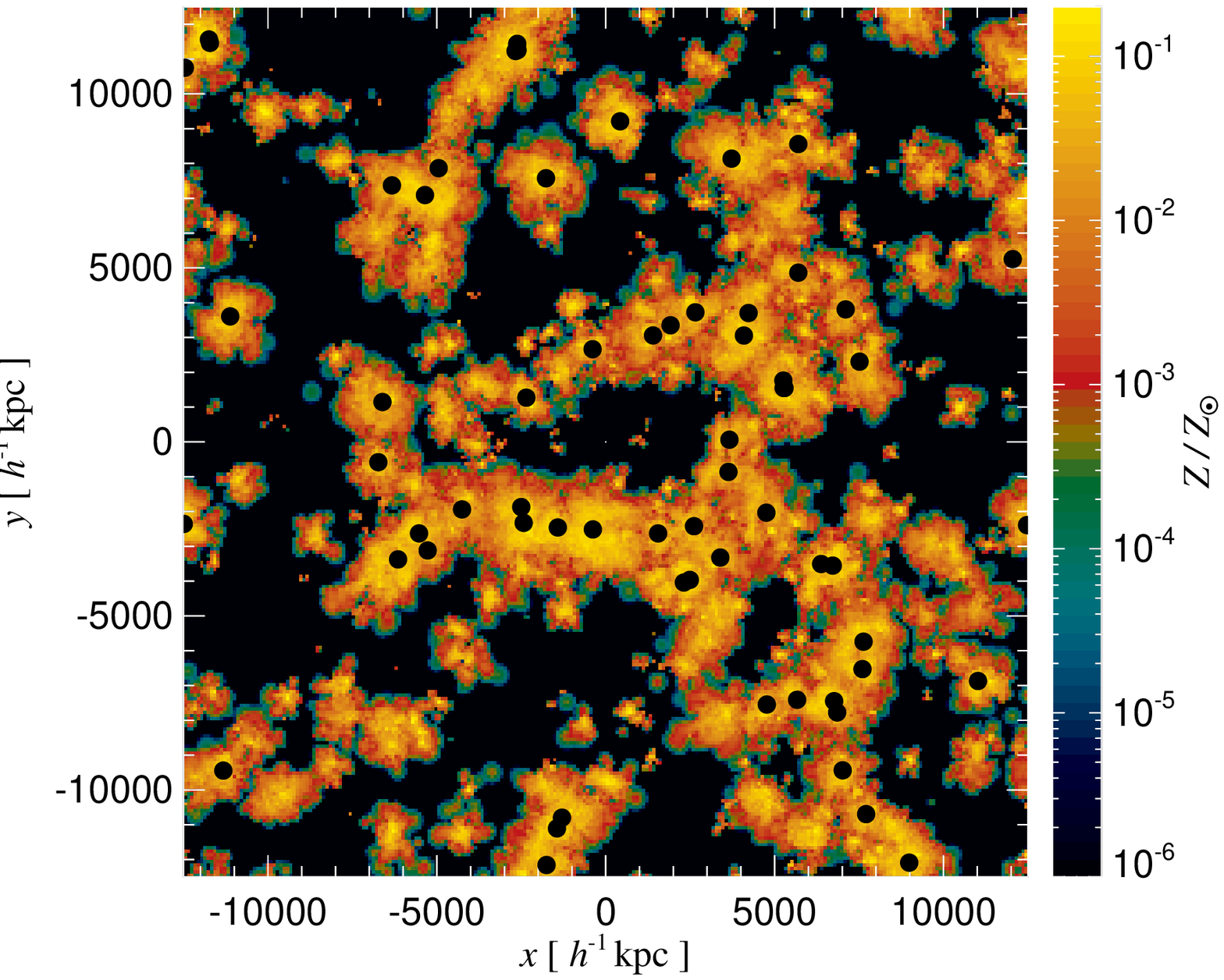,width=8.5truecm,height=8.truecm}
}
\hbox{
\vspace{-0.5truecm}
\psfig{file=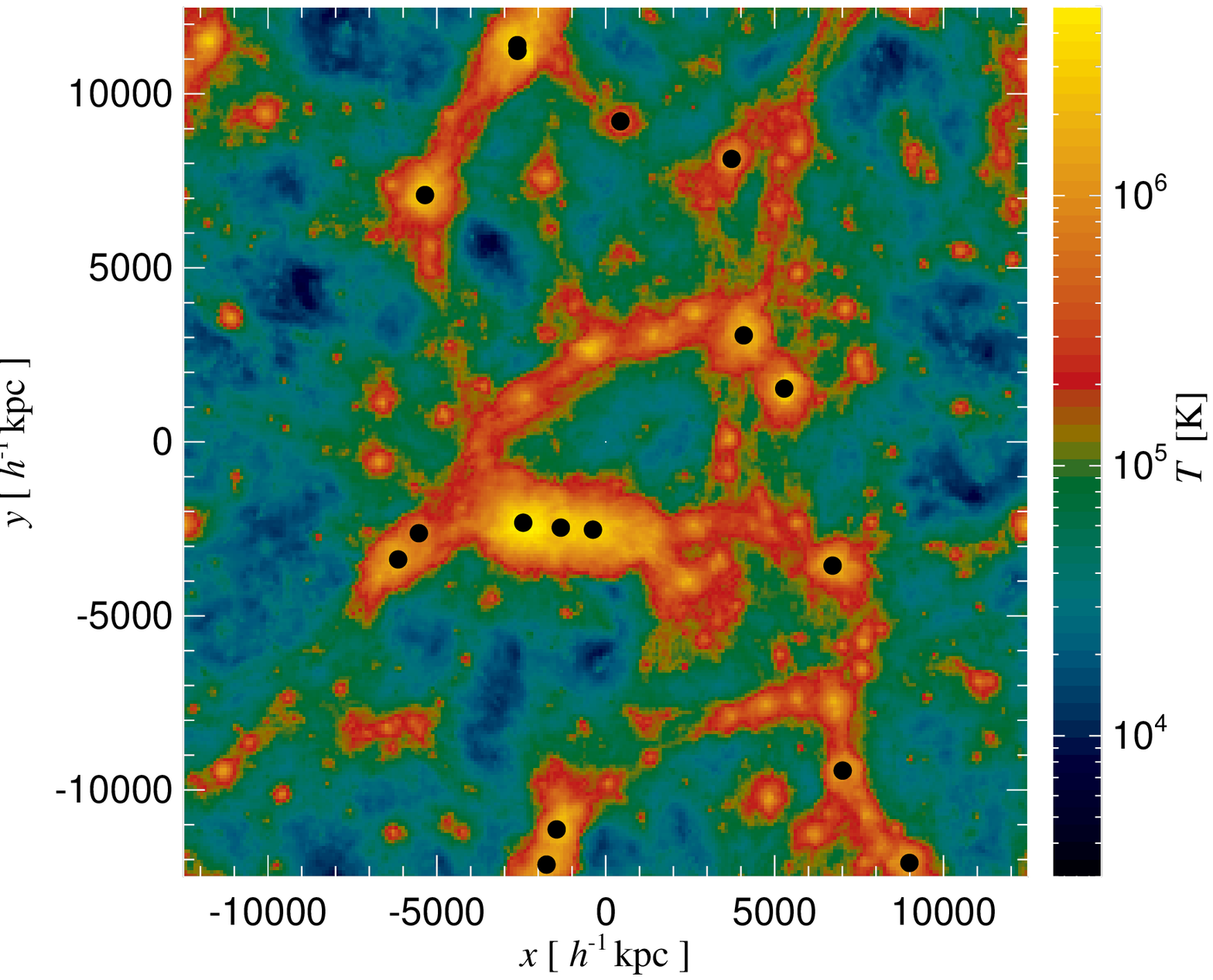,width=8.5truecm,height=8.0truecm}
\psfig{file=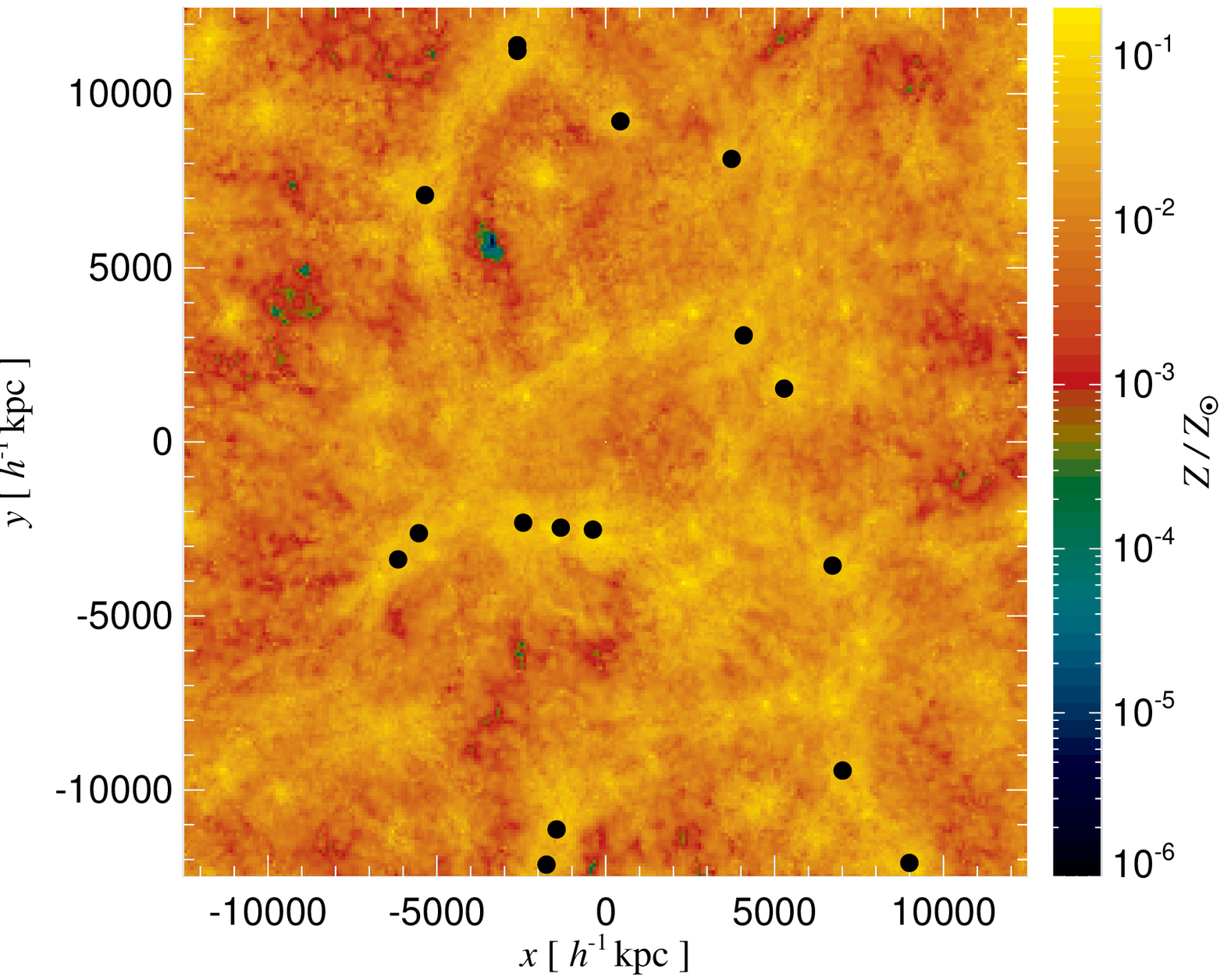,width=8.5truecm,height=8.0truecm}
}
}}
\caption{Projected mass-weighted temperature maps (left-hand panels)
  and gas metallicity maps (right-hand panels) at $z=1$ of the R2
  run. Top row: simulation with cooling and star formation only;
  middle row: simulation with additional BH model; bottom row: run
  with BH model and galactic winds ``switched-on''. The positions of
  BH particles more massive than $2\times10^7h^{-1}{\rm M}_\odot$ are
  marked with black dots.}
\label{temp_box25_256}
\end{figure*}

In Figure~\ref{temp_box25_256}, we show projected mass-weighted
temperature and metallicity maps of our cosmological box at $z=1$. In
the upper row, the run with cooling and star formation only is
plotted, while the other two rows give results for simulations where the AGN
feedback was included. In the lower row, feedback by galactic winds
was additionally included. The positions of BHs more massive than
$2\times10^7h^{-1}{\rm M}_\odot$ are marked with black dots. It can be
seen that galactic winds modify the gas temperature and metallicity
properties more significantly than the BH feedback alone. As expected,
galactic winds are especially important in low mass systems, given
that they can expel gas from objects with small escape speeds
relatively easily due to their shallow potential wells. Galactic winds
can therefore pollute a sizable fraction of the volume of the
intergalactic medium (IGM) with metals, to a level of
$10^{-3}-10^{-2}$ of the Solar metallicity, in agreement with a number
of observational findings \citep[e.g.][]{Rauch1997, Cowie1998,
Schaye2000, Ellison2000}.

However, as can be seen from Figure~\ref{temp_box25_256}, AGN
contribute to the process of metal pollution of the low density gas,
albeit with lower efficiency. If only AGN feedback is considered, we
find that metals can be transported up to a distance of $\sim 1.5\,
h^{-1}{\rm Mpc}$ from a galactic centre, but the large-scale
metallicity distribution remains extremely patchy. Thus, the main
difference relative to enrichment by galactic winds lies not only in
the level of IGM metal enrichment, but in the volume filling factor of
the metal-enriched regions.  Nevertheless, by redshift $z=3$, AGN
feedback is making an important contribution to the enrichment of the
IGM, where the characteristic size of high metallicity regions is of
the order of $\sim 1\, h^{-1}{\rm Mpc}$. However, we should point out
that the AGN impact on the metal distribution found in our simulations
is probably a lower limit, due to the fact that we cannot account for
the contribution of very bright quasars powered by massive BHs at high
redshifts.

Interestingly, even though AGN heating is affecting the gas
metallicity of the IGM, it hardly affects its mean temperature,
which is probably a reflection of the small volume filling factor of
AGN-driven outflows. In contrast, galactic winds raise the
intergalactic temperature significantly. Since this can delay the
formation of small galaxies, the number and the mass of galaxies at a
given epoch is reduced as well, which in turn slows the growth of the
BH mass density.

\begin{figure*}
\centerline{ \vbox{
\hbox{
\vspace{-0.7truecm}
\psfig{file=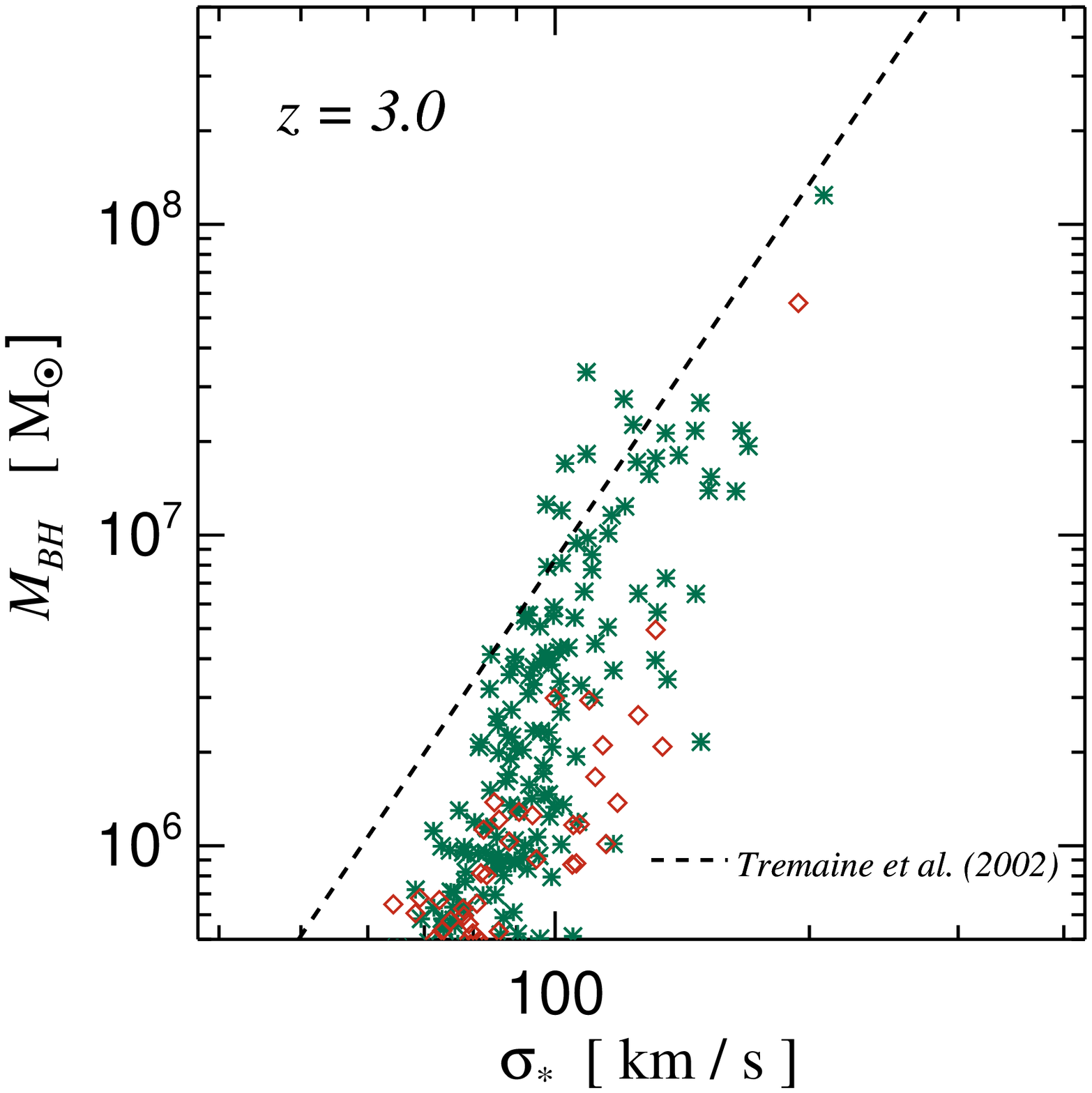,width=8.5truecm,height=8.truecm}
\psfig{file=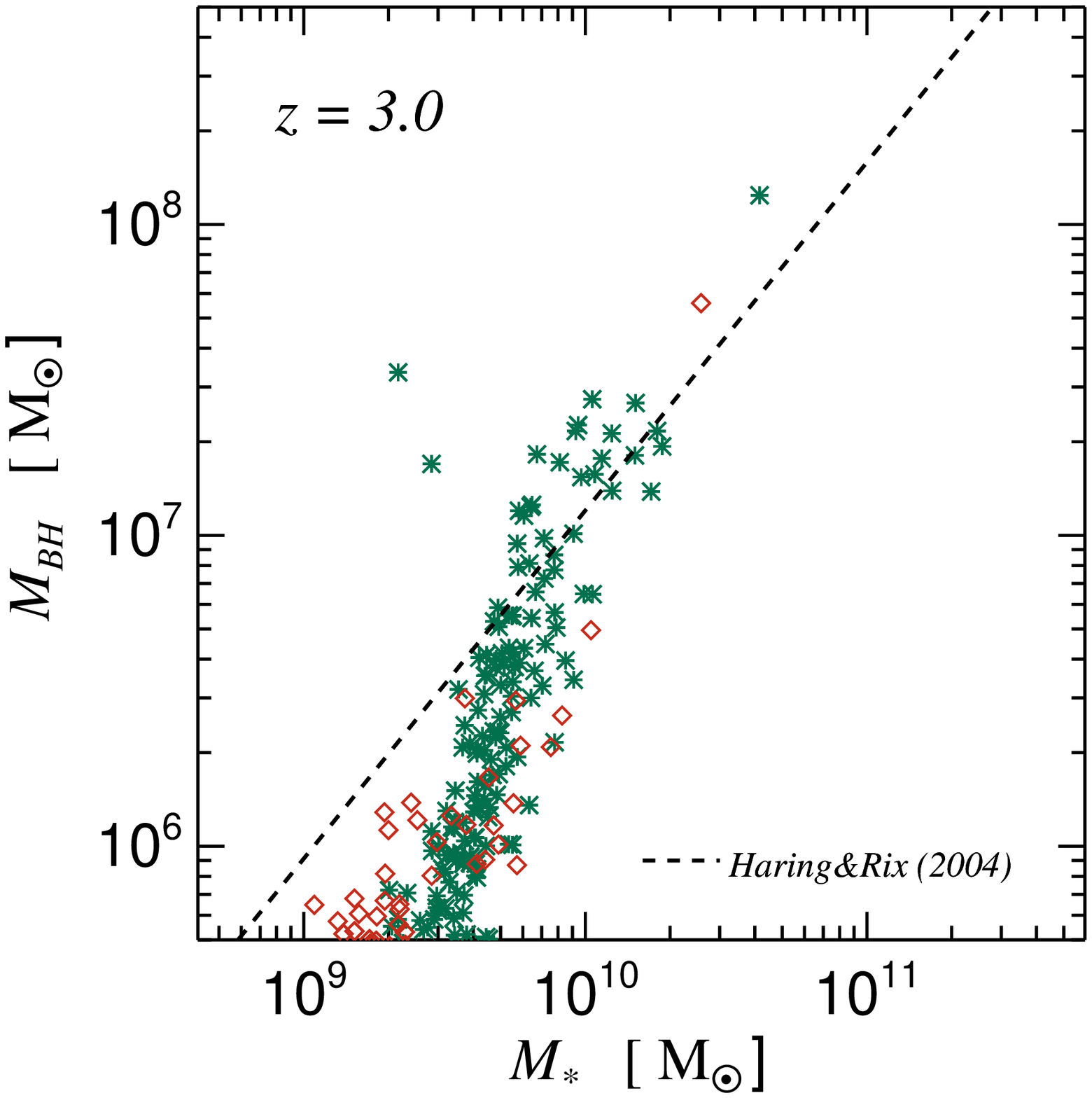,width=8.5truecm,height=8.truecm}
}
\hbox{
\vspace{-0.7truecm}
\psfig{file=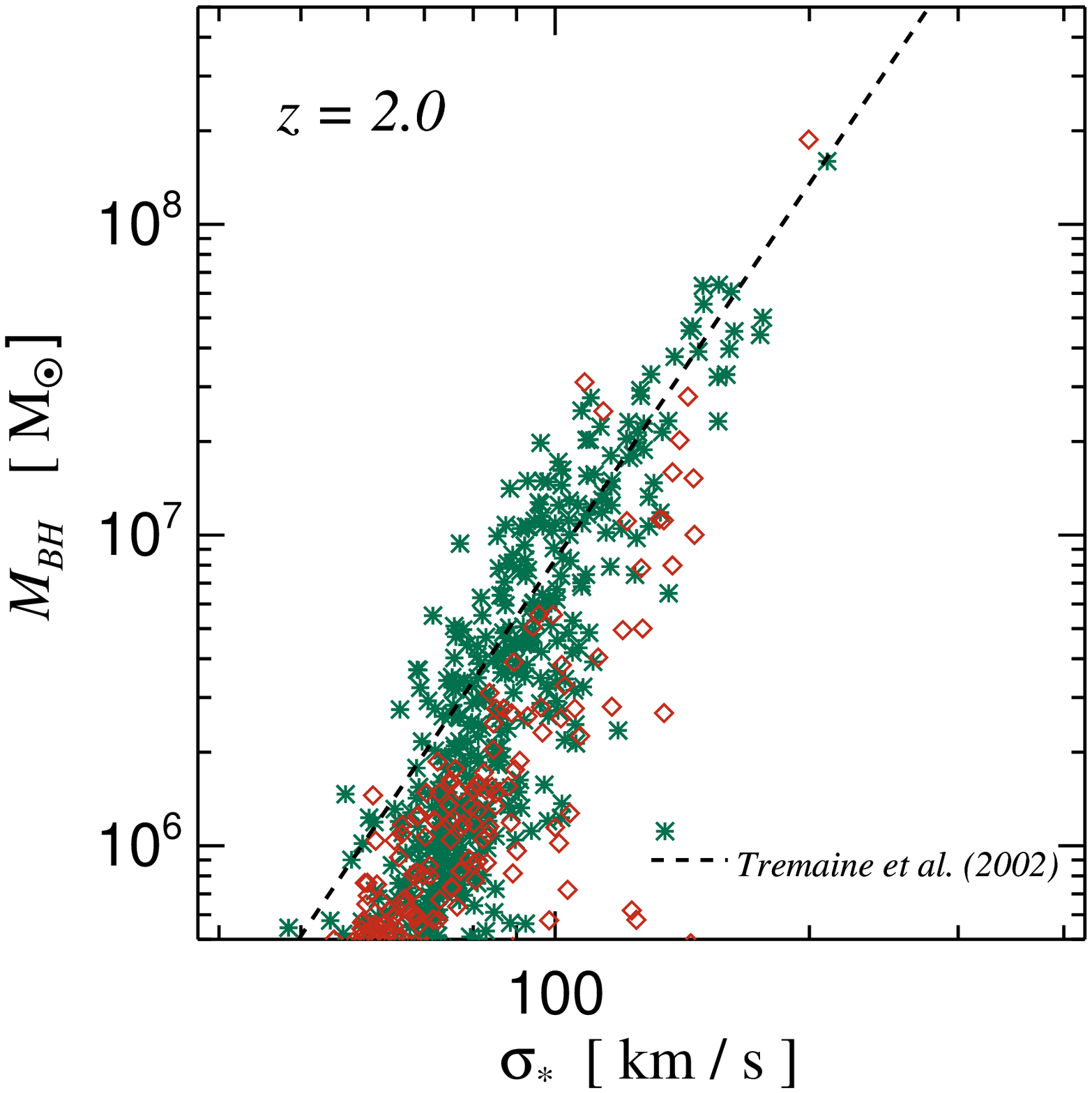,width=8.5truecm,height=8.truecm}
\psfig{file=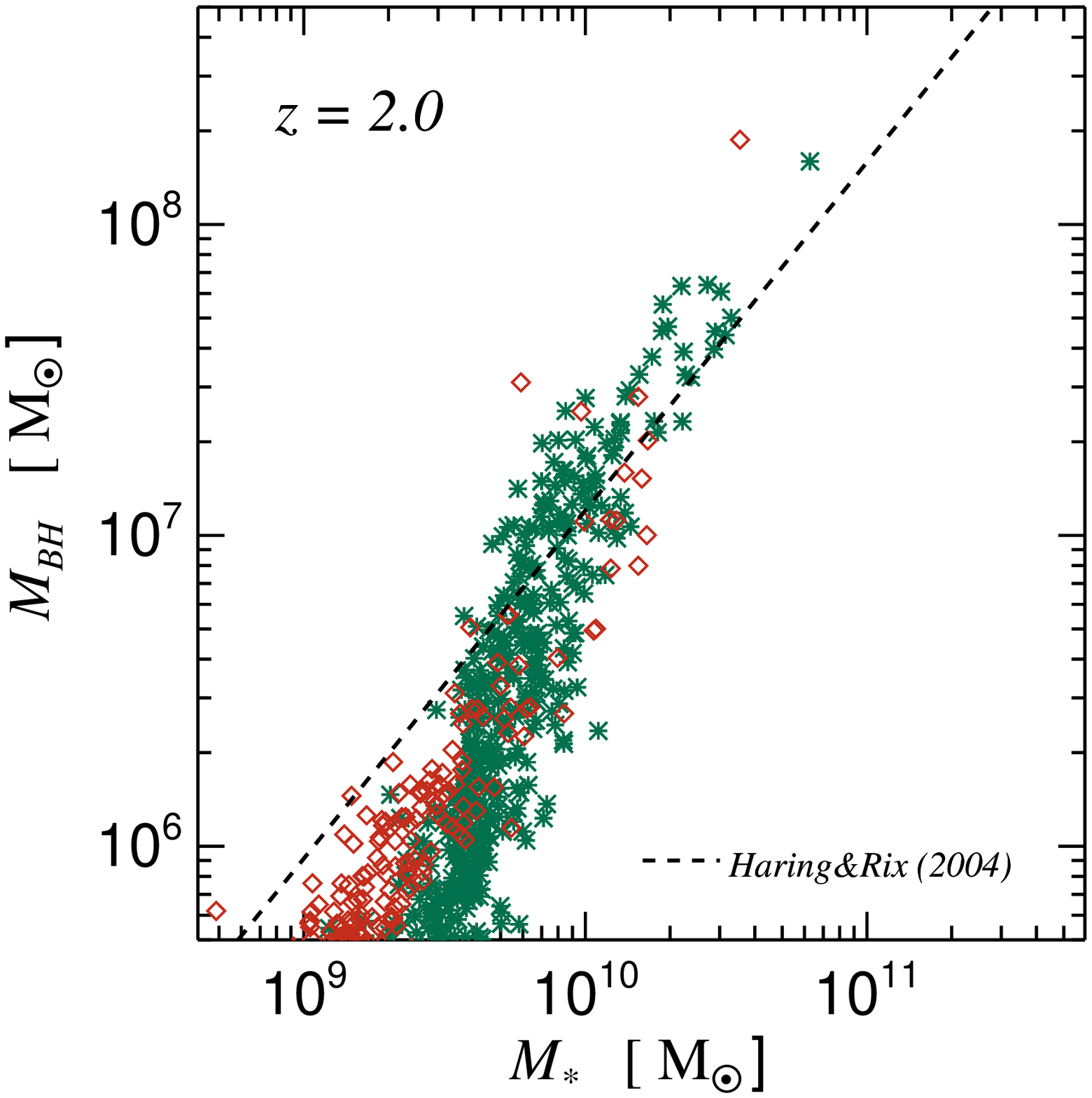,width=8.5truecm,height=8.truecm}
}
\hbox{
\vspace{-0.5truecm}
\psfig{file=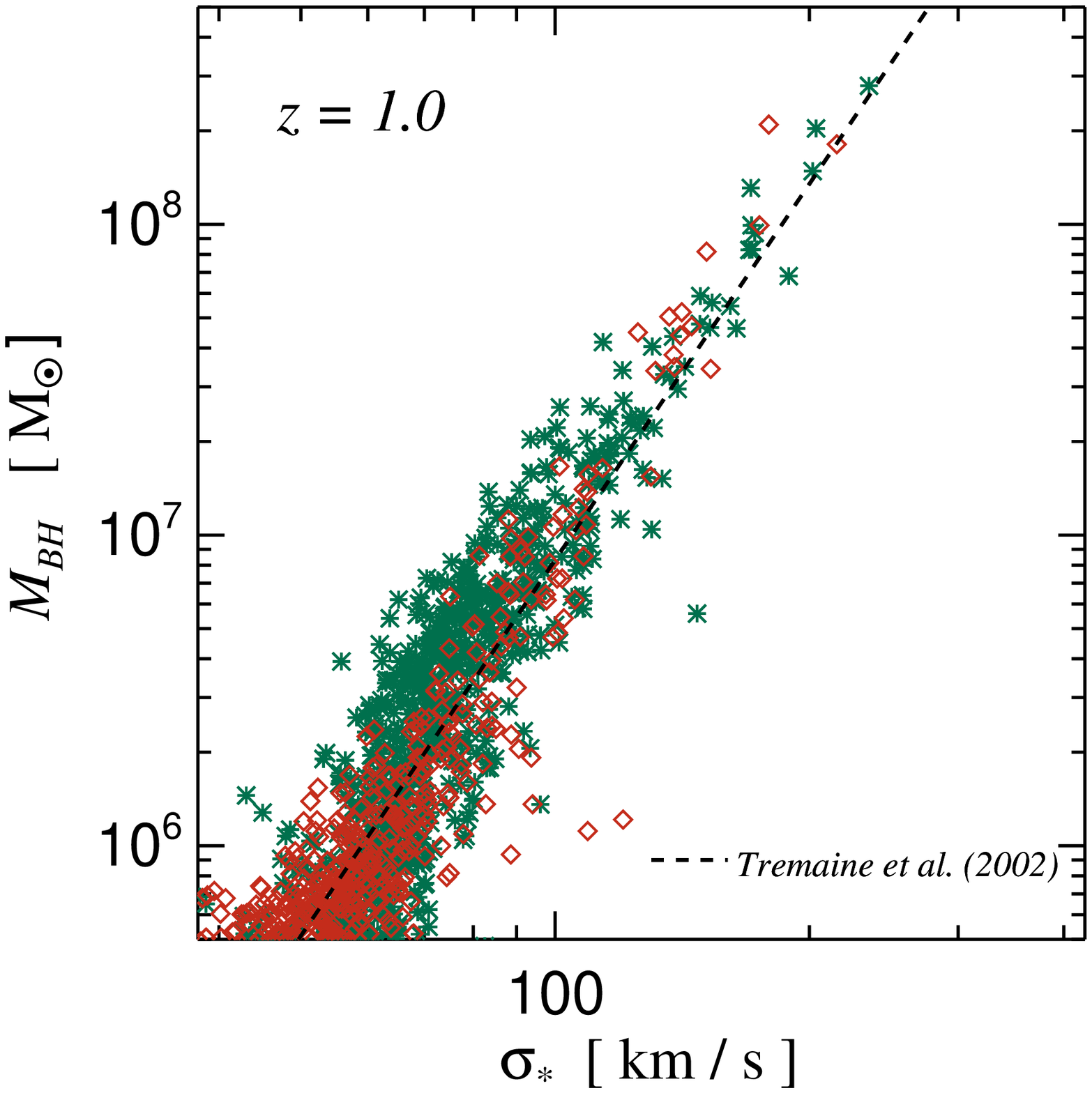,width=8.5truecm,height=8.0truecm}
\psfig{file=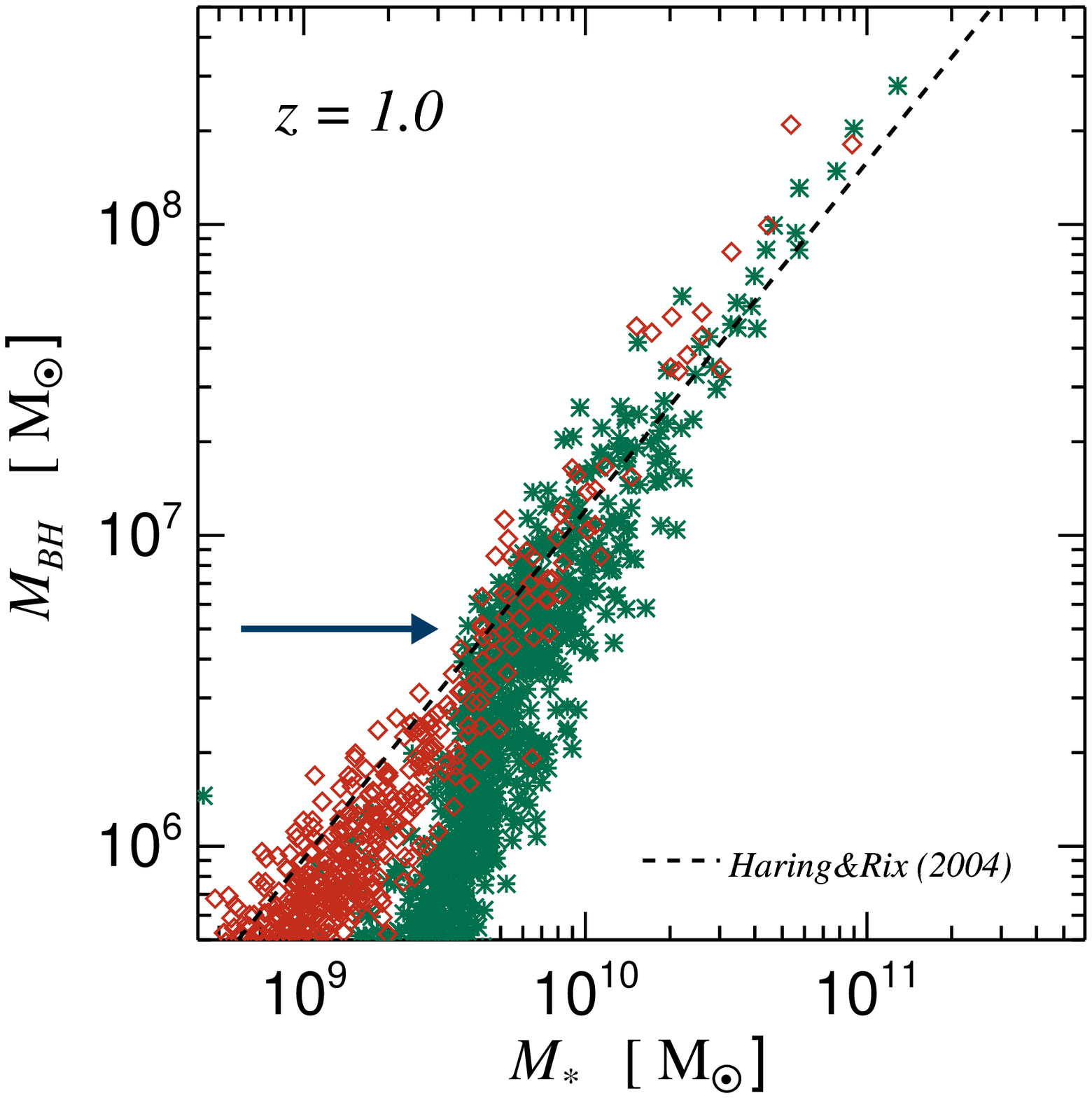,width=8.5truecm,height=8.0truecm}
}
}}
\caption{BH mass--stellar velocity dispersion relation (left-hand
  panels) and BH mass--stellar mass relation (right-hand panels), at
  redshifts $z=1,\,2$ and $3$. Green star symbols denote the run
  without galactic winds, while red diamonds are for a run where
  galactic winds have been included as well. Both simulations have
  been performed at the resolution of the R2 cosmological box. The
  dashed lines give the locally observed relationships between the
  considered quantities, as determined by \cite{Tremaine2002} and
  \cite{Haring2004}.}
\label{mbh_relations}
\end{figure*}

\begin{figure*}
\centerline{
\hbox{
\psfig{file=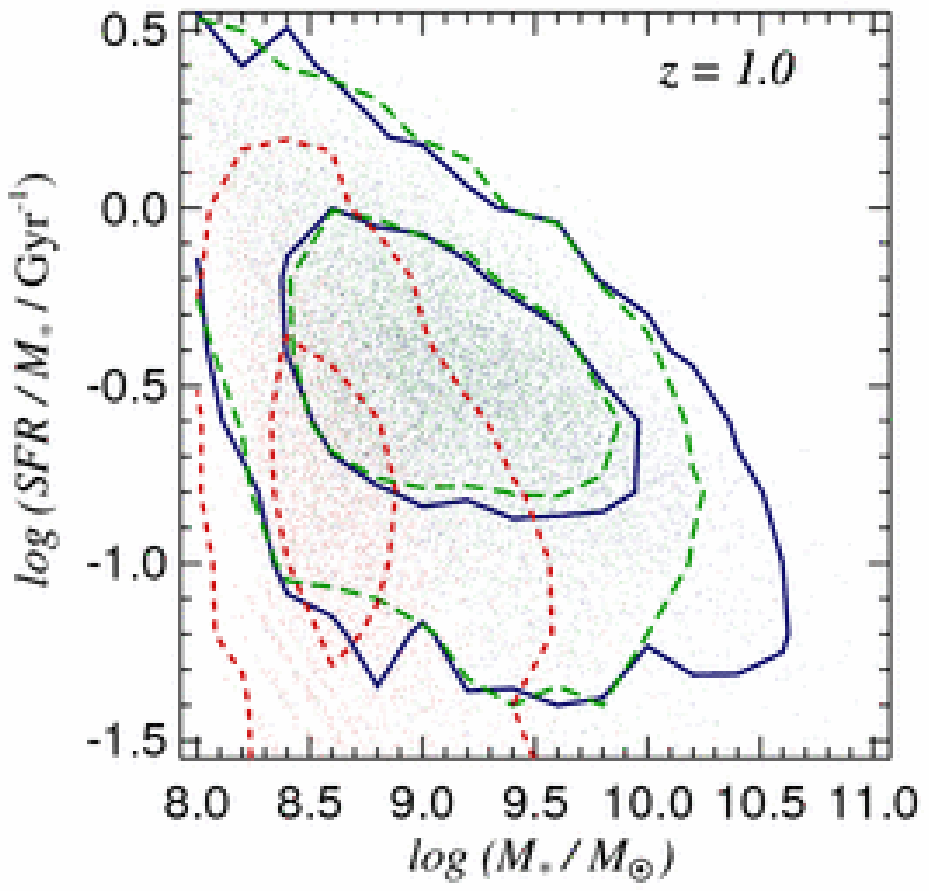,width=8.5truecm,height=8.truecm}
\psfig{file=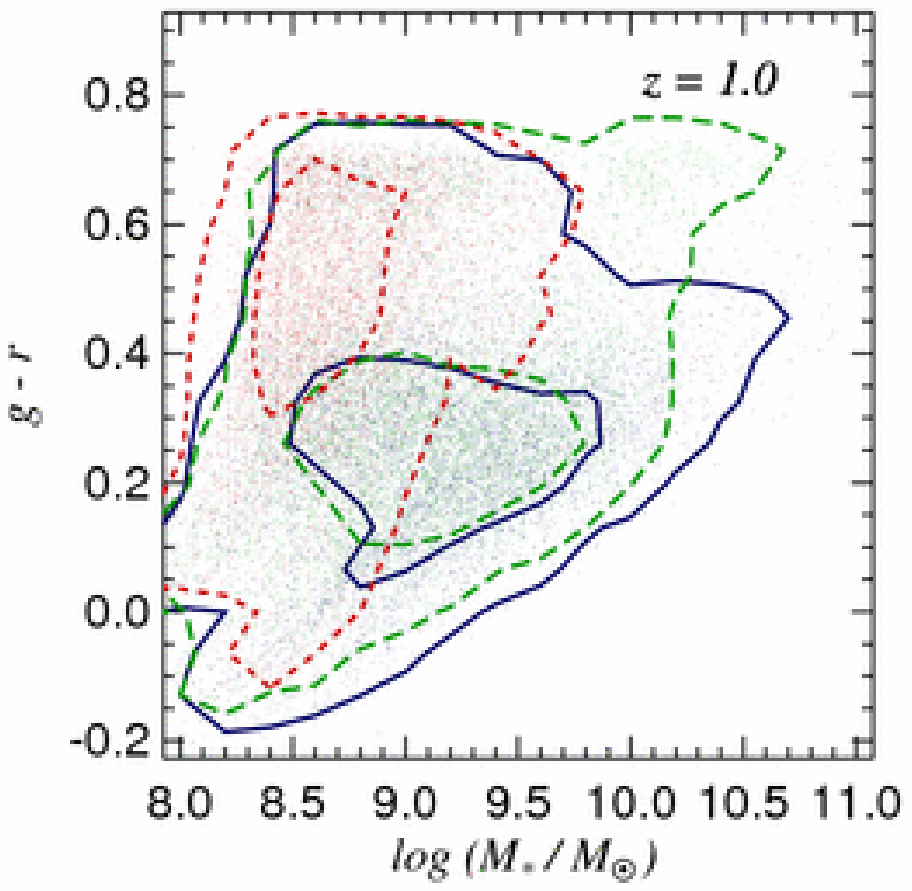,width=8.5truecm,height=8.truecm}
}}
\caption{Specific SFR (left-hand panel) and $g-r$ colour (right-hand panel)
  for simulated galaxies as a function of stellar mass at $z=1$. Blue dots and
  continuous contours are for the run with cooling and star formation only.
  Green dots and long-dashed contours indicate the case where our BH model was
  included, while red dots and short-dashed contours are for the simulation
  which in addition contains galactic winds.}
\label{gal_prop}
\end{figure*}

\subsection{Galaxy properties and evolution}

We now study the question how the properties of galaxies are affected
by BH feedback, and how the BH masses relate to their hosts.  For this
purpose, we identify the galaxies present in our cosmological box at
different redshifts using a special group finder, and compute a number
of properties for them, e.g.~their stellar mass $M_{*}$, stellar
velocity dispersion $\sigma_*$, their total SFR, and their colours in
the rest-frame SDSS bands. This also allows us to look for
correlations with their central BH masses. Our scheme for identifying
simulated galaxies as isolated groups of stars is based on a
simplified variant of the {\small SUBFIND} algorithm
\citep{SpringelSubfind2001}. We first compute an adaptively smoothed
baryonic density field for all stars and gas particles, allowing us to
robustly identify centres of individual galaxies as isolated density
peaks. Using {\small SUBFIND}'s approach for growing these centres by
attaching nearby particles of ever lower density, we then obtain a list of
cleanly separated galaxies. By restricting the set of gas particles processed
in this way to those that have at least a density of 0.01 times the threshold
density for star formation, we avoid that spatially separated galaxies are
linked together, while all the stars present in the simulation are assigned to
the different galaxies. A gravitational unbinding procedure is then not
necessary in this procedure. Note that we are not interested in the gaseous
components of galaxies here. We only include the gas particles in the galaxy
finding procedure because the dense, cold gas that most galaxies contain makes
the method more robust.

In Figure~\ref{mbh_relations}, we show the relation between BH mass
and stellar velocity dispersion $\sigma_*$ (left-hand panels), and
between BH mass and $M_{*}$ (right-hand panels), at three different
redshifts, $z=1,\,2$ and $3$. Here we evaluate $M_{*}$ and $\sigma_*$
within the effective radius, $R_{e}$, chosen as the half-mass radius.
Green star symbols are for the run
without galactic winds, while red diamonds are for the simulation that
also includes galactic winds. The results shown are from our
intermediate resolution box, but we have verified good convergence in
these quantities by comparing with the R3 box.  With the dashed lines
we overplot the locally observed $M_{\rm BH}-\sigma_*$ and $M_{\rm
BH}-M_*$ relations, as determined by \cite{Tremaine2002} and
\cite{Haring2004}, respectively. A number of interesting features are
noteworthy in this figure:

\begin{enumerate}
\item Both relations show some evolution with redshift that however seems
most prominent at the low BH mass end and for the run with galactic
winds. The most massive objects found at every epoch stay very close
to the local relation over the redshift interval considered.

\item At high redshifts, galactic winds give rise to less massive
galaxies with somewhat lower $\sigma_*$, and for a given $M_{*}$ or $\sigma_*$,
BHs are less massive when the winds are ``switched-on''. However, by
redshift $z=1$ and for $M_{\rm BH} > 5\times10^6 {\rm M}_\odot$, the central
BH masses are comparable for a given stellar mass of a galaxy,
regardless of the presence of winds. This suggests that the feedback
by galactic winds is delaying BH growth by expelling significant
amounts of gas at high redshift.  However, this gas becomes available
for BH accretion at latter epochs, when it has been reincorporated from
the IGM back into the galaxies.

\item For BH masses smaller than $5\times10^6\, {\rm M}_\odot$
(as indicated with the arrow in Figure~\ref{mbh_relations}) there is a
tilt in both relations, especially noticeable for the $M_{\rm BH}-M_*$
relation. This tilt is significantly reduced in the run with galactic
winds. This can be understood by considering that the AGN feedback in
these small galaxies is very modest, and does not modify the
properties of the hosts significantly.  On the other hand, galactic
winds are very efficient for these low mass objects, reducing their
stellar mass and bringing them into better agreement with the expected
relation. However, we caution that this result of our modelling may be
sensitive to our prescription for BH seeding and the early initial growth,
which could be affected by numerical resolution effects. If small mass
galaxies could grow somewhat bigger BHs in their centre, they would
also be more affected by BH feedback so that it may not be necessary
to invoke galactic winds to match the low mass end of the $M_{\rm
BH}-\sigma$ relation. In any case, it is interesting that galactic
winds shift these smaller objects in the desired direction while at
the same time more massive galaxies are unaffected and still lie on
the local relation.

\item Finally, given that the simulated $M_{\rm BH}-M_*$ and $M_{\rm
  BH}-\sigma_*$ relations at $z=1$ match the locally observed relations well it
  would be interesting to explore how the model will evolve further
  until $z=0$. It seems quite likely, especially for more massive
  galaxies, that not much evolution will occur for $z<1$, given that
  these galaxies have a reduced star formation and depleted gas
  content due to the AGN feedback. However, considering the complex interplay
  between BHs and galactic winds, a reliable answer of this question
  will require explicit numerical simulations of a bigger cosmological
  box at comparable or better resolution, which we will tackle in
  future work.
\end{enumerate}

In \citet{DiMatteo2007}, a measurement of the $M_{\rm BH}-M_*$ and
$M_{\rm BH}-\sigma_*$ relations is presented as well, for an
overlapping redshift range. Their results extend to somewhat higher BH
mass, due to their larger box-size and their higher normalization
$\sigma_8$ of the power spectrum. In general, we find a reassuring
agreement with the findings of \citet{DiMatteo2007}, indicating that
the $M_{\rm BH}-M_*$ and $M_{\rm BH}-\sigma_*$ relations are not
significantly affected by the bubble feedback or the underlying
cosmology, as we will discuss latter on in more detail. Only for the
$M_{\rm BH}-\sigma_*$ relations at $z=3$, our measurements lie
somewhat below those of \citet{DiMatteo2007}, a trend that we
attribute to the fact that we required dark matter halos to be five
times more massive than \citet{DiMatteo2007} before we endowed them
with a seed black hole, which may have delayed their early growth.

\begin{figure*}
\centerline{
\hbox{
\psfig{file=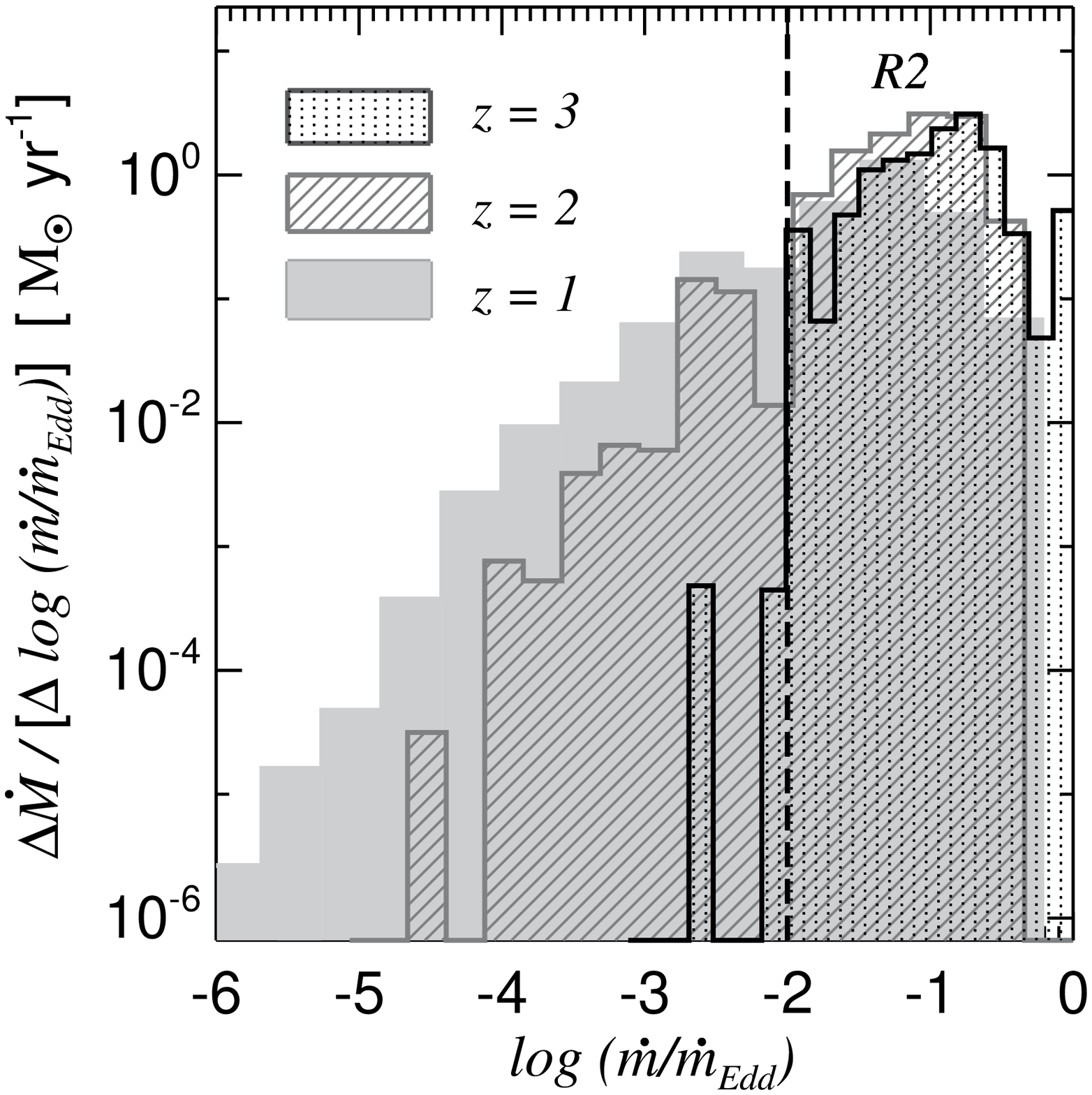,width=8.5truecm,height=8.truecm}
\psfig{file=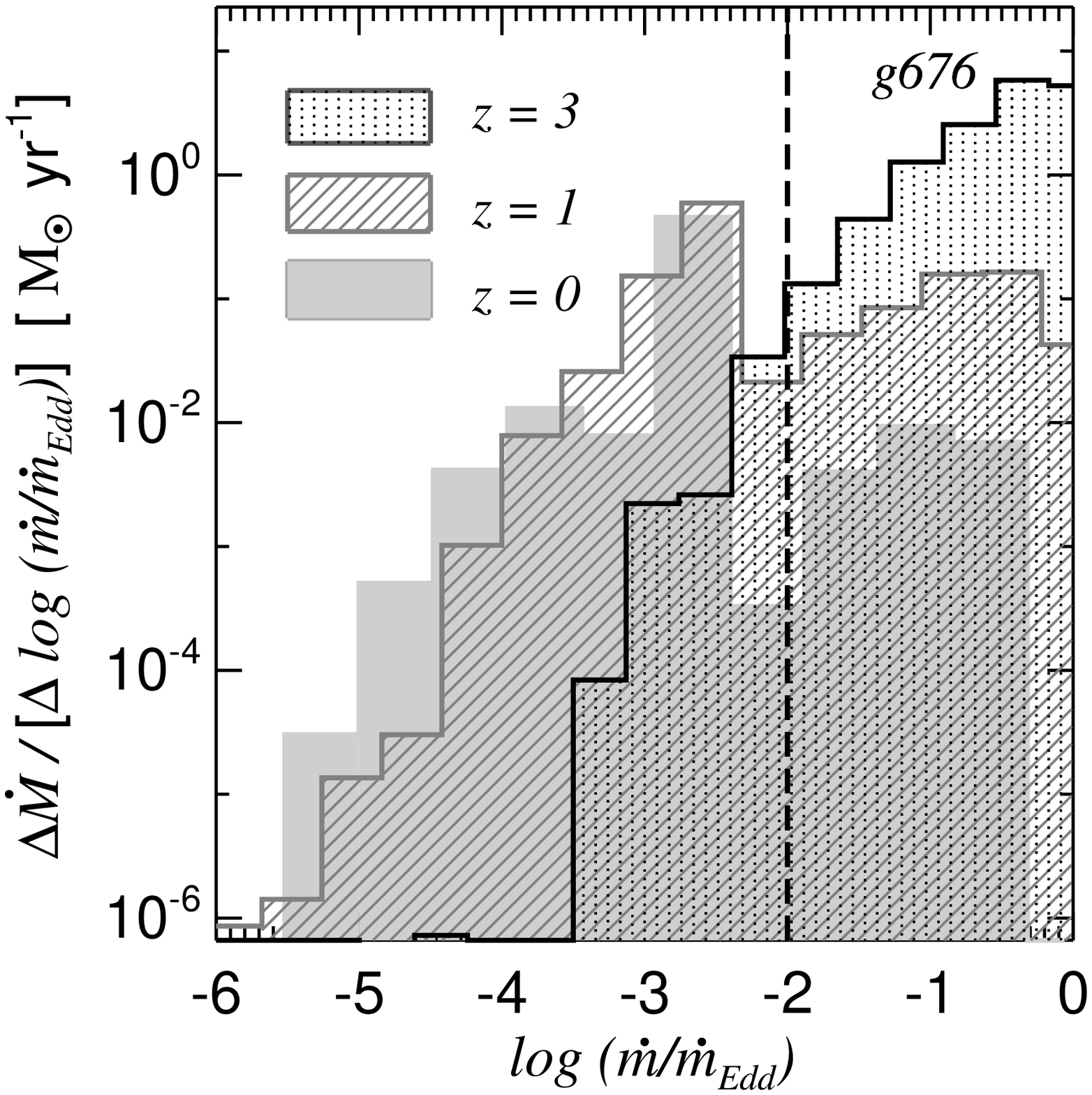,width=8.5truecm,height=8.truecm}
}}
\caption{Distribution of the BHAR, expressed in ${\rm M}_\odot/{\rm
  yr}$, as a function of accretion rate in Eddington units, for three
  different redshifts, as indicated on the panels. The left-hand panel
  shows our measurements for the cosmological box simulation at the
  intermediate resolution (R2), while the right-hand panel gives
  analogous results for the g676 galaxy cluster simulation. The
  vertical dashed line denotes the transition adopted in our model
  between BHs accreting in the ``quasar'' mode and those accreting in
  the radiatively inefficient ``radio'' mode.}
\label{histbhacc}
\end{figure*}

In Figure~\ref{gal_prop}, we show the specific SFRs and the $g-r$
colours of our simulated galaxies at $z=1$ as a function of their
stellar mass. Blue dots and blue continuous contours show the results
when only cooling and star formation is considered. Green dots and
long-dashed contours give the results for the run with AGN feedback,
while red dots and short-dashed contours represent the case with
additional galactic winds.  While AGN feedback does not affect the
bulk of the population significantly, it is evident that the stellar
masses and the SFRs of most massive galaxies are reduced. When
feedback by galactic winds is included as well this trend is even more
pronounced, and extends to the lower mass systems. Note also that the
galaxy colours change in our BH model -- a number of galaxies become
much redder, forming a red sequence, as it is shown in the right-hand
plot of Figure~\ref{gal_prop}. However, if the galactic wind model is
included, the clearly defined red sequence disappears. Instead, all
galaxies are in general less massive and redder. The fact that the red
sequence becomes largely unpopulated in the run with galactic winds is
not necessarily problematic, however. A larger cosmological box may
again fill this region of the diagram with a population of more
massive galaxies.

Finally, in order to assess whether the colours of our galaxies in the
AGN feedback model are realistic, we have computed the mean $u-r$
colour of the red sequence at $z=1$, obtaining $u-r \sim 2.2$. We can
compare this result to the work of \cite{Bell2004} on the COMBO-17
survey. Using the conversion between $U-V$ and $u-r$ colours suggested
by these authors, we find that the colour of our simulated galaxies
along the red sequence is in good agreement with the early-type
galaxies from the COMBO-17 survey.

\subsection{Dependence on cosmological parameters}

In order to evaluate how sensitively our BH model depends on the underlying
cosmological model, we have performed two additional simulations at
intermediate resolution (R2) with exactly identical BH parameters. In one case
we have adopted cosmological parameters consistent with the WMAP 1st-year data
analysis, while for the other run we have selected the updated parameters of
WMAP's 3rd year data release. The main difference between the corresponding
cosmological models lies in a reduction of the normalization parameter
$\sigma_8$ ($0.9 \rightarrow 0.75$), a lowering of the matter density
$\Omega_m$ ($0.3 \rightarrow 0.26$) and in the introduction of a tilt in the
primordial power spectrum slope $n_s$ ($1.0 \rightarrow 0.938$).

Our results show that BHs form earlier in the WMAP-1 cosmology, as
expected from the fact that host halos above a given mass threshold
appear earlier in this model due to the higher normalization, and are
therefore seeded with a BH at higher $z$. Also, the number density of
BHs is higher and there are more high mass BHs at lower redshift with
respect to the WMAP-3 run.  Consequently, the comoving BH mass density
is somewhat higher in the WMAP-1 cosmology. At $z=1$, it is
$3.9\times10^5 {\rm M}_\odot {\rm Mpc}^{-3}$, while for WMAP-3 we
obtain $2.2\times10^5 {\rm M}_\odot {\rm Mpc}^{-3}$ at the same
numerical resolution. The BHAR density is higher in the WMAP-1 model
as well, with its peak shifted to a slightly earlier redshift of $\sim
3.5$.  However, for lower redshifts this is compensated by a steeper
decline, such that at $z=1$ the BHAR densities have a very similar
value in both cosmologies.  These trends can simply be understood as a
result of ``delayed'' structure formation in the WMAP-3
cosmology. Increasing $\sigma_8$ will let the BHs form earlier, evolve
faster, and reach their peak activity at higher redshift.
Interestingly, within the scatter, the ${M}_{\rm BH}-\sigma_*$ and ${
M}_{\rm BH}-M_*$ relations are however not significantly affected by
the change in the cosmological parameters.

\section{Relevance of radio versus quasar feedback mode}\label{Radio}

In this section, we explore the relative importance of our two
different modes of black hole growth and quasar feedback for the build-up
of the cosmological black hole mass density. For this purpose, we
show in Figure~\ref{histbhacc} the distribution function of the net
black hole mass accretion rate in ${\rm M}_\odot/{\rm yr}$, as a
function of the Eddington-normalized accretion rate. In the left-hand
panel, we give results for our cosmological box R2 at redshifts $z=1$,
$2$ and $3$, while on the right-hand side we show results for the g676
galaxy cluster simulation at different redshifts.

It can be seen that at $z=3$ the bulk of the BH growth occurs in the
``quasar'' mode, where the accretion rate is more than 0.01 times the
Eddington rate. In fact, the ``radio'' mode contributes less than $2\%$
to the total BHAR at this epoch, and for the g676 cluster simulation,
the contribution from growth in this radiatively inefficient mode is
similarly small ($\sim 1\%$).  At lower redshifts, BHs accreting at
low rates make up for an increasingly higher fraction of the total
BHAR density (which however declines with time). For the R2 simulation
at $z=2$ they contribute $\sim 13\%$, and at $z=1$ they are at $\sim
37\%$. Instead, for the g676 galaxy cluster, these numbers are somewhat
higher, with $65\%$ at $z=1$ and $96\%$ at $z=0$ coming from BHs in a
low accretion state. Thus, while in the cosmological box BHs in the
quasar regime are the main channel of BH growth at all epochs
considered, for the galaxy cluster run there appears to be a
transition epoch at $z \sim 1$, below which BHs in the ``radio'' mode
are driving the BH growth at low redshifts. This difference between
the BHAR distributions for the R2 run and the g676 cluster simulation
can be explained by the presence of a dominating, very massive BH in
the centre of the main progenitor of the cluster simulation, while the
cosmological box contains a much fairer sample of the black hole mass
function.

However, integrating the BHAR over cosmic time for all BHs in the
simulated volumes, we find that the bulk of BH mass is always grown
during phases of quasar activity, i.e.~in the high accretion rate
regime. In fact, BH accreting in the ``radio'' mode contribute less than
$\sim 5\%$ to the integrated black hole mass density in the
cosmological box, while for the g676 cluster simulation this number is
$\sim 20\%$. Performing the same calculations for our g1 galaxy
cluster simulation we obtain a very similar result ($\sim 15\%$),
further strengthening this point. Thus, we conclude that most of the
BH mass is assembled during periods of rapid growth in the radiatively
efficient ``quasar'' mode. This implies that our models are consistent
with the Soltan argument \citep{Soltan1982, Yu2002, Marconi2004,
Merloni2004, HopkinsNarayan2006}, and shows that merger-driven quasar
activity is still dominating the black hole assembly in our unified
model for AGN feedback.

Furthermore, by repeating our cosmological box simulations without
``radio'' mode feedback we have explicitly checked that the inclusion of
bubble feedback has a negligible influence on the $M_{\rm BH}-M_*$ and
$M_{\rm BH}-\sigma_*$ relations. Only at $z<2$, the ``radio'' mode makes
BHs slightly less massive due to the somewhat more efficient heating
by the bubbles.  Thus, quasar activity triggered by the merging of
host galaxies is the primary process responsible for establishing the
scaling relationships between the central supermassive BHs and their
host galaxies, and this process remains intact and is essentially
unaffected if an additional ``radio'' mode feedback is invoked in the
low accretion state of BHs.
 
On the other hand, our simulations also show that the ``radio'' mode
feedback in the form of AGN-driven bubbles is essential in massive
groups and clusters. In order to demonstrate this point directly, we
have rerun our g676 galaxy cluster simulation by switching off the
radio channel of feedback mode, only allowing the ordinary quasar
activity to proceed until $z=0$. This results in simulated ICM
properties that are in disagreement with observations. In particular,
the gas temperature profile keeps rising towards the very centre. Thus
the intermittent nature of our bubble feedback appears necessary in
this respect, as well of course for explaining the observed phenomena
of radio galaxies and X-ray cavities in clusters of galaxies.

\section{Discussion and Conclusions} \label{Discussion} 

In this study, we have proposed a new model for following BH growth and
feedback in cosmological simulations of structure formation. Within
our prescription, BH seeds are introduced at early cosmic times in the
centres of all halos once their mass exceeds a certain threshold value
for the first time.  The BH seeding is accomplished by
frequently running a FoF algorithm on the fly in our simulation code,
which identifies newly formed halos and their properties. The BHs
themselves are represented as collisionless sink particles that can
grow their mass via two channels, by (1) gas accretion with an
accretion rate estimated by a simple Bondi formulae, and (2) via
mergers with other BHs that are close enough. Motivated by the
observed phenomenology of radio galaxies and quasars, and by analogy
with X--ray binaries, we have assumed that the AGN feedback that
accompanies gas accretion can be decomposed into two physically
distinct modes: BHs accreting at high accretion rates in terms of
their Eddington limit are operating in a radiatively efficient
``quasar'' regime, while BHs accreting at much lower rates live in a
radiatively inefficient ``radio'' mode, which however produces
significant mechanical luminosity that gives rise to jet-inflated
bubbles.

In our initial tests of the model, we have applied it to simulations
of isolated galaxy clusters consisting of a static dark matter
potential and gas in hydrostatic equilibrium. Considering a range of
cluster masses from $10^{13} h^{-1}{\rm M}_\odot$ to $10^{15}
h^{-1}{\rm M}_\odot$, we have found that the model leads to a
self-regulated AGN feedback that produces realistic ICM properties and
prevents the overcooling problem in the central cluster regions. At
the same time, it yields reasonable BH masses and accretion rates.

We have then applied our model to fully self-consistent cosmological
simulations of galaxy cluster formation, where the seeding of BHs and
their subsequent growth is followed self-consistently. Together with
our work in \citet{DiMatteo2007}, these simulations are the first
cosmological hydrodynamical models that simultaneously follow the
growth of structure in the dark matter as well as the baryonic physics
of star formation and AGN feedback.  These simulations provide
information on how the BHs are spatially distributed in the simulated
volume, how they grow with cosmic time as a function of their
environment, and how they influence the surrounding medium and their
host galaxies.

Our analysis of these calculations has focused on the history of the BH in
the central cD galaxy. We have shown that while essentially all the mass in
supermassive BHs originates in gas accretion, a large fraction of the
mass of the BH in the cD grows from mergers with other BHs, as opposed to
being all accreted by a single massive progenitor. This suggests that mergers
of supermassive BHs are an important factor for building up very
massive BHs.  In line with our findings for the isolated cluster
simulations we found that feedback from BHs is preventing the formation of
excessive cooling flows out of hot cluster atmospheres, and is changing the
central ICM properties substantially in the process.  These changes are all
acting to bring the simulated temperature and entropy profiles into much
better agreement with observational findings of cool core clusters
\citep[e.g.][]{DeGrandi2002, Pratt2003, Ponman2003, Vikhlinin2005, Bauer2005,
  Dunn2006, Sanderson2006, Pratt2007}. Moreover, the stellar properties of the
central galaxy are affected as well, with the forming cD galaxies being less
massive, and having older and redder stellar populations which are consistent
with the results found in recent studies of a large sample of BCGs in the SDSS
survey \citep{Linden2006}.

We have also studied our unified AGN feedback model in simulations of
homogeneously sampled cosmological volumes, thereby investigating how
successful it is on the scale of galaxies, and with respect to the
mean properties of the cosmic population of BHs. In our simulation box
of size $25\, h^{-1}{\rm Mpc}$ on a side, we were able to resolve
galaxies with stellar masses greater than $\sim 10^8 {\rm
M}_\odot$. We have found that our model produces a comoving BH mass
density equal to $2.8\times10^5 {\rm M}_\odot {\rm Mpc}^{-3}$, which
is in very good agreement with observational estimates
\citep{Fabian1999, Merritt2001,Yu2002,Cowie2003}.  Moreover, the AGN
feedback influences the properties and the formation of the host
galaxies and vice versa, and this mutual coupling establishes $M_{\rm
BH}-\sigma_*$ and $M_{\rm BH}-M_*$ relationships already at early
epochs.  Again, these relations are in broad agreement with
observations, at least for the more massive and better resolved
systems.  Interestingly, a model that besides BH feedback includes
feedback from supernova-driven galactic winds produces a better match
to the expected relationships at the low mass end. At the same time,
this model also leads to a much more widespread enrichment of the IGM,
which appears required by the data on quasar absorption line
systems. Additionally, AGN heating reduces the stellar mass of the
most luminous galaxies identified in the simulated volume, and it
affects their SFR and colours as well. The galaxies in simulations
with AGN feedback show a clearly defined red sequence already at
$z=1$.

Finally, we have explored the relative importance of ``radio'' versus
``quasar'' mode.  We have found that the bulk of the BH growth occurs at
high accretion rates, corresponding to radiatively efficient AGN
activity.  While the relative importance of the ``radio'' mode grows
towards late times, and becomes large in clusters of galaxies at $z<
1$, this mode of accretion is unimportant for the total black hole
mass density today.  This also implies that our model is consistent
with the Soltan argument. We also found that the feedback of the
quasar regime is the key factor for establishing the $M_{\rm
BH}-\sigma_*$ and $M_{\rm BH}-M_*$ relationships, a process that
appears unaffected by the bubble feedback.  On the other hand, the
mechanical luminosity of AGN-driven bubbles in groups and cluster of
galaxies is necessary in order to successfully reproduce the observed
ICM properties.

These results represent highly encouraging successes for
hydrodynamical models of hierarchical galaxy formation in $\Lambda$CDM
cosmologies, and provide support for the theoretical conjecture that
galaxy formation and supermassive BH growth are intimately linked.  In
spite of these successes we need to emphasize that our BH model
clearly represents a drastically simplified picture of BH physics. In
part, the simplifications are driven by numerical limitations, because
even state-of-the-art simulations of galaxy formation in
cosmologically representative regions of the Universe cannot directly
resolve the gravitational sphere of influence of individual BHs. We
are therefore forced to adopt a subresolution approach for the
representation of BH accretion, and also for the modelling of bubble
feedback where the initial stages of bubble inflation by a jet are not
resolved.  However, this subresolution approach can still capture the
essential parts of the physics relevant for the coupling of the BHs to
their larger-scale environment, and this is what allows us to study
their cosmological significance for galaxy formation.

While we have tried in this study to shed some light onto the entangled
histories of galaxies and their central BHs, a number of important questions
remain open for future work, and our model appears suitable to help answering
them. They include: (i) The early growth of BHs - can the observed
`downsizing' of BH growth be reproduced by our model? (ii) Do the galaxy
cluster scaling relations change as a result of the AGN feedback? (iii) Can a
galaxy luminosity function with a bright end consistent with observational
constraints be reproduced in cosmological simulations, and what is the
respective AGN luminosity function? (iv) Do the quasar clustering properties
come out right?

Indeed, AGN appear to be a key ingredient of structure formation, and future
numerical simulations should be a very helpful theoretical tool to unveil the
intriguing and complex picture of their interactions and growth.

\section*{Acknowledgements}
We are grateful to Simon White, Eugene Churazov, Andrea Merloni and
Martin Haehnelt for very stimulating discussions and useful comments
on the manuscript. We thank Klaus Dolag for proving us with galaxy
cluster initial conditions used in this study. DS acknowledges the PhD
fellowship of the International Max Planck Research School in
Astrophysics, and the support received from a Marie Curie Host
Fellowship for Early Stage Research Training.

\bibliographystyle{mnras}

\bibliography{paper}

\end{document}